# Second Order Bi-Scalar-Tensor Field Equations in a Space of Four-Dimensions

by

Gregory W. Horndeski
Adjunct Associate Professor of Applied Mathematics
University of Waterloo
200 University Avenue
Waterloo, Ontario
Canada
N2L 3G1

email:
horndeskimath@gmail.com
ghorndeski@uwaterloo.ca

August 1, 2025

## ABSTRACT

Lagrange scalar densities which are concomitants of two scalar fields, a pseudo-Riemannian metric tensor, and their derivatives of arbitrary differential order are investigated in a space of four-dimensions. I construct the most general second-order Euler-Lagrange tensor densities derivable from such a Lagrangian. It is demonstrated that all such second-order Euler-Lagrange tensor densities can be derived from a set of eleven Lagrangians which are at most of second-order. Of these eleven Lagrangians six do not involve the second derivatives of the metric tensor, and are algebraically at most of first degree in the second derivatives of the scalar fields. Each of the eleven Lagrangians will have a scalar coefficient which is a concomitant of five variables: the two scalar-fields, and the three inner products of the gradients of the two scalar fields. Of these eleven coefficient functions only one is arbitrary, while the other ten must satisfy linear partial differential equations. Surprisingly these partial differential equations are related to the wave equation in 2+1-dimensional Minkowski space. The paper concludes with a few observations on the form that a Lagrangian which yields the most general second-order, multi-scalar-tensor field equations in a space of arbitrary dimension can assume.

## Section 1: Introduction and Statement of the Theorem

In Horndeski [1] I constructed in a space of four-dimensions all of the second-order scalar-tensor field equations which could be derived from a Lagrange scalar density which was a concomitant of a pseudo-Riemannian metric tensor, scalar field and their derivatives of arbitrary order. Applications of these equations are discussed in an essay by Horndeski and Silvestri [2]. Fairly recently Ohashi, *et al.* [3] attempted to extend my result for one scalar field to include a second scalar field. Why one might be interested in adding a second field is briefly discussed in Ohashi, *et al.* [3]. To their reasons I would like to add that it might be possible to use two scalar fields to represent dark energy and dark matter in the context of a metric theory of gravity. Another possible reason for investigating bi-scalar-tensor theories is that since the Higgs field is a complex valued scalar field its real and imaginary parts would provide us with two scalar fields. Thus a second-order, bi-scalar-tensor theory would provide us with equations for coupling the Higgs field with the gravitational field.

By generalizing the techniques I developed in my Ph.D thesis, Horndeski [4], Ohashi, *et al.* [3], constructed the form that the second-order, bi-scalar-tensor field equations must have, and made a great deal of progress toward constructing a Lagrangian that could yield those field equations**.** My goal in this paper is to complete the construction of a Lagrangian that yields the most general second-order bi-scalar-tensor field equations in a space of four-dimensions. To do that I shall often draw upon the work of Ohashi, *et al.,* [3] and answer some of the questions they raise in their paper concerning the construction of Lagrangians.

Before stating the main result of this paper let me introduce some terminology.

Our considerations will be of a purely local nature. If x is an arbitrary chart in a four-dimensional manifold M with domain U, then I let $g_{ij}$ denote the x components of a pseudo-Riemannian metric tensor of fixed signature. Small Latin indices will run from 1 to 4 and obey the Einstein summation convention. All fields and tensors will be assumed to be of class $C^\infty$, unless stated to the contrary. In order to allow the possibility of tensorial concomitants involving the Levi-Civita symbol, I shall require permissible coordinate transformations to have positive Jacobian.

The components $\Gamma^h_{ij}$ of the Christoffel connection are defined by

$$\Gamma^h_{ij} := \tfrac{1}{2} g^{hm}(g_{mj,i} + g_{im,j} - g_{ij,m})$$

where "$_{,j}$" denotes a partial derivative with respect to $x^j$, $g^{hm}$ is the matrix inverse of $g_{hm}$. If $Y^h$ denotes the local components of a contravariant vector field then I define the components $R_h{}^i{}_{jk}$ of the Riemann-Christoffel curvature tensor by $Y^i{}_{|jk} - Y^i{}_{|kj} = Y^h R_h{}^i{}_{jk}$, where a vertical bar preceding Latin indices denotes covariant differentiation. However, for the two scalar fields $\varphi$ and $\xi$ used in this paper I shall drop the vertical bar for covariant differentiation. So, *e.g.,* $\varphi_{abc\ldots} := \varphi_{|abc\ldots}$, with $\Box\varphi := \varphi^a{}_a$. The last geometric quantities I need to define are the Ricci tensor, scalar curvature and Einstein tensor; *viz.,*

$$R_{hj} := R_h{}^i{}_{ji}\,;\ R := g^{hj} R_{hj} \text{ and } G_{hj} := R_{hj} - \tfrac{1}{2} g_{hj} R\,.$$

The gradient of the two scalar fields $\varphi$ and $\xi$ generate three scalar fields:

$$X := g^{ab}\varphi_{,a}\varphi_{,b}\,;\ Y := g^{ab}\xi_{,a}\xi_{,b} \text{ and } Z := g^{ab}\varphi_{,a}\xi_{,b}\,. \tag{1.1}$$

During the course of our investigations we shall encounter a plethora of scalar functions of the form $F=F(\varphi,\xi,X,Y,Z)$. Such a function will be said to be a BST (:=bi-scalar-tensor) function if

$$F_{XY} = \tfrac{1}{4} F_{ZZ} \tag{1.2a}$$

where the subscripts $_{X,\,Y,\,Z}$ denote partial derivatives with respect to X, Y and Z. An (ordered) pair of functions $(F_1,F_2)$ will be referred to as a BST conjugate pair (=:BSTCP) of functions if

$$F_{1Z} = 2F_{2X} \text{ and } F_{2Z} = 2F_{1Y}\,. \tag{1.2b}$$

If $(F_1,F_2)$ is a BSTCP then (1.2b) implies that $F_1$ and $F_2$ satisfy (1.2a), and hence are actually BST

functions.

If F is a BST function then the pairs $(2F_X,F_Z)$ and $(F_Z,2F_Y)$ are each BSTCPs. We shall call F the generator of each of these pairs. It is easy to see that if one has a BSTCP, then (at least formally) there exists generating functions for that pair–in fact many, since generating functions for BSTCPs are not unique.

There is an interesting relationship between the BST functions and solutions to the wave equation in 2+1 dimensional Minkowski Space, $M_3$. To see the connection I define three new variables λ, μ and ν by

$$\lambda := \tfrac{1}{2}(X+Y),\ \mu := \tfrac{1}{2}(X-Y),\ \nu := Z, \text{ and hence } X = \lambda+\mu,\ Y = \lambda-\mu\ ,\ Z=\nu. \tag{1.3}$$

If F is a BST function then the function $\psi=\psi(\varphi,\xi,\lambda,\mu,\nu):=F(\varphi,\xi,X,Y.Z)$ satisfies

$$-\psi_{\lambda\lambda} + \psi_{\mu\mu} + \psi_{\nu\nu} = 0,$$

which is the wave equation in $M_3$ (where I am regarding φ and ξ as fixed parameters in this equation). This shows us how we can obtain BST functions.

Throughout this paper we shall be concerned with Lagrange scalar densities which are concomitants of the components of a pseudo-Riemannian metric tensor, $g_{ij}$, two scalar fields, φ and ξ, along with their derivatives of arbitrary, but finite order. I express this by writing

$$L = L(g_{ij};\ g_{ij,k};\ \ldots;\ \varphi;\ \varphi_{,k};\ \ldots;\ \xi;\ \xi_{,k};\ \ldots)\quad . \tag{1.4}$$

The Euler-Lagrange tensors associated with these Lagrangians are:

$$E^{ij}(L) \equiv \frac{\delta L}{\delta g_{ij}} := \frac{\partial L}{\partial g_{ij}} - \frac{d}{dx^k}\frac{\partial L}{\partial g_{ij,k}} + \ldots \tag{1.5}$$

$$E_{\varphi}(L) \equiv \frac{\delta L}{\delta \varphi} := \frac{\partial L}{\partial \varphi} - \frac{d}{dx^k}\frac{\partial L}{\partial \varphi_{,k}} + \ldots \tag{1.6}$$

and

$$E_{\xi}(L) \equiv \frac{\delta L}{\delta \xi} := \frac{\partial L}{\partial \xi} - \frac{d}{dx^k}\frac{\partial L}{\partial \xi_{,k}} + \ldots \tag{1.7}$$

(Note that these Euler-Lagrange tensor densities are the negatives of those I used in Horndeski [1]

and [4].) For those of you who are not familiar with tensorial concomitants or how one goes about differentiating them with respect to their various arguments I suggest that you read Appendix A in Horndeski [5].

My goal in this paper is to determine in a space of four-dimensions, the form of a Lagrangian L which yields the most general second-order bi-scalar-tensor Euler-Lagrange tensor densities. To that end note that if the Lagrangian L presented in (1.4) is of $p^{th}$ order in $g_{ab}$, of $q^{th}$ order in $\varphi$, and of $r^{th}$ order in $\xi$, then in general $E^{ij}(L)$ will be of $2p^{th}$ order in $g_{ab}$, $(p+q)^{th}$ order in $\varphi$, and $(p+r)^{th}$ order in $\xi$. So it is evident that in general to get the Euler-Lagrange tensor densities of L to be of second-order will require severe restrictions on the form of L. Fortunately the Euler-Lagrange tensor densities associated with L are not independent. For a simple generalization of the work I did in Horndeski [4] (see, Theorem I.5.1 on page 63) shows that in general

$$E^{ij}(L)_{|j} = \tfrac{1}{2}\varphi^{i}E_{\varphi}(L) + \tfrac{1}{2}\xi^{i}E_{\xi}(L) \,. \tag{1.8}$$

(This result is also derived in Ohashi, *et al.*[3].) Consequently we can immediately deduce that when the Euler-Lagrange tensor densities are of second-order the divergence of $E^{ij}(L)$ must also be of second-order. Employing this observation along with numerous other properties of Lagrange scalar densities I shall prove the following theorem in the next section:

**Theorem:** In a space of four-dimensions the most general second-order bi-scalar-tensor field equations can be derived from the Lagrangian

$$\mathcal{L} = \mathcal{L}_0 + \mathcal{L}_1 + \mathcal{L}_2 + \mathcal{L}_3 \tag{1.9}$$

where

$$\mathcal{L}_0 := g^{\frac{1}{2}}S_0 \tag{1.10}$$

$$\mathcal{L}_1 := L_{11} + L_{12} + L_{13} + L_{14} + L_{15} \tag{1.11}$$

with

$$L_{11} := g^{\frac{1}{2}}\{\lambda_{11}\Box\varphi + \lambda_{12}\Box\xi\} \tag{1.12a}$$

$$L_{12} := g^{\frac{1}{2}}\{\lambda_{21}(X\Box\varphi - \varphi^{a}\varphi^{b}\varphi_{ab}) + \lambda_{22}(X\Box\xi - \varphi^{a}\varphi^{b}\xi_{ab})\} \tag{1.12b}$$

$$L_{13} := g^{1/2}\{\lambda_{31}(Y\Box\varphi - \xi^a\xi^b\varphi_{ab}) + \lambda_{32}(Y\Box\xi - \xi^a\xi^b\xi_{ab})\} \tag{1.12c}$$

$$L_{14} := g^{1/2}\{\lambda_{41}(Z\Box\varphi - \varphi^a\xi^b\varphi_{ab}) + \lambda_{42}(Z\Box\xi - \varphi^a\xi^b\xi_{ab})\} \tag{1.12d}$$

$$L_{15} := g^{1/2}\{\lambda_{51}[(XY - Z^2)\Box\varphi - X\xi^a\xi^b\varphi_{ab} - Y\varphi^a\varphi^b\varphi_{ab} + 2Z\varphi^a\xi^b\varphi_{ab}] +$$

$$+ \lambda_{52}[(XY - Z^2)\Box\xi - X\xi^a\xi^b\xi_{ab} - Y\varphi^a\varphi^b\xi_{ab} + 2Z\varphi^a\xi^b\xi_{ab}]\}\ , \tag{1.12e}$$

$$\mathcal{L}_2 := L_{21} + L_{22} + L_{23} + L_{24} \tag{1.13}$$

with

$$L_{21} := g^{1/2}\{S_{21}R - 2S_{21X}[(\Box\varphi)^2 - \varphi^{ab}\varphi_{ab}] - 2S_{21Y}[(\Box\xi)^2 - \xi^{ab}\xi_{ab}] - 2S_{21Z}[\Box\varphi\Box\xi - \varphi^{ab}\xi_{ab}]\} \tag{1.14a}$$

$$L_{22} := g^{1/2}\{S_{22}\varphi^a\varphi^b G_{ab} + S_{22X}[X((\Box\varphi)^2 - \varphi^{ab}\varphi_{ab}) - 2\varphi^a\varphi^b\varphi_{ab}\Box\varphi + 2\varphi^a\varphi^b\varphi_{ac}\varphi_b{}^c] +$$

$$+ S_{22Z}[X(\Box\varphi\Box\xi - \varphi^{ab}\xi_{ab}) - \varphi^a\varphi^b\varphi_{ab}\Box\xi - \varphi^a\varphi^b\xi_{ab}\Box\varphi + 2\varphi^a\varphi^b\varphi_{ac}\xi_b{}^c] +$$

$$+ S_{22Y}[X((\Box\xi)^2 - \xi^{ab}\xi_{ab}) - 2\varphi^a\varphi^b\xi_{ab}\Box\xi + 2\varphi^a\varphi^b\xi_{ac}\xi_b{}^c]\}\ , \tag{1.14b}$$

$$L_{23} := g^{1/2}\{S_{23}\xi^a\xi^b G_{ab} + S_{23X}[Y((\Box\varphi)^2 - \varphi^{ab}\varphi_{ab}) - 2\xi^a\xi^b\varphi_{ab}\Box\varphi + 2\xi^a\xi^b\varphi_{ac}\varphi_b{}^c] +$$

$$+ S_{23Z}[Y(\Box\varphi\Box\xi - \varphi^{ab}\xi_{ab}) - \xi^a\xi^b\varphi_{ab}\Box\xi - \xi^a\xi^b\xi_{ab}\Box\varphi + 2\xi^a\xi^b\varphi_{ac}\xi_b{}^c] +$$

$$+ S_{23Y}[Y((\Box\xi)^2 - \xi^{ab}\xi_{ab}) - 2\xi^a\xi^b\xi_{ab}\Box\xi + 2\xi^a\xi^b\xi_{ac}\xi_b{}^c]\}\ , \tag{1.14c}$$

$$L_{24} := g^{1/2}\{S_{24}\varphi^a\xi^b G_{ab} + S_{24X}[Z((\Box\varphi)^2 - \varphi^{ab}\varphi_{ab}) - 2\varphi^a\xi^b\varphi_{ab}\Box\varphi + 2\varphi^a\xi^b\varphi_{ac}\varphi_b{}^c] +$$

$$+ S_{24Z}[Z(\Box\varphi\Box\xi - \varphi^{ab}\xi_{ab}) - \varphi^a\xi^b\varphi_{ab}\Box\xi - \varphi^a\xi^b\xi_{ab}\Box\varphi + \varphi^a\xi^b\varphi_{ac}\xi_b{}^c + \xi^a\varphi^b\varphi_{ac}\xi_b{}^c] +$$

$$+ S_{24Y}[Z((\Box\xi)^2 - \xi^{ab}\xi_{ab}) - 2\varphi^a\xi^b\xi_{ab}\Box\xi + 2\varphi^a\xi^b\xi_{ac}\xi_b{}^c]\}\ , \tag{1.14d}$$

and

$$\mathcal{L}_3 := L_{31} + L_{32} \tag{1.15}$$

with

$$L_{31} := g^{1/2}\{S_{31}\varphi_{ab}G^{ab} + \tfrac{1}{3}S_{31X}[(\Box\varphi)^3 - 3\Box\varphi\varphi^{ab}\varphi_{ab} + 2\varphi^a{}_b\varphi^b{}_c\varphi^c{}_a] +$$

$$+ \tfrac{1}{2}S_{31Z}[(\Box\varphi)^2\Box\xi - 2\Box\varphi\varphi^{ab}\xi_{ab} - \varphi^{ab}\varphi_{ab}\Box\xi + 2\varphi^a{}_b\varphi^b{}_c\xi^c{}_a]\} \tag{1.16a}$$

and

$$L_{32} := g^{1/2}\{S_{32}\xi_{ab}G^{ab} + \tfrac{1}{3}S_{32Y}[(\Box\xi)^3 - 3\Box\xi\xi^{ab}\xi_{ab} + 2\xi^a{}_b\xi^b{}_c\xi^{ca}] +$$

$$+ \tfrac{1}{2}S_{32Z}[(\Box\xi)^2\Box\varphi - 2\Box\xi\varphi^{ab}\xi_{ab} - \xi^{ab}\xi_{ab}\Box\varphi + 2\xi^a{}_b\xi^b{}_c\varphi^c{}_a]\}\ . \tag{1.16b}$$

In the above Lagrangians $S_0 = S_0(\varphi,\xi,X,Y,Z)$ is arbitrary, while $(\lambda_{11},\lambda_{12})$, $(\lambda_{21},\lambda_{22})$, $(\lambda_{31},\lambda_{32})$, $(\lambda_{41},\lambda_{42})$, $(\lambda_{51},\lambda_{52})$ and $(S_{31},S_{32})$ are BSTCPs with $S_{21}$, $S_{22}$, $S_{23}$ and $S_{24}$ being BST functions. ■

For the case in which there is only one scalar field, $\varphi$, the above theorem reduces to the usual result for second-order scalar-tensor field equations in a space of four-dimensions Horndeski [1], Deffayet, *et al.,* [6] and Kobayashi *et al.,* [7], where the last paper reconciles the results of [1] and [6].

One remark about the notation used in the statement of the **Theorem.** You should note that each of the four Lagrangians can be viewed as a polynomial in the second derivatives of $\varphi$ and $\xi$. As such $\mathcal{L}_0$ is of degree zero, $\mathcal{L}_1$ is of degree one, $\mathcal{L}_2$ is of degree two, and $\mathcal{L}_3$ is of degree three. There is no relation between the names I have chosen and those used for scalar-tensor theories where the numerical subscripts on the Lagrangians bear a connection to Galileons (*see,* Deffayet, *et al.,*[6]).

The Lagrangians $\mathcal{L}_2$ and $\mathcal{L}_3$ presented in the **Theorem** can be expressed more succinctly using generalized Kronecker deltas. Such expressions can be found using (2.29) and (2.40).

At first sight there appears to be seventeen coefficient functions appearing in the **Theorem**. However, due to my earlier remarks, we know that each of the six BSTCPs occurring in the **Theorem** has a generating function, and so these pairs really only correspond to six BST functions. In view of these remarks on generating functions, we see that there are actually only eleven coefficient functions appearing in the **Theorem.**

Another thing that one should note is that $L_{15}$ in (1.12e) can be expressed in terms of $L_{11}$,..., $L_{14}$ in the following way:

$$L_{15} = YL_{12} + XL_{13} - 2ZL_{14} - (XY - Z^2)L_{11} \tag{1.17}$$

when we take $(\lambda_{11},\lambda_{12})=(\lambda_{21},\lambda_{22})=(\lambda_{31},\lambda_{32})=(\lambda_{41},\lambda_{42})=(\lambda_{51},\lambda_{52})$. However when you multiply a BSTCP by a function of $\varphi,\xi,X,Y,Z$ that actually involves at least one of X, Y, or Z, then that new pair is no longer a BSTCP. That is why each of the separate Lagrangians on the right-hand side of (1.17) is not in the $\mathfrak{L}_1$ class. Hence $L_{15}$ can not be dropped from $\mathfrak{L}_1$.

The Euler-Lagrange tensor densities of many of the Lagrangians appearing in the **Theorem** can be found in Appendix C of Ohashi, *et al.*[3].

My proof of this theorem is quite arduous and, like the work presented in Ohashi, *et al.*, [3], begins by generalizing the construction used in Horndeski [1], [4], dealing with the scalar-tensor case, to the present situation. So at the outset I shall be duplicating some of the work of Ohashi, *et al.*, but I have included it so as to make the paper self-contained. However, where I shall depart from their treatment of the problem is in my construction of the Lagrangian required for the field equations, which they were unable to obtain. This construction will differ markedly from what was done in [1] and [4], and it will take quite a bit of effort.

So let's now get to work.

**Section 2: A Proof of the Theorem**

The proof will proceed by means of a series of lemmas, some of which present interesting results in their own right.

Due to (1.8) we know that in a space of four-dimensions, all of the Euler-Lagrange tensor densities we seek will have $E^{ij}(L)$ contained in the set of symmetric (2,0) tensor density concomitants $A^{ij}$ which are such that:

(i) $A^{ij}$ is symmetric with the functional form:

$$A^{ij} = A^{ij}(g_{hk}; g_{hk,p}; g_{hk,pq}; \varphi; \varphi_{,p}; \varphi_{,pq}; \xi; \xi_{,p}; \xi_{,pq}) ; \tag{2.1}$$

and

(ii) there exist scalar density concomitants B and C, which are at most second-order bi-scalar-tensor concomitants like $A^{ij}$ and for which

$$A^{ij}{}_{|j} = \tfrac{1}{2}\varphi^{i}B + \tfrac{1}{2}\xi^{i}C. \tag{2.2}$$

Consequently $A^{ij}{}_{|j}$ must be at most of second-order.

Owing to (2.1) we know that in general $A^{ij}{}_{|j}$ would be of third-order, but (2.2) says it isn't. As a result we have the following three conditions on $A^{ij}$:

$$\frac{\partial A^{ij}{}_{|j}}{\partial g_{rs,tuv}} = 0; \quad \frac{\partial A^{ij}{}_{|j}}{\partial \varphi_{,tuv}} = 0 \text{ ; and } \frac{\partial A^{ij}{}_{|j}}{\partial \xi_{,tuv}} = 0 \quad . \tag{2.3}$$

Recall that whenever you differentiate a bi-scalar-tensor tensorial concomitant with respect to the highest occurring derivatives of $g_{hk}$, $\varphi$ and $\xi$ in that concomitant the result is a tensorial concomitant. Hence each term in (2.3) is a tensorial concomitant.

To simplify the form of the ensuing calculations I shall introduce the following notation: if $T^{\cdots}$ denotes the components of a second-order bi-scalar-tensor concomitant, let

$$T^{\cdots;ab} := \frac{\partial T^{\cdots}}{\partial g_{ab}}; \; T^{\cdots;ab,c} := \frac{\partial T^{\cdots}}{\partial g_{ab,c}} \; ; \; T^{\cdots;ab,cd} := \frac{\partial T^{\cdots}}{\partial g_{ab,cd}} \quad ; \tag{2.4a}$$

$$T_{\varphi}{}^{\cdots} := \frac{\partial T^{\cdots}}{\partial \varphi} \; ; \; T_{\varphi}{}^{\cdots;c} := \frac{\partial T^{\cdots}}{\partial \varphi_{,c}} \; ; \; T_{\varphi}{}^{\cdots;cd} := \frac{\partial T^{\cdots}}{\partial \varphi_{,cd}} \; ; \; T_{\xi}{}^{\cdots} := \frac{\partial T^{\cdots}}{\partial \xi} \; ; \; T_{\xi}{}^{\cdots;c} := \frac{\partial T^{\cdots}}{\partial \xi_{,c}} \; ; \; T_{\xi}{}^{\cdots;cd} := \frac{\partial T^{\cdots}}{\partial \xi_{,cd}} . \tag{2.4b}$$

One should note that $T^{\cdots;ab,cd}$ is symmetric in a,b and in c,d. While $T_{\varphi}{}^{\cdots;cd}$ and $T_{\xi}{}^{\cdots;cd}$ are symmetric in c,d.

Since the only third-order terms in $A^{ij}{}_{|j}$ occur in $A^{ij}{}_{,j}$ we easily find that (2.3) implies that

$$3A^{i(u;|rs|,tv)} \equiv A^{iu;rs,tv} + A^{it;rs,vu} + A^{iv;rs,ut} = 0 \, , \tag{2.5a}$$

$$3A_{\varphi}{}^{i(t;uv)} \equiv A_{\varphi}{}^{it;uv} + A_{\varphi}{}^{iu;vt} + A_{\varphi}{}^{iv;tu} = 0, \tag{2.5b}$$

$$3A_{\xi}{}^{i(t;uv)} \equiv A_{\xi}{}^{it;uv} + A_{\xi}{}^{iu;vt} + A_{\xi}{}^{iv;tu} = 0 \, , \tag{2.5c}$$

where parentheses around a group of indices denotes symmetrization over those indices, except for those between vertical bars.

To proceed further we require an invariance identity (*see,* Rund [8], [9], du Plessis [10], Horndeski and Lovelock [11], Horndeski [4], Lovelock and Rund [12]), which can be derived as follows. Suppose that x and x' are charts at an arbitrary point P. So that on a neighborhood of P we can functionally write $x^r = x^r(x'^h)$ and $x'^r = x'^r(x^h)$. Then at P the x and x' components of the metric tensor are related by

$$g'_{hk} = g_{rs}B^r_{\ h}B^s_{\ k} \tag{2.6a}$$

where I shall let

$$B^r_{\ h} := \frac{\partial x^r}{\partial x'^h}\,,\ B^r_{\ hk} := \frac{\partial^2 x^r}{\partial x'^h \partial x'^k} \quad \text{and} \quad B^r_{\ hkl} := \frac{\partial^3 x^r}{\partial x'^h \partial x'^k \partial x'^l}\quad . \tag{2.6b}$$

Consequently at P we can write

$$g'_{hk,pq} = g_{rs}(B^r_{\ hpq}B^s_{\ k} + B^r_{\ h}B^s_{\ kpq}) + \text{lower order terms}$$

where the "lower order terms" are those terms that do not involve $B^r_{\ hpq}$. Since $A^{ij}$ is a tensor density its components in the primed and unprimed indices at P are related by

$$\begin{aligned} &A^{ij}(g'_{hk};\, g'_{hk,p};\, g'_{hk,pq};\, \varphi';\, \varphi'_{,p};\, \varphi'_{,pq};\, \xi';\, \xi'_{,p};\, \xi'_{,pq}) = \\ &\quad = \det(B^l_{\ m})B'^i_{\ e}B'^j_{\ f}A^{ef}(g_{rs};\, g_{rs,t};\, g_{rs,tu};\, \varphi;\, \varphi_{,t};\, \varphi_{,tu};\, \xi;\, \xi_{,t};\, \xi_{,tu})\ , \end{aligned} \tag{2.7}$$

where $B'^i_{\ e}$ is the derivative of $x'^i$ with respect to $x^e$, and is the matrix inverse of $B^i_{\ e}$. If we differentiate (2.7) with respect to $B^a_{\ bcd}$ and evaluate the result for the identity coordinate transformation at P we discover that the tensor density $A^{ij;ab,cd}$ must satisfy

$$A^{ij;a(b,cd)} = 0\ . \tag{2.8}$$

It is easily shown that the results presented in (2.5) and (2.8) imply that

$$A^{ij;ab,cd} = A^{ij;cd,ab} = A^{cd;ab,ij}\ ;\ A^{i(j;ab),cd} = 0;\ A_\varphi^{\ ij;ab} = A_\varphi^{\ ab;ij}\ ;\ A_\xi^{\ ij;ab} = A_\xi^{\ ab;ij}\ . \tag{2.9}$$

Throughout our construction of $A^{ij}$ we shall encounter numerous quantities which possess the symmetries evinced in (2.5), (2.8) and (2.9). Such quantities were first encountered in Lovelock [13], and following his nomenclature we shall say a quantity $Q = Q[i_1 i_2; \ldots ; i_{2p-1} i_{2p}]$ has Property S if it satisfies the following three conditions:

(i) It is symmetric in the indicies $i_{2h-1} i_{2h}$ for h=1,...,p;

(ii) It is symmetric under the interchange of the pair $(i_1 i_2)$ with the pair $(i_{2h-1} i_{2h})$, for h=2,...,p; and

(iii) It satisfies the cyclic identity involving any three of the four indicies $(i_1i_2)$, $(i_{2h-1}i_{2h})$, h=2,...,p;*e.g.,*

$$Q[i_1(i_2;|...i_{2h-2}|;i_{2h-1}i_{2h});. . . ;i_{2p-1}i_{2p}] = 0.$$

A quantity $Q[i_1i_2]$ is said to have Property S if it is symmetric in $i_1$, $i_2$. Due to (2.5), (2.8) and (2.9) we have established the following:

**Lemma 1:** If the tensor density $A^{ij}$ is a second-order bi-scalar-tensor concomitant with the property that $A^{ij}{}_{|j}$ is also second-order then all partial derivatives of $A^{ij}$ with respect to any combination of $g_{ab,cd}$, $\varphi_{,cd}$ and $\xi_{,cd}$ will be tensorial concomitants with property S.■

The significance of **Lemma 1** lies in the fact that Lovelock [13] proved that in a four-dimensional space any 10 index quantity with Property S must vanish. Hence the lemma tells us that under its assumptions $A^{ij}$ must be a polynomial of at most first degree in $g_{ab,cd}$ and at most of third degree in the second derivatives of $\varphi$ and $\xi$. This polynomial can also involve terms of the form $g_{ab,cd}\varphi_{,ef}$ and $g_{ab,cd}\xi_{,ef}$ but there can be no terms like $g_{ab,cd}\varphi_{,ef}\xi_{,gh}$ . Consequently when $A^{ij}$ satisfies the conditions of **Lemma 1** we can express it as follows:

$$A^{ij} = \alpha_1{}^{ijabcdef}g_{ab,cd}\varphi_{,ef} + \alpha_2{}^{ijabcdef}g_{ab,cd}\xi_{,ef} + \alpha_3{}^{ijabcd}g_{ab,cd} + \beta_1{}^{ijabcdef}\varphi_{,ab}\varphi_{,cd}\varphi_{,ef} + \beta_2{}^{ijabcdef}\varphi_{,ab}\varphi_{,cd}\xi_{,ef}+ +\beta_3{}^{ijabcdef}\varphi_{,ab}\xi_{,cd}\xi_{,ef}+ \beta_4{}^{ijabcdef}\xi_{,ab}\xi_{,cd}\xi_{,ef} + \beta_5{}^{ijabcd}\varphi_{,ab}\varphi_{,cd} + \beta_6{}^{ijabcd}\varphi_{,ab}\xi_{,cd} + \beta_7{}^{ijabcd}\xi_{,ab}\xi_{,cd} + \beta_8{}^{ijab}\varphi_{,ab} + \beta_9{}^{ijab}\xi_{,ab}+ +\beta_{10}{}^{ab}, \qquad (2.10)$$

where each of the $\alpha$ and $\beta$ coefficient functions are concomitants of $g_{ij}$, $g_{ij,h}$, $\varphi$, $\varphi_{,h}$, $\xi$ and $\xi_{,h}$ which enjoy Property S. It is easy to show that $\alpha_1$, $\alpha_2$, $\beta_1$, $\beta_2$, $\beta_3$ and $\beta_4$ are tensorial concomitants, but at present it is not clear if that is so for $\alpha_3$ and the remaining $\beta$ coefficients. In order to write $A^{ij}$ in a manifestly tensorial form we need Thomas's Replacement Theorem [14] which I state as

**Lemma 2: (Thomas's Replacement Theorem for Third-Order Bi-Scalar-Tensor Concomitants)**

If $T^{\cdot\cdot} = T^{\cdot\cdot}(g_{ab}; g_{ab,c}; g_{ab,cd}; g_{ab,cde}; \varphi; \varphi_{,c}; \varphi_{,cd}; \varphi_{,cde}; \xi; \xi_{,c}; \xi_{,cd}; \xi_{,cde})$ denotes the components of a

tensorial concomitant of the indicated form then its value is unaffected when $g_{ab,c}$ is replaced by 0; $g_{ab,cd}$ is replaced by $\frac{1}{3}(R_{acdb} + R_{adcb})$; $g_{ab,cde}$ is replaced by $\frac{1}{3}(R_{a(cd)b|e} + R_{a(de)b|c} + R_{a(ec)b|d})$; $\varphi_{,cd}$ is replaced by $\varphi_{cd}$ ; $\varphi_{,cde}$ is replaced by $\varphi_{(cde)}$; $\xi_{,ab}$ is replaced by $\xi_{ab}$; and $\xi_{,cde}$ is replaced by $\xi_{(cde)}$.■

I prove this result for third-order scalar-tensor concomitants in Lemma 2 of Horndeski [5] and that proof goes over virtually unchanged for the bi-scalar-tensor case. As an immediate consequence of **Lemma 2** and (2.10) we have

**Lemma 3:** If in a space of four-dimensions the tensor density $A^{ij}$ is a symmetric, second-order bi-scalar-tensor concomitant which is such that $A^{ij}{}_{|j}$ is of second-order then

$$A^{ij} = \alpha_1{}^{ijabcdef}R_{acdb}\varphi_{ef} + \alpha_2{}^{ijabcdef}R_{acdb}\xi_{ef} + \alpha_3{}^{ijabcd}R_{acdb} + \beta_1{}^{ijabcdef}\varphi_{ab}\varphi_{cd}\varphi_{ef} + \beta_2{}^{ijabcdef}\varphi_{ab}\varphi_{cd}\xi_{ef} + \beta_3{}^{ijabcdef}\varphi_{ab}\xi_{cd}\xi_{ef} + \\ + \beta_4{}^{ijabcdef}\xi_{ab}\xi_{cd}\xi_{ef} + \beta_5{}^{ijabcd}\varphi_{ab}\varphi_{cd} + \beta_6{}^{ijabcd}\varphi_{ab}\xi_{cd} + \beta_7{}^{ijabcd}\xi_{ab}\xi_{cd} + \beta_8{}^{ijab}\varphi_{ab} + \beta_9{}^{ijab}\xi_{ab} + \beta_{10}{}^{ab} \qquad (2.11)$$

where the $\alpha$'s and $\beta$'s are tensorial concomitants of $g_{ab}$; $\varphi$; $\varphi_{,a}$; $\xi$; and $\xi_{,a}$ with Property S.■

Strictly speaking I should have used different symbols for the $\alpha$'s and $\beta$'s in (2.10) and (2.11), but that is really unnecessary.

In **Appendix B**, I construct the most general 2, 4, 6 and 8 index tensorial concomitants of $g_{ij}$; $\varphi$; $\varphi_{,i}$; $\xi$; and $\xi_{,i}$ which have property S in a four-dimensional space. This is also done in Ohashi, *et al.,* [3], by assuming that the concomitants in question must be polynomials, and then appealing to a result in Weyl [15] which says that they must be built from all possible products of $g_{ij}$, $\varphi_{,i}$, $\xi_{,i}$ and $\varepsilon^{abcd}$ with the appropriate symmetries, where $\varepsilon^{abcd}$ is the contravariant form of the Levi-Civita symbol. In **Appendix A** and **B**, I present a formal way of building these concomitants based on Lovelock's [13] approach to building metric concomitants with property S, which does not require the polynomial assumption. I also show that there is a "duality" between the 2 index and 6 index concomitants with property S that enables one to construct all of the 6 index concomitants if you know all of the 2 index concomitants. This duality becomes useful when trying to generalize the **Theorem** to even dimensional spaces of dimension greater than four.

Using the results of **Appendix A** and **B** it can be shown, with some effort, that (2.11) gives us

**Lemma 4:** If the tensor density $A^{ij}$ is a symmetric tensorial concomitant of the form

$$A^{ij} = A^{ij}(g_{hk};\, g_{hk,p};\, g_{hk,pq};\, \varphi;\, \varphi_{,p};\, \varphi_{,pq};\, \xi;\, \xi_{,p};\, \xi_{,pq})$$

which is such that $A^{ij}{}_{|j}$ is also of second-order in $g_{hk}$, $\varphi$ and $\xi$, then in a space of four-dimensions

$$\begin{aligned}
A^{ij} = g^{1/2}\{ & \alpha_1 \delta^{abei}_{cdfq}\, g^{qj}R_{ab}{}^{cd}\varphi_e{}^f + \alpha_2 \delta^{abei}_{cdfq}\, g^{qj}R_{ab}{}^{cd}\xi_e{}^f + \alpha_{31} \delta^{abi}_{cdq}\, g^{qj}R_{ab}{}^{cd} + \alpha_{32} \delta^{abei}_{cdfq}\, g^{qj}\varphi_e\varphi^f R_{ab}{}^{cd} + \\
& + \alpha_{33} \delta^{abei}_{cdfq}\, g^{qj}\xi_e\xi^f R_{ab}{}^{cd} + \alpha_{34} \delta^{abei}_{cdfq}\, g^{qj}(\varphi_e\xi^f + \xi_e\varphi^f)R_{ab}{}^{cd} + \beta_1 \delta^{acei}_{bdfq}\, g^{qj}\varphi_a{}^b\varphi_c{}^d\varphi_e{}^f + \beta_2 \delta^{acei}_{bdfq}\, g^{qj}\varphi_a{}^b\varphi_c{}^d\xi_e{}^f + \\
& + \beta_3 \delta^{acei}_{bdfq}\, g^{qj}\varphi_a{}^b\xi_c{}^d\xi_e{}^f + \beta_4 \delta^{acei}_{bdfq}\, g^{qj}\xi_a{}^b\xi_c{}^d\xi_e{}^f + \beta_{51} \delta^{aci}_{bdq}\, g^{qj}\varphi_a{}^b\varphi_c{}^d + \beta_{52} \delta^{acei}_{bdfq}\, g^{qj}\varphi_a\varphi^b\varphi_c{}^d\varphi_e{}^f + \\
& + \beta_{53} \delta^{acei}_{bdfq}\, g^{qj}\xi_a\xi^b\varphi_c{}^d\varphi_e{}^f + \beta_{54} \delta^{acei}_{bdfq}\, g^{qj}(\varphi_a\xi^b + \xi_a\varphi^b)\varphi_c{}^d\varphi_e{}^f + \beta_{61} \delta^{aci}_{bdq}\, g^{qj}\varphi_a{}^b\xi_c{}^d + \beta_{62} \delta^{acei}_{bdfq}\, g^{qj}\varphi_a\varphi^b\varphi_c{}^d\xi_e{}^f + \\
& + \beta_{63} \delta^{acei}_{bdfq}\, g^{qj}\xi_a\xi^b\varphi_c{}^d\xi_e{}^f + \beta_{64} \delta^{acei}_{bdfq}\, g^{qj}(\varphi_a\xi^b + \xi_a\varphi^b)\varphi_c{}^d\xi_e{}^f + \beta_{71} \delta^{aci}_{bdq}\, g^{qj}\xi_a{}^b\xi_c{}^d + \beta_{72} \delta^{acei}_{bdfq}\, g^{qj}\varphi_a\varphi^b\xi_c{}^d\xi_e{}^f + \\
& + \beta_{73} \delta^{acei}_{bdfq}\, g^{qj}\xi_a\xi^b\xi_c{}^d\xi_e{}^f + \beta_{74} \delta^{acei}_{bdfq}\, g^{qj}(\varphi_a\xi^b + \xi_a\varphi^b)\xi_c{}^d\xi_e{}^f + \beta_{81}(\square\varphi g^{ij} - \varphi^{ij}) + \beta_{82}(\square\varphi\varphi^i\varphi^j + g^{ij}\varphi_a\varphi_b\varphi^{ab} + \\
& - \varphi^i\varphi_a\varphi^{ja} - \varphi^j\varphi_a\varphi^{ia}) + \beta_{83}(\square\varphi\xi^i\xi^j + g^{ij}\xi_a\xi_b\varphi^{ab} - \xi^i\xi_a\varphi^{ja} - \xi^j\xi_a\varphi^{ia}) + \beta_{84}(\square\varphi\varphi^i\xi^j + \square\varphi\varphi^j\xi^i + 2g^{ij}\varphi_a\xi_b\varphi^{ab} + \\
& - \xi^i\varphi_a\varphi^{ja} - \xi^j\varphi_a\varphi^{ia} - \varphi^i\xi_a\varphi^{ja} - \varphi^j\xi_a\varphi^{ia}) + \beta_{85}(\varphi_a\varphi_b\varphi^{ab}\xi^i\xi^j + \xi_a\xi_b\varphi^{ab}\varphi^i\varphi^j - \varphi_a\xi_b\varphi^{ab}\xi^i\varphi^j - \varphi_a\xi_b\varphi^{ab}\xi^j\varphi^i) + \\
& + \beta_{86}(*\omega^{ia}\varphi^b\xi^j\varphi_{ab} + *\omega^{ja}\varphi^b\xi^i\varphi_{ab} - *\omega^{ia}\varphi^j\xi^b\varphi_{ab} - *\omega^{ja}\varphi^i\xi^b\varphi_{ab}) + \beta_{91}(\square\xi g^{ij} - \xi^{ij}) + \beta_{92}(\square\xi\varphi^i\varphi^j + g^{ij}\varphi_a\varphi_b\xi^{ab} + \\
& - \varphi^i\varphi_a\xi^{ja} - \varphi^j\varphi_a\xi^{ia}) + \beta_{93}(\square\xi\xi^i\xi^j + g^{ij}\xi_a\xi_b\xi^{ab} - \xi^i\xi_a\xi^{ja} - \xi^j\xi_a\xi^{ia}) + \beta_{94}(\square\xi\varphi^i\xi^j + \square\xi\varphi^j\xi^i + 2g^{ij}\varphi_a\xi_b\xi^{ab} + \\
& - \xi^i\varphi_a\xi^{ja} - \xi^j\varphi_a\xi^{ia} - \varphi^i\xi_a\xi^{ja} - \varphi^j\xi_a\xi^{ia}) + \beta_{95}(\varphi_a\varphi_b\xi^{ab}\xi^i\xi^j + \xi_a\xi_b\xi^{ab}\varphi^i\varphi^j - \varphi_a\xi_b\xi^{ab}\xi^i\varphi^j - \varphi_a\xi_b\xi^{ab}\xi^j\varphi^i) + \\
& + \beta_{96}(*\omega^{ia}\varphi^b\xi^j\xi_{ab} + *\omega^{ja}\varphi^b\xi^i\xi_{ab} - *\omega^{ia}\varphi^j\xi^b\xi_{ab} - *\omega^{ja}\varphi^i\xi^b\xi_{ab}) + \gamma_1 g^{ij} + \gamma_2\varphi^i\varphi^j + \gamma_3\xi^i\xi^j + \gamma_4(\varphi^i\xi^j + \xi^i\varphi^j)\}, \qquad (2.12)
\end{aligned}$$

where the $\alpha_{..}$, $\beta_{..}$ and $\gamma_{.}$ coefficients are functions of $\varphi$, $\xi$, X, Y and Z, $*\omega^{ij} := \tfrac{1}{2}g^{-1/2}\varepsilon^{ijab}\varphi_a\xi_b$, $\varepsilon^{ijab}$ is the Levi-Civita tensor density, and $\delta^{\cdots}_{\cdots}$ denotes the generalized Kronecker delta .■

So far Ohashi, *et al.*[3] and I have been following the proof I laid out in Horndeski [1], [4]. I shall now start to diverge from that program, while Ohashi, *et al*, continue to follow it.

From (1.8) we know that if $A^{ij}$ is to arise from a variational principle then not only must $A^{ij}{}_{|j}$

be of second-order, but it must satisfy (2.2). So if L were a Lagrangian for which $E^{ij}(L) = A^{ij}$, B := $E_{\varphi}(L)$, and C := $E_{\xi}(L)$, then $A^{ij}{}_{|j}$ lies in the plane spanned by $\{\varphi^i, \xi^i\}$. This property of $A^{ij}{}_{|j}$ can be expressed by writing:

$$A^{ij}{}_{|j} \equiv 0 \bmod(\varphi^i,\xi^i), \tag{2.13}$$

(read: $A^{ij}{}_{|j}$ is congruent to 0 modulo $(\varphi^i,\xi^i)$). So we shall now determine the conditions that the coefficients in $A^{ij}$ must satisfy if (2.13) is to hold. This is done by first computing $A^{ij}{}_{|j}$ and while doing so discard all terms that involve a scalar density times either $\varphi^i$ or $\xi^i$. Ohashi, *et al.* [3] keep track of all the terms we are throwing out, since they also want to know the form of B and C. I feel that trying to keep track of all those terms just complicates the calculations, while B and C can be computed later if one wishes. However, there is another good reason to keep track of these terms, which I shall discuss later.

When $A^{ij}$ is given by (2.12), the expression for $A^{ij}{}_{|j}$ is monstrously long (over 100 feet), even when working $\bmod(\varphi^i, \xi^i)$. There is no need for me to reproduce it here. Analysis of that equation is discussed in Ohashi, *et al.*[3], along with **Appendix C** here, and leads to

**Lemma 5:** If $A^{ij}$ satisfies the conditions of **Lemma 4** along with (2.13) then the coefficients appearing in $A^{ij}$ must satisfy the following conditions:

$$4\alpha_{1X} = -3\beta_1;\ 4\alpha_{1Y} = -\beta_3;\ 2\alpha_{1Z} = -\beta_2;\ 4\alpha_{2X} = -\beta_2;\ 4\alpha_{2Y} = -3\beta_4;\ 2\alpha_{2Z} = -\beta_3; \tag{2.14}$$

$$6\beta_{1Y} = \beta_{2Z};\ 3\beta_{1Z} = 2\beta_{2X};\ 2\beta_{2Y} = \beta_{3Z};\ \beta_{2Z} = 2\beta_{3X};\ 2\beta_{3Y} = 3\beta_{4Z};\ \beta_{3Z} = 6\beta_{4X}; \tag{2.15}$$

$$\beta_{51} = -4\alpha_{31X};\ \beta_{52} = -4\alpha_{32X};\ \beta_{53} = -4\alpha_{33X};\ \beta_{54} = -4\alpha_{34X};\ \beta_{61} = -4\alpha_{31Z};\ \beta_{62} = -4\alpha_{32Z};\ \beta_{63} = -4\alpha_{33Z}; \tag{2.16a}$$

$$\beta_{64} = -4\alpha_{34Z};\ \beta_{71} = -4\alpha_{31Y};\ \beta_{72} = -4\alpha_{32Y};\ \beta_{73} = -4\alpha_{33Y};\ \beta_{74} = -4\alpha_{34Y}; \tag{2.16b}$$

$$-\alpha_{1\varphi}+\alpha_{32}+2\alpha_{31X}+2X\alpha_{32X}+Y\alpha_{34Z}+Z(2\alpha_{34X}+\alpha_{32Z})=0; \tag{2.17a}$$

$$-\alpha_{1\xi}+\alpha_{34}+\alpha_{31Z}+2X\alpha_{34X}+Y\alpha_{33Z}+Z(\alpha_{34Z}+2\alpha_{33X})=0; \tag{2.17b}$$

$$-\alpha_{2\xi}+\alpha_{33}+2\alpha_{31Y}+X\alpha_{34Z}+2Y\alpha_{33Y}+Z(2\alpha_{34Y}+\alpha_{33Z})=0; \tag{2.18a}$$

$$-\alpha_{2\varphi}+\alpha_{34}+\alpha_{31Z}+2Y\alpha_{34Y}+X\alpha_{32Z}+Z(\alpha_{34Z}+2\alpha_{32Y}) = 0; \tag{2.18b}$$

$$\alpha_{2\varphi} - \alpha_{1\xi} + X(2\alpha_{34X} - \alpha_{32Z}) + Y(\alpha_{33Z} - 2\alpha_{34Y}) + 2Z(\alpha_{33X} - \alpha_{32Y}) = 0; \tag{2.19}$$

$$\beta_{86}=\beta_{96}=0;\ 4(\alpha_{32\xi}-\alpha_{34\varphi}) = \beta_{92}-\beta_{84};\ 4(\alpha_{33\varphi}-\alpha_{34\xi}) = \beta_{83}-\beta_{94}; \tag{2.20}$$

$$\beta_{81}=4[Z(\alpha_{32\xi}-\alpha_{34\varphi}) + Y(\alpha_{34\xi}-\alpha_{33\varphi}) - \alpha_{31\varphi}];\ \beta_{91}=4[Z(\alpha_{33\varphi}-\alpha_{34\xi}) + X(\alpha_{34\varphi}-\alpha_{32\xi}) - \alpha_{31\xi}]\,. \tag{2.21}$$

Consequently $\alpha_1$, $\alpha_2$, $\alpha_{31}$, $\alpha_{32}$, $\alpha_{33}$ and $\alpha_{34}$ determine $\beta_1$, $\beta_2$, $\beta_3$, $\beta_4$, $\beta_{51}$, $\beta_{52}$, $\beta_{53}$, $\beta_{54}$, $\beta_{61}$, $\beta_{62}$, $\beta_{63}$, $\beta_{64}$, $\beta_{71}$, $\beta_{72}$, $\beta_{73}$ and $\beta_{74}$. (2.14) and (2.15) imply that $\alpha_1$, $\alpha_2$, $\beta_1,\ldots, \beta_4$ are BST functions, with $(\alpha_1,\alpha_2)$ being a BSTCP. When $\alpha_1=\alpha_2=0$, then $\alpha_{31}$, $\alpha_{32}$, $\alpha_{33}$ and $\alpha_{34}$ turn out to be BST functions, and hence due to (2.16), $\beta_{51}$, $\beta_{52}$, $\beta_{53}$, $\beta_{54}$, $\beta_{61}$, $\beta_{62}$, $\beta_{63}$, $\beta_{64}$, $\beta_{71}$, $\beta_{72}$, $\beta_{73}$ and $\beta_{74}$ are BST functions when $\alpha_1=\alpha_2=0$.■

The conditions presented in the above lemma are not exhaustive of all the implications of $A^{ij}{}_{|j}\equiv 0 \bmod(\varphi^i,\xi^i)$, but they are more than enough for our present needs. We shall determine more later.

Using the results of **Lemma 5** enables us to rewrite the expression for $A^{ij}$ given in Eq.2.12 as follows:

$$A^{ij} = A_3{}^{ij} + A_2{}^{ij} + A_1{}^{ij} + A_0{}^{ij} \tag{2.22a}$$

where

$$\begin{aligned} A_3{}^{ij} := g^{1/2}\{&\alpha_1\, \delta^{abei}_{cdfq}\, g^{qj}R_{ab}{}^{cd}\varphi_e{}^f + \alpha_2\, \delta^{abei}_{cdfq}\, g^{qj}R_{ab}{}^{cd}\xi_e{}^f - 4/3\alpha_{1X}\, \delta^{acei}_{bdfq}\, g^{qj}\varphi_a{}^b\varphi_c{}^d\varphi_e{}^f - 4\alpha_{2X}\, \delta^{acei}_{bdfq}\, g^{qj}\varphi_a{}^b\varphi_c{}^d\xi_e{}^f + \\ &-4\alpha_{1Y}\, \delta^{acei}_{bdfq}\, g^{qj}\varphi_a{}^b\xi_c{}^d\xi_e{}^f - 4/3\alpha_{2Y}\, \delta^{acei}_{bdfq}\, g^{qj}\xi_a{}^b\xi_c{}^d\xi_e{}^f\}\,, \end{aligned} \tag{2.22b}$$

$$\begin{aligned} A_2{}^{ij} := g^{1/2}\{&\alpha_{31}\, \delta^{abi}_{cdq}\, g^{qj}R_{ab}{}^{cd} + \alpha_{32}\, \delta^{abei}_{cdfq}\, g^{qj}\varphi_e\varphi^f R_{ab}{}^{cd} + \alpha_{33}\, \delta^{abei}_{cdfq}\, g^{qj}\xi_e\xi^f R_{ab}{}^{cd} + \alpha_{34}\, \delta^{abei}_{cdfq}\, g^{qj}(\varphi_e\xi^f+\xi_e\varphi^f)R_{ab}{}^{cd} + \\ &-4\alpha_{31X}\, \delta^{aci}_{bdq}\, g^{qj}\varphi_a{}^b\varphi_c{}^d - 4\alpha_{32X}\, \delta^{acei}_{bdfq}\, g^{qj}\varphi_a\varphi^b\varphi_c{}^d\varphi_e{}^f - 4\alpha_{33X}\, \delta^{acei}_{bdfq}\, g^{qj}\xi_a\xi^b\varphi_c{}^d\varphi_e{}^f + \\ &-4\alpha_{34X}\, \delta^{acei}_{bdfq}\, g^{qj}(\varphi_a\xi^b+\xi_a\varphi^b)\varphi_c{}^d\varphi_e{}^f - 4\alpha_{31Z}\, \delta^{aci}_{bdq}\, g^{qj}\varphi_a{}^b\xi_c{}^d - 4\alpha_{32Z}\, \delta^{acei}_{bdfq}\, g^{qj}\varphi_a\varphi^b\varphi_c{}^d\xi_e{}^f - 4\alpha_{33Z}\, \delta^{acei}_{bdfq}\, g^{qj}\xi_a\xi^b\varphi_c{}^d\xi_e{}^f \\ &-4\alpha_{34Z}\, \delta^{acei}_{bdfq}\, g^{qj}(\varphi_a\xi^b+\xi_a\varphi^b)\varphi_c{}^d\xi_e{}^f - 4\alpha_{31Y}\, \delta^{aci}_{bdq}\, g^{qj}\xi_a{}^b\xi_c{}^d - 4\alpha_{32Y}\, \delta^{acei}_{bdfq}\, g^{qj}\varphi_a\varphi^b\xi_c{}^d\xi_e{}^f - 4\alpha_{33Y}\, \delta^{acei}_{bdfq}\, g^{qj}\xi_a\xi^b\xi_c{}^d\xi_e{}^f + \end{aligned}$$

$$-4\alpha_{34Y}\,\delta^{acei}_{bdfq}\, g^{qj}(\varphi_a\xi^b + \xi_a\varphi^b)\xi_c{}^d\xi_e{}^f\}, \tag{2.22c}$$

$$A_1{}^{ij} := g^{1/2}\{\beta_{81}(\Box\varphi g^{ij} - \varphi^{ij}) + \beta_{82}(\Box\varphi\varphi^i\varphi^j + g^{ij}\varphi_a\varphi_b\varphi^{ab} - \varphi^i\varphi_a\varphi^{ja} - \varphi^j\varphi_a\varphi^{ia}) + \beta_{83}(\Box\varphi\xi^i\xi^j + g^{ij}\xi_a\xi_b\varphi^{ab} - \xi^i\xi_a\varphi^{ja}$$

$$- \xi^j\xi_a\varphi^{ia}) + \beta_{84}(\Box\varphi\varphi^i\xi^j + \Box\varphi\varphi^j\xi^i + 2g^{ij}\varphi_a\xi_b\varphi^{ab} - \xi^i\varphi_a\varphi^{ja} - \xi^j\varphi_a\varphi^{ia} - \varphi^i\xi_a\varphi^{ja} - \varphi^j\xi_a\varphi^{ia}) + \beta_{85}(\varphi_a\varphi_b\varphi^{ab}\xi^i\xi^j +$$

$$+\xi_a\xi_b\varphi^{ab}\varphi^i\varphi^j - \varphi_a\xi_b\varphi^{ab}\xi^i\varphi^j - \varphi_a\xi_b\varphi^{ab}\xi^j\varphi^i) + \beta_{91}(\Box\xi g^{ij} - \xi^{ij}) + \beta_{92}(\Box\xi\varphi^i\varphi^j + g^{ij}\varphi_a\varphi_b\xi^{ab} - \varphi^i\varphi_a\xi^{ja} - \varphi^j\varphi_a\xi^{ia})$$

$$+ \beta_{93}(\Box\xi\xi^i\xi^j + g^{ij}\xi_a\xi_b\xi^{ab} - \xi^i\xi_a\xi^{ja} - \xi^j\xi_a\xi^{ia}) + \beta_{94}(\Box\xi\varphi^i\xi^j + \Box\xi\varphi^j\xi^i + 2g^{ij}\varphi_a\xi_b\xi^{ab} - \xi^i\varphi_a\xi^{ja} - \xi^j\varphi_a\xi^{ia} - \varphi^i\xi_a\xi^{ja}$$

$$- \varphi^j\xi_a\xi^{ia}) + \beta_{95}(\varphi_a\varphi_b\xi^{ab}\xi^i\xi^j + \xi_a\xi_b\xi^{ab}\varphi^i\varphi^j - \varphi_a\xi_b\xi^{ab}\xi^i\varphi^j - \varphi_a\xi_b\xi^{ab}\xi^j\varphi^i)\}, \tag{2.22d}$$

$$A_0{}^{ij} := g^{1/2}\{\gamma_1 g^{ij} + \gamma_2\varphi^i\varphi^j + \gamma_3\xi^i\xi^j + \gamma_4(\varphi^i\xi^j + \xi^i\varphi^j)\}\ . \tag{2.22e}$$

One should note that the terms in the tensor densities $A_\mu$ ($\mu$=0,1,2,3) are of $\mu$th degree in the second partial derivatives of the scalar fields.

Before I begin our construction of Lagrangians that can be used to realize $A^{ij}$, I need to remark on other possible constraints that $A^{ij}$ satisfies, even though at first it may not be obvious why.

In Ohashi, *et al.* [3] they discuss what they call, following the terminology of Deffayet, *et al.* [16], "integrability conditions," that the field tensor candidates must satisfy in order to arise from a variational principle (*see,* (5.1)-(5.3) in Ohashi *et al,* [3]). In terms of my notation these conditions would amount to the following equations:

$$E^{hk}(A^{ij}) = E^{ij}(A^{hk});\ E_\varphi(A^{ij}) = E^{ij}(B);\ E_\xi(A^{ij}) = E^{ij}(C);\ \text{and}\ E_\varphi(C) = E_\xi(B)\ . \tag{2.23}$$

I hesitate to call these equations "integrability conditions" since that suggests that if the triple ($A^{ij}$, B, C) were to satisfy (2.23), then there would exist a bi-scalar-tensor Lagrangian that realizes them as its Euler-Lagrange tensor densities. I prefer to call them consistency conditions, since if the triple ($A^{ij}$,B,C) is to be consistent with the notion that they are derivable from a variational principle, they must satisfy the conditions presented in (2.23). Since I have not been keeping track of B and C, the only one of the four conditions presented in (2.23) that we could check out is the $E^{hk}(A^{ij}) = E^{ij}(A^{hk})$ condition, and that would be ridiculously complicated. Nevertheless it is possible to obtain information from (2.23) which could be relevant to the form of $A^{ij}$ without really computing the Euler-Lagrange tensors. For example if L is the Lagrangian which generates $A^{ij}$ then

$$E^{hk}(A^{ij}) = E^{hk}\left(\frac{\partial L}{\partial g_{ij}} + \text{a divergence}\right).$$

Since $E^{hk}$ is a linear operator that commutes with $\frac{\partial}{\partial g_{ij}}$ and annihilates divergences (*see*, Lovelock[17] for a direct proof of this), the above equation tells us that

$$E^{hk}(A^{ij}) = A^{hk;ij} \quad . \tag{2.24}$$

Hence $E^{hk}(A^{ij})$ must be of second-order, and not of fourth order. This in turn will lead to conditions on the derivatives of $A^{ij}$, just like the demand that $A^{ij}{}_{|j}$ must be of second-order. Similarly the second and third parts of (2.22) tell us that $E_{\varphi}(A^{ij})$ and $E_{\xi}(A^{ij})$ must be of second-order giving us more constraints on the partial derivatives of $A^{ij}$. I shall not bother to write these constraints down. It suffices to say that I checked them out, and after lengthy calculations they turned out to provide us with no more information than what could be obtained from $A^{ij}{}_{|j} \equiv 0 \bmod(\varphi^{i}, \xi^{i})$. This led me to start thinking that perhaps the fundamental condition, (2.2) is not only necessary for the triple $(A^{ij}, B, C)$ to arise from a variational principle but possibly also sufficient. This is suggested by the next amazing

**Lemma 6A:** Suppose that L is a bi-scalar-tensor Lagrangian of arbitrary differential order in a space of any dimension, and let $A^{ij} := E^{ij}(L)$, $B := E_{\varphi}(L)$, $C := E_{\xi}(L)$. Then $E^{ij}(B) = A^{ij}{}_{\varphi}$, $E_{\varphi}(B) = B_{\varphi}$, $E_{\xi}(B) = C_{\varphi}$ and $E^{ij}(C) = A^{ij}{}_{\xi}$, $E_{\varphi}(C) = B_{\xi}$ , $E_{\xi}(C) = C_{\xi}$.

**Proof:** Since $B = E_{\varphi}(L)$ we can write

$$E^{ij}(B) = E^{ij}\left(\frac{\partial L}{\partial \varphi} - \text{a divergence}\right) . \tag{2.25}$$

Since all of the Euler-Lagrange operators annihilate divergences and commute with the partial differential operators $\frac{\partial}{\partial g_{ij}}$ , $\frac{\partial}{\partial \varphi}$ , $\frac{\partial}{\partial \xi}$ , and with no other partial derivatives with respect to the arguments of L, (2.25) tells us that $E^{ij}(B) = \frac{\partial A^{ij}}{\partial \varphi}$ , as desired.

In virtually the same way it is possible to establish all of the other relations presented in the statement of **Lemma 6A**.■

The following Lemma should now be obvious:

**Lemma 6B:** Suppose that in an n-dimensional space, L is a multi-scalar-tensor Lagrange scalar density of arbitrary differential order built from the scalar fields $\varphi_I$, I=1,...,N$\leq$n, and a pseudo-Riemannian metric tensor $g_{ij}$. Let $A^{ij} := E^{ij}(L)$, $B_I \equiv E_I(L) := \frac{\delta L}{\delta \varphi_I}$. Then each of the Lagrangians $B_I$ generates the same Euler-Lagrange tensor densities as does L, except they are differentiated with respect to $\varphi_I$.■

It should now be clear how to attempt to construct $k^{th}$ order multi-scalar-tensor field theories in a space of n-dimensions. We begin with the identity

$$E^{ij}(L)_{|j} = \tfrac{1}{2}\{\varphi_1{}^{i}E_1(L) + \ldots + \varphi_N{}^{i}E_N(L)\}$$

(N$\leq$n) and search for all $k^{th}$ order multi-scalar-tensor density concomitants ($A^{ij}$, $B_1$,. . . , $B_N$) which are such that $A^{ij}$ is symmetric and

$$A^{ij}{}_{|j} = \tfrac{1}{2}\{\varphi_1{}^{i}B_1 + \ldots +\varphi_N{}^{i}B_N\} \ . \tag{2.26}$$

(Note this decomposition can only be unique if N$\leq$n.) Once we have determined all of the conditions on $A^{ij}$ which guarantee that $A^{ij}{}_{|j}$ is of $k^{th}$ order and expressible as in (2.26), we know from **Lemma 6B** that the Lagrangians $B_I$ will yield $A^{ij}$, $B_1$, . . . , $B_N$ differentiated with respect to $\varphi_I$ as their Euler-Lagrange tensor densities, if $A^{ij}$, $B_1$,..., $B_N$ in fact arise from a variational principle. I shall now explain how to modify any one of the $B_I$ Lagrangians to yield a Lagrangian that yields $A^{ij}$, $B_1$,..., $B_N$ as its variational derivative.

Let's consider the second-order bi-scalar tensor problem we have been studying. Take the scalar density B. From (2.25) we see that if L generates B then B is the partial of L with respect to $\varphi$ plus a divergence. So in practice one would not want to use B as a Lagrangian since it has a large divergence sitting in it. Thus we define a new scalar density B' to be what we obtain from B when we drop every group of terms that clearly constitute a divergence and every term in which the coefficient function is not differentiated with respect to $\varphi$. We then define the Lagrangian $L_B$ to be the scalar density we get from B' when we replace every coefficient function in B' by the obvious indefinite integral of that function with respect to $\varphi$. So if (say) $\alpha_{\varphi\varphi}$ were a coefficient function in B' we would replace it by $\alpha_\varphi$ in $L_B$. This Lagrangian should yield $A^{ij}$, B and C as its Euler Lagrange tensor densities, if they in fact arise from a variational principle. I tried this out for the second-order scalar-tensor case and it worked yielding the usual Lagrangian of that theory.

However, there seems to be a "minor" faux pas in the construction I just described. What if the triple ($A^{ij}$,B,C) which arise from a variational principle is actually independent of explicit $\varphi$ and $\xi$ dependence. Then due to the argument presented in the proof of **Lemma 6A** all the variational derivatives of B would a vanish! But don't let that worry you. The argument I presented above is meant to be employed for the case of the most general second-order bi-scalar-tensor triple ($A^{ij}$,B,C) satisfying (2.2). And *if* the Lagrangian $L_B$ you obtain using it yields $A^{ij}$, B and C for the general case you are fine.

In view of what I just said you might expect me to now go to the B and C associated with the $A^{ij}$ of (2.22) to build the Lagrangians we need to finish the proof of the **Theorem**. But, alas, I have not been keeping track of B and C in my construction of $A^{ij}$. In Ohashi, *et al.*[3] they have been keeping track of B and C. They are the $Q_1$ and $Q_2$ appearing in (B.1) of their paper. I leave it to them, or to others, to use the method I just outlined above to provide a second proof of the **Theorem**.

I shall take a more "traditional" approach to constructing the Lagrangians we need, which proceeds as follows. In the purely metric case treated by Lovelock [13] he constructed all of the second-order, symmetric, divergence-free tensorial concomitants $A^{ij}$. Then to show that they arose from a variational principle he considered the Lagrangian $A^{ij}g_{ij}$, which worked perfectly. Why this Lagrangian should work was a mystery. Then when I was dealing with the case of second-order scalar-tensor field equations in a four-dimensional space, I too considered the Lagrangian $A^{ij}g_{ij}$, and with some effort, it also worked. So logically, Ohashi, *et al.*[3] also tried to use the same type of Lagrangian to finish their proof of the bi-scalar-tensor **Theorem**, and ran into numerous difficulties, expressing wonder why the Lagrangian $A^{ij}g_{ij}$ should even work. I shall now prove a lemma which explains why the Lagrangian $A^{ij}g_{ij}$ produces second-order field equations, and then go onto explain how we can use that knowledge to finish the proof of the **Theorem**.

**Lemma 7:** If the second-order, symmetric, bi-scalar-tensor tensor density concomitant $A^{ij}$ arises from a variational principal, and $A^{ij}{}_{|j}$ is of second-order, then the Euler-Lagrange tensor densities of $L_A := A^{ij}g_{ij}$ are of second-order. The same is also true for the Lagrangians $L_{A\varphi\varphi} := A^{ij}\varphi_i\varphi_j$; $L_{A\xi\xi} := A^{ij}\xi_i\xi_j$ and $L_{A\varphi\xi} := A^{ij}\varphi_i\xi_j$.

**Proof:** Let L be the Lagrangian that generates $A^{ij}$, so $E^{ij}(L) = A^{ij}$. To prove the **Lemma** I shall just compute what $E^{hk}(L_A)$ is equal to, and we shall see that it is the sum of second-order quantities. In a similar way $E_\varphi(L_A)$ and $E_\xi(L_A)$ can be shown to be such a sum.

To begin, since $L_A$ is of second-order

$$E^{hk}(L_A) = E^{hk}(g_{ij}A^{ij}) = \frac{\partial}{\partial g_{hk}}(g_{ij}A^{ij}) - \frac{d}{dx^l}\frac{\partial}{\partial g_{hk,l}}(g_{ij}A^{ij}) + \frac{d^2}{dx^m dx^l}\frac{\partial}{\partial g_{hk,lm}}(g_{ij}A^{ij})$$

$$= A^{hk} + g_{ij}A^{ij;hk} - g_{ij,l}A^{ij;hk,l} - g_{ij}\frac{d}{dx^l}A^{ij;hk,l} + g_{ij,lm}A^{ij;hk,lm} + 2g_{ij,l}\frac{d}{dx^m}A^{ij;hk,lm} + g_{ij}\frac{d^2}{dx^m dx^l}A^{ij;hk,lm} \qquad (2.27)$$

where I am using the notation introduced in (2.4). Upon regrouping terms in (2.27) we find

$$E^{hk}(L_A) = A^{hk} + g_{ij}E^{hk}(A^{ij}) - g_{ij,l}A^{ij;hk,l} + g_{ij,lm}A^{ij;hk,lm} + 2g_{ij,l}\frac{d}{dx^m}A^{ij;hk,lm} \quad . \qquad (2.28)$$

Thomas's Replacement Theorem, **Lemma 2**, along with (2.24), permit us to rewrite (2.28) as

$$E^{hk}(L_A) = A^{hk} + g_{ij}A^{hk;ij} + \tfrac{2}{3}R_{ilmj}A^{ij;hk,lm} ,$$

which clearly shows why $E^{hk}(L_A)$ is second-order. In this expression $A^{hk;ij}$ and $A^{ij;hk,lm}$ are computed by taking the usual partial derivative of $A^{hk}$ with respect to $g_{ij}$ and $g_{hk,lm}$, and then evaluating the result using **Lemma 2**.

It should be noted that in deriving this result I did not use the fact that $A^{ij}{}_{|j}$ is second-order, but could have to show that the last term on the right-hand side of (2.28) is actually of second-order, and not third due to Property S, and thereby obtain another proof of the desired result. However, to show that $E_\varphi(L_A)$ and $E_\xi(L_A)$ are second-order, the fact that $A^{ij}{}_{|j}$ is of second-order is required. I leave that simple proof to you.

The proof that $L_{A\varphi\varphi}$, $L_{A\xi\xi}$ and $L_{A\varphi\xi}$ yield second-order variational derivatives is similar.■

Ohashi, *et al.* [3] had hoped to use the Lagrangian $A^{ij}g_{ij}$, where $A^{ij}$ has the form given in **Lemma 4** with the coefficients determined by their version of **Lemma 5**, to finish the proof of the **Theorem**. I feel that such a "frontal assault" to obtain the coveted Lagrangian is "mathematical suicide." I shall use a different approach which I call "peeling the layers off of an onion." I introduced a version of this approach (although I did not give it that name) in Horndeski [5], where I solved the problem of constructing all of the conformally invariant ("flat space compatible") scalar-tensor fields theories in a space of 4-dimensions. There I first showed that all such field theories must have differential order less than 5. I then constructed the 4th order part of the field equations, followed by the 3rd order part and lastly the 2nd order part. We can do something analogous to that here.

(2.22) shows us that $A^{ij}$ is the sum of four blocks of terms. The first block $A_3{}^{ij}$ involves $\varphi_{ab}R_{cdef}$, $\xi_{ab}R_{cdef}$ and terms that are purely of third degree in the second derivatives of $\varphi$ and $\xi$. (Throughout what follows if I refer to the degree of a term I mean its algebraic degree in the second derivatives of $\varphi$ and $\xi$.) So suppose that I can construct a Lagrangian $L_3$ that yields precisely $A_3^{ij}$ along with some additional terms that fall into $A_2^{ij}$, $A_1^{ij}$ and $A_0^{ij}$. Then if L is the Lagrangian that yields $A^{ij}$, the Lagrangian $L - L_3$ will yield terms which are only of the form $A_2^{ij}+A_1^{ij}+A_0^{ij}$. Next suppose that I can produce a Lagrangian, $L_2$, that yields precisely $A_2^{ij}$, along with some terms that fall into $A_1^{ij}$ and $A_0^{ij}$. If that can be accomplished then the Lagrangian $L-L_3-L_2$ will produce an $A^{ij}$ that is at most of first degree and has the form of $A_1^{ij}+A_0^{ij}$. I shall then construct a Lagrangian $L_1$ that yields precisely $A_1^{ij}$, so that the Lagrangian $L-L_3-L_2-L_1$ yields that part of $A^{ij}$ that is of zeroth order. A Lagrangian $L_0$ will then be found to yield these zeroth degree terms so that the Lagrangian $L_3+L_2+L_1+L_0$ will be equivalent to the Lagrangian L that yields the most general $A^{ij}$ with the required properties. This will complete the proof of the **Theorem**. Without mincing any words, the calculations that we are about to perform "stink," so you can see why I call this approach "peeling the layers off of an onion."

Motivated by **Lemma 7** and the above remarks let us begin our quest for a Lagrangian by considering the Lagrangian $L_3$ obtained from (2.22b) by contracting $A_3^{ij}$ with $g_{ij}$. Doing so leads us to consider the Lagrangian

$$L_3 := g^{1/2}\{S_{31}\,\delta^{abe}_{cdf}\,R_{ab}{}^{cd}\varphi_e{}^f + S_{32}\,\delta^{abe}_{cdf}\,R_{ab}{}^{cd}\xi_e{}^f - {}^4/_3 S_{31X}\,\delta^{ace}_{bdf}\,\varphi_a{}^b\varphi_c{}^d\varphi_e{}^f - 2S_{31Z}\,\delta^{ace}_{bdf}\,\varphi_a{}^b\varphi_c{}^d\xi_e{}^f +$$

$$- 2S_{32Z}\,\delta^{ace}_{bdf}\,\varphi_a{}^b\xi_c{}^d\xi_e{}^f - {}^4/_3 S_{32Y}\,\delta^{ace}_{bdf}\,\xi_a{}^b\xi_c{}^d\xi_e{}^f\}\,, \qquad (2.29)$$

where $S_{31}$ and $S_{32}$ are scalar functions of $\varphi$, $\xi$, X, Y and Z such that $S_{31Z} = 2S_{32X}$ and $S_{32Z} = 2S_{31Y}$ so that the pair ($S_{31}$, $S_{32}$) forms a BSTCP of functions (*see,* (1.3)).

At this juncture one should note that the Lagrangian $L_3$ is invariant under the transformation:

$$\varphi_i \leftrightarrow \xi_i,\ \varphi_{ij} \leftrightarrow \xi_{ij},\ S_{31} \leftrightarrow S_{32},\ \partial_X \leftrightarrow \partial_Y,\ \partial_Z \leftrightarrow \partial_Z\,.$$

This observation will prove to be very useful when computing partial derivatives and Euler-Lagrange tensors densities. Similar remarks will apply to various other Lagrangians that will arise throughout this paper.

We desire to compute $E^{ij}(L_3)$. To that end we can use the work in my thesis (*see*, page 116 of [4]) to conclude that if L is a second-order bi-scalar-tensor Lagrangian then its Euler Lagrange tensor densities are given by:

$$E^{ij}(L) = \Pi^{ij}(L) - \Pi^{ij,k}(L)_{|k} + L^{;ij,hk}{}_{|hk}\,;\; E_\varphi(L) = L_\varphi - \Phi^i(L)_{|i} + L_\varphi{}^{;ij}{}_{|ij} \text{ and } E_\xi(L) = L_\xi - \Xi^i(L)_{|i} + L_\xi{}^{;ij}{}_{|ij} \tag{2.30}$$

where $\Pi^{ij}(L)$, $\Pi^{ij,k}(L)$, $\Phi^i(L)$ and $\Xi^i(L)$ are the tensorial derivatives of L with respect to $g_{ij}$, $g_{ij,k}$, $\varphi_i$ and $\xi_i$ respectively. If $T^{\cdots}$ is a second-order bi-scalar tensor concomitant then $\Phi^i(T^{\cdots})$ and $\Xi^i(T^{\cdots})$ are defined by

$$\Phi^i(T^{\cdots}) := T^{\cdots}{}_\varphi{}^{;i} + \Gamma^i{}_{ab}T^{\cdots}{}_\varphi{}^{;ab} \text{ and } \Xi^i(T^{\cdots}) := T^{\cdots}{}_\xi{}^{;i} + \Gamma^i{}_{ab}T^{\cdots}{}_\xi{}^{;ab}\ . \tag{2.31}$$

$\Phi^i$ and $\Xi^i$ satisfy the usual rules obeyed by differential operators and have the useful property that $\Phi^i(\varphi_{ab}) = \Phi^i(\xi_{ab}) = \Xi^i(\varphi_{ab}) = \Xi^i(\xi_{ab}) = 0$, which is quite handy when computing $\Phi^i(L_3)$ and $\Xi^i(L_3)$. I shall not bother to state the formal definitions of $\Pi^{ij}(T^{\cdots})$ and $\Pi^{ij,k}(T^{\cdots})$. It suffices to say that using the results in Sections 3 and 4 of Part I in [4] it can be shown that

$$\Pi^{ij}(L) = \tfrac{1}{3}R_k{}^j{}_{mh}L^{;hk,im} - R_k{}^i{}_{mh}L^{;hk,jm} - \tfrac{1}{2}\varphi^i\Phi^j(L) - \varphi^i{}_h L_\varphi{}^{;jh} - \tfrac{1}{2}\xi^i\Xi^j(L) - \xi^i{}_h L_\xi{}^{;jh} + \tfrac{1}{2}g^{ij}L \tag{2.32a}$$

$$\Pi^{ij,k}(L) = \tfrac{1}{2}(L_\varphi{}^{;ij}\varphi^k - L_\varphi{}^{;kj}\varphi^i - L_\varphi{}^{;ki}\varphi^j + L_\xi{}^{;ij}\xi^k - L_\xi{}^{;kj}\xi^i - L_\xi{}^{;ki}\xi^j)\ . \tag{2.32b}$$

Right now we do not know if the Euler-Lagrange tensors of $L_3$ are of second-order. So we shall begin by computing the HOT (:=higher order terms) of $E^{ij}(L_3)$, which are the terms of order $\geq 2$. To that end we require

$$L_3{}^{;ij,hk} = g^{1/2}\{S_{31}\,\delta^{abe}_{cdf}\,\frac{\partial R_{abpq}}{\partial g_{ij,hk}}\,g^{pc}g^{qd}\,\varphi_e{}^f + S_{32}\,\delta^{abe}_{cdf}\,\frac{\partial R_{abpq}}{\partial g_{ij,hk}}\,g^{pc}g^{qd}\xi_e{}^f\}\ , \tag{2.33}$$

where

$$R_{abpq} = \tfrac{1}{2}(g_{bp,aq} + g_{aq,bp} - g_{ap,bq} - g_{bq,ap}) + 1^{st} \text{ order terms}\ . \tag{2.34}$$

Using (2.33) and (2.34) we find that

$$L_3{}^{;ij,hk} = g^{1/2}\{S_{31}[\,\delta^{hie}_{cdf}\,g^{jc}g^{kd}\varphi_e{}^f + \delta^{hje}_{cdf}\,g^{ic}g^{kd}\varphi_e{}^f] + S_{32}[\,\delta^{hie}_{cdf}\,g^{jc}g^{kd}\xi_e{}^f + \delta^{hje}_{cdf}\,g^{ic}g^{kd}\xi_e{}^f]\}\ . \tag{2.35}$$

Similarly it can be shown that

$$L_{3\varphi}{}^{;ij} = g^{1/2}\{S_{31}\,\delta^{abi}_{cdf}\,g^{fj}R_{ab}{}^{cd} - 4S_{31X}\,\delta^{ice}_{bdf}\,g^{bj}\varphi_c{}^d\varphi_e{}^f - 4S_{31Z}\,\delta^{ice}_{bdf}\,g^{bj}\varphi_c{}^d\xi_e{}^f - 2S_{32Z}\,\delta^{ice}_{bdf}\,g^{bj}\xi_c{}^d\xi_e{}^f\} \tag{2.36a}$$

$$L_{3\xi}{}^{;ij} = g^{1/2}\{S_{32}\,\delta^{abi}_{cdf}\,g^{fj}R_{ab}{}^{cd} - 2S_{31Z}\,\delta^{ice}_{bdf}\,g^{bj}\varphi_c{}^d\varphi_e{}^f - 4S_{32Z}\,\delta^{ice}_{bdf}\,g^{bj}\varphi_c{}^d\xi_e{}^f - 4S_{32Y}\,\delta^{ice}_{bdf}\,g^{bj}\xi_c{}^d\xi_e{}^f\} \tag{2.36b}$$

$$\begin{aligned}\Phi^i(L_3) = g^{1/2}\{&2S_{31X}\varphi^i\,\delta^{abe}_{cdf}\,R_{ab}{}^{cd}\varphi_e{}^f + S_{31Z}\xi^i\,\delta^{abe}_{cdf}\,R_{ab}{}^{cd}\varphi_e{}^f + 2S_{32X}\varphi^i\,\delta^{abe}_{cdf}\,R_{ab}{}^{cd}\xi_e{}^f + S_{32Z}\xi^i\,\delta^{abe}_{cdf}\,R_{ab}{}^{cd}\xi_e{}^f + \\ &- {}^8/_3S_{31XX}\varphi^i\,\delta^{ace}_{bdf}\,\varphi_a{}^b\varphi_c{}^d\varphi_e{}^f - {}^4/_3S_{31XZ}\xi^i\,\delta^{ace}_{bdf}\,\varphi_a{}^b\varphi_c{}^d\varphi_e{}^f - 4S_{31ZX}\varphi^i\,\delta^{ace}_{bdf}\,\varphi_a{}^b\varphi_c{}^d\xi_e{}^f - 2S_{31ZZ}\xi^i\,\delta^{ace}_{bdf}\,\varphi_a{}^b\varphi_c{}^d\xi_e{}^f \\ &- 4S_{32ZX}\varphi^i\,\delta^{ace}_{bdf}\,\varphi_a{}^b\xi_c{}^d\xi_e{}^f - 2S_{32ZZ}\xi^i\,\delta^{ace}_{bdf}\,\varphi_a{}^b\xi_c{}^d\xi_e{}^f - {}^8/_3S_{32YX}\varphi^i\,\delta^{ace}_{bdf}\,\xi_a{}^b\xi_c{}^d\xi_e{}^f + \\ &- {}^4/_3S_{32YZ}\xi^i\,\delta^{ace}_{bdf}\,\xi_a{}^b\xi_c{}^d\xi_e{}^f\}\end{aligned} \tag{2.37a}$$

$$\begin{aligned}\Xi^i(L_3) = g^{1/2}\{&2S_{31Y}\xi^i\,\delta^{abe}_{cdf}\,R_{ab}{}^{cd}\varphi_e{}^f + S_{31Z}\varphi^i\,\delta^{abe}_{cdf}\,R_{ab}{}^{cd}\varphi_e{}^f + 2S_{32Y}\xi^i\,\delta^{abe}_{cdf}\,R_{ab}{}^{cd}\xi_e{}^f + S_{32Z}\varphi^i\,\delta^{abe}_{cdf}\,R_{ab}{}^{cd}\xi_e{}^f + \\ &- {}^8/_3S_{31XY}\xi^i\,\delta^{ace}_{bdf}\,\varphi_a{}^b\varphi_c{}^d\varphi_e{}^f - {}^4/_3S_{31XZ}\varphi^i\,\delta^{ace}_{bdf}\,\varphi_a{}^b\varphi_c{}^d\varphi_e{}^f - 4S_{31ZY}\xi^i\,\delta^{ace}_{bdf}\,\varphi_a{}^b\varphi_c{}^d\xi_e{}^f - 2S_{31ZZ}\varphi^i\,\delta^{ace}_{bdf}\,\varphi_a{}^b\varphi_c{}^d\xi_e{}^f \\ &- 4S_{32ZY}\xi^i\,\delta^{ace}_{bdf}\,\varphi_a{}^b\xi_c{}^d\xi_e{}^f - 2S_{32ZZ}\varphi^i\,\delta^{ace}_{bdf}\,\varphi_a{}^b\xi_c{}^d\xi_e{}^f - {}^8/_3S_{32YY}\xi^i\,\delta^{ace}_{bdf}\,\xi_a{}^b\xi_c{}^d\xi_e{}^f + \\ &- {}^4/_3S_{32YZ}\varphi^i\,\delta^{ace}_{bdf}\,\xi_a{}^b\xi_c{}^d\xi_e{}^f\}\ .\end{aligned} \tag{2.37b}$$

Using (2.30), (2.35) and (2.36) it is easy to see that the Euler-Lagrange tensor densities of $L_3$ have no fourth-order terms. With a bit more work it can be seen that the third-order terms involving $g_{ij}$ in $E^{ij}(L_3)$ vanish, while the terms that are of third-order in $\varphi$ generate second-order terms in $E^{ij}(L_3)$. So we know that $E^{ij}(L_3)$ is at most of second-order. A similar result applies to $E_\varphi(L_3)$ and $E_\xi(L_3)$. Consequently the triple $(E^{ij}(L_3), E_\varphi(L_3), E_\xi(L_3))$ satisfies the assumptions of **Lemma 4**, in which case $E^{ij}(L_3)$ must have the form presented in (2.22). Thus when computing $E^{ij}(L_3)$ we need only keep track of the terms involving $\varphi_{ab}R_{cdef}$ and $\xi_{ab}R_{cdef}$ since knowledge of them will provide us with the form of the $A_3{}^{ij}$ part of $E^{ij}(L_3)$. A lengthy calculation, involving the use of some dimensionally dependent identities Lovelock [18], such as

$$0 = \delta^{iabcd}_{lmnpq}\,g^{lj}\varphi_a\varphi^m\varphi_b{}^nR_{cd}{}^{pq},$$

shows that the $\varphi_{ab}R_{cdef}$ and $\xi_{ab}R_{cdef}$ terms in $E^{ij}(L_3)$ have the form presented in (2.22b) if $S_{31}$ and $S_{32}$ are chosen so that

$$\alpha_1 = -XS_{31X} - YS_{31Y} - ZS_{31Z} \quad \text{and} \quad \alpha_2 = -XS_{32X} - YS_{32Y} - ZS_{32Z} \ . \tag{2.38}$$

Note that since the pair $(S_{31},S_{32})$ is assumed to be a BST conjugate pair the pair $(\alpha_1,\alpha_2)$ given in (2.38) is also a BST conjugate pair, as it must be. The question now is given $\alpha_1$ and $\alpha_2$ can you solve the partial differential equations presented in (2.38) for $S_{31}$ and $S_{32}$? To see how this can be done I need to introduce some new notation. If $f=f(\varphi,\xi,X,Y,Z)$, then I shall define $f(t) := f(\varphi,\xi,tX,tY,tZ)$, where t is a real variable defined in a neighborhood of t=1. Note that

$$\frac{df(t)}{dt} = Xf_X(t) + Yf_Y(t) + Zf_Z(t).$$

Thus (2.38) implies that $\alpha_1$, $\alpha_2$, $S_{31}$ and $S_{32}$ are related by

$$\alpha_1(t) = -t\frac{dS_{31}(t)}{dt} \quad \text{and} \quad \alpha_2(t) = -t\frac{dS_{32}(t)}{dt} \ .$$

Upon taking an indefinite integral of these equations with respect to t and evaluating the result when t=1 we obtain

$$S_{31} = -\int t^{-1}\alpha_1(t)dt\Big|_{t=1} \quad \text{and} \quad S_{32} = -\int t^{-1}\alpha_2(t)dt\Big|_{t=1} \ . \tag{2.39}$$

Thus we have established

**Lemma 8:** The variational derivative with respect to $g_{ij}$ of the Lagrangian $L_3$ defined in (2.29), with the coefficient functions $S_{31}$ and $S_{32}$ being the BST conjugate pair $(S_{31},S_{32})$ given by (2.39), will yield the $A_3^{ij}$ term in the $A^{ij}$ of (2.22a), along other terms of the form $A_2^{ij}$, $A_1^{ij}$ and $A_0^{ij}$. ■

Consequently I have shown that if L is the Lagrangian that yields the $A^{ij}$ given in (2.22a), then $L - L_3$ will be devoid of third degree terms in $\varphi$ and $\xi$, and also of terms involving $\varphi_{ab}R_{cdef}$ and $\xi_{ab}R_{cdef}$. Thus upon varying $g_{ij}$ in $L-L_3$ only terms of the form appearing $A_2^{ij}$, $A_1^{ij}$ and $A_0^{ij}$ will appear.

Since the Lagrangian $\mathfrak{L}_3$ in (1.13) and the Lagrangian $L_3$ just constructed are such that $\mathfrak{L}_3 = -\frac{1}{4}L_3$, they are essentially the same Lagrangian for the purpose of the **Theorem**.

Our next challenge is to construct a Lagrangian that yields the $A_2^{ij}$ term appearing in (2.22c). For this case we are investigating the situation in which $\alpha_1 = \alpha_2 = 0$. So from **Lemma 5** we know that $\alpha_{31}$, $\alpha_{32}$, $\alpha_{33}$ and $\alpha_{34}$ are BST functions.

Upon examining the $g_{ij}$ contraction of the $A_2^{ij}$ term given in (2.22c) I concluded that a suitable scalar density for us to consider is

$$L_2 := g^{1/2}\{2S_{21}\,\delta^{ab}_{cd}\,R_{ab}{}^{cd} + S_{22}\,\delta^{abe}_{cdf}\,\varphi_e\varphi^f R_{ab}{}^{cd} + S_{23}\,\delta^{abe}_{cdf}\,\xi_e\xi^f R_{ab}{}^{cd} + 2S_{24}\,\delta^{abe}_{cdf}\,\varphi_e\xi^f R_{ab}{}^{cd} - 8S_{21X}\,\delta^{ac}_{bd}\,\varphi_a{}^b\varphi_c{}^d$$

$$- 4S_{22X}\,\delta^{ace}_{bdf}\,\varphi_a\varphi^b\varphi_c{}^d\varphi_e{}^f - 4S_{23X}\,\delta^{ace}_{bdf}\,\xi_a\xi^b\varphi_c{}^d\varphi_e{}^f - 4S_{24X}\,\delta^{ace}_{bdf}\,\varphi_a\xi^b\varphi_c{}^d\varphi_e{}^f - 4S_{24X}\,\delta^{ace}_{bdf}\,\xi_a\varphi^b\varphi_c{}^d\varphi_e{}^f +$$

$$- 8S_{21Z}\,\delta^{ac}_{bd}\,\varphi_a{}^b\xi_c{}^d - 4S_{22Z}\,\delta^{ace}_{bdf}\,\varphi_a\varphi^b\varphi_c{}^d\xi_e{}^f - 4S_{23Z}\,\delta^{ace}_{bdf}\,\xi_a\xi^b\varphi_c{}^d\xi_e{}^f - 4S_{24Z}\,\delta^{ace}_{bdf}\,\varphi_a\xi^b\varphi_c{}^d\xi_e{}^f +$$

$$- 4S_{24Z}\,\delta^{ace}_{bdf}\,\xi_a\varphi^b\varphi_c{}^d\xi_e{}^f - 8S_{21Y}\,\delta^{ac}_{bd}\,\xi_a{}^b\xi_c{}^d - 4S_{22Y}\,\delta^{ace}_{bdf}\,\varphi_a\varphi^b\xi_c{}^d\xi_e{}^f - 4S_{23Y}\,\delta^{ace}_{bdf}\,\xi_a\xi^b\xi_c{}^d\xi_e{}^f +$$

$$- 4S_{24Y}\,\delta^{ace}_{bdf}\,\varphi_a\xi^b\xi_c{}^d\xi_e{}^f - 4S_{24Y}\,\delta^{ace}_{bdf}\,\xi_a\varphi^b\xi_c{}^d\xi_e{}^f\} \qquad (2.40)$$

where $S_{21}$, $S_{22}$, $S_{23}$ and $S_{24}$ are BST functions.

So I want to demonstrate that the higher order terms (=:HOT) involving $R_{abcd}$ in $E^{ij}(L_2)$ are expressible in the form

$$\mathrm{HOT}[E^{ij}(L_2)] =$$

$$= g^{1/2}\{\alpha_{31}\,\delta^{abi}_{cdq}\,g^{qj}R_{ab}{}^{cd} + \alpha_{32}\,\delta^{abei}_{cdfq}\,g^{qj}\varphi_e\varphi^f R_{ab}{}^{cd} + \alpha_{33}\,\delta^{abei}_{cdfq}\,g^{qj}\xi_e\xi^f R_{ab}{}^{cd} + \alpha_{34}\,\delta^{abei}_{cdfq}\,g^{qj}(\varphi_e\xi^f + \xi_e\varphi^f)\,R_{ab}{}^{cd}\}, \qquad (2.41)$$

for a suitable choice of the $S_{..}$'s, where $\alpha_{31}$, $\alpha_{32}$, $\alpha_{33}$ and $\alpha_{34}$ satisfy the conditions presented in **Lemma 5** when $\alpha_1 = \alpha_2 = 0$.

Using (2.34) and the properties of the tensorial derivative operators $\Phi^i$ and $\Xi^i$ stated after (2.31) we find that

$$L_2{}^{;ij,hk} = g^{1/2}\{S_{21}[\,\delta^{ih}_{qr}\,g^{qk}g^{rj} + \delta^{jh}_{qr}\,g^{qk}g^{ri} + \delta^{ik}_{qr}\,g^{qh}g^{rj} + \delta^{jk}_{qr}\,g^{qh}g^{ri}] +$$

$$+ \tfrac{1}{2}S_{22}\,[\,\delta^{aih}_{pqr}\,\varphi_a\varphi^p g^{qk}g^{rj} + \delta^{ajh}_{pqr}\,\varphi_a\varphi^p g^{qk}g^{ri} + \delta^{aik}_{pqr}\,\varphi_a\varphi^p g^{qh}g^{rj} + \delta^{ajk}_{pqr}\,\varphi_a\varphi^p g^{qh}g^{ri}] +$$

$$+ \tfrac{1}{2}S_{23}\,[\,\delta^{aih}_{pqr}\,\xi_a\xi^p g^{qk}g^{rj} + \delta^{ajh}_{pqr}\,\xi_a\xi^p g^{qk}g^{ri} + \delta^{aik}_{pqr}\,\xi_a\xi^p g^{qh}g^{rj} + \delta^{ajk}_{pqr}\,\xi_a\xi^p g^{qh}g^{ri}] +$$

$$+ S_{24}\,[\,\delta^{aih}_{pqr}\,\varphi_a\xi^p g^{qk}g^{rj} + \delta^{ajh}_{pqr}\,\varphi_a\xi^p g^{qk}g^{ri} + \delta^{aik}_{pqr}\,\varphi_a\xi^p g^{qh}g^{rj} + \delta^{ajk}_{pqr}\,\varphi_a\xi^p g^{qh}g^{ri}]\}\,, \qquad (2.42)$$

$$L_{2\varphi}{}^{;ij} = -4g^{\frac{1}{2}}\{4S_{21X}[\Box\varphi g^{ij} - \varphi^{ij}] + 2S_{22X}\,\delta^{aie}_{bdf}\,\varphi_a\varphi^b g^{dj}\varphi_e{}^f + 2S_{23X}\,\delta^{aie}_{bdf}\,\xi_a\xi^b g^{dj}\varphi_e{}^f + 2S_{24X}\,\delta^{aie}_{bdf}\,\varphi_a\xi^b g^{dj}\varphi_e{}^f +$$

$$+ 2S_{24X}\,\delta^{aie}_{bdf}\,\xi_a\varphi^b g^{dj}\varphi_e{}^f + S_{21Z}[\Box\xi g^{ij} - \xi^{ij}] + S_{22Z}\,\delta^{aie}_{bdf}\,\varphi_a\varphi^b g^{dj}\xi_e{}^f + S_{23Z}\,\delta^{aie}_{bdf}\,\xi_a\xi^b g^{dj}\varphi_e{}^f +$$

$$+ S_{24Z}\,\delta^{aie}_{bdf}\,\varphi_a\xi^b g^{dj}\varphi_e{}^f + S_{24Z}\,\delta^{aie}_{bdf}\,\xi_a\varphi^b g^{dj}\varphi_e{}^f\,\} \qquad (2.43)$$

$$L_{2\xi}{}^{;ij} = -4g^{\frac{1}{2}}\{4S_{21y}[\Box\xi g^{ij} - \xi^{ij}] + 2S_{22Y}\,\delta^{aie}_{bdf}\,\varphi_a\varphi^b g^{dj}\xi_e{}^f + 2S_{23Y}\,\delta^{aie}_{bdf}\,\xi_a\xi^b g^{dj}\xi_e{}^f ++ 2S_{24Y}\,\delta^{aie}_{bdf}\,\varphi_a\xi^b g^{dj}\xi_e{}^f +$$

$$+ 2S_{24Y}\,\delta^{aie}_{bdf}\,\xi_a\varphi^b g^{dj}\xi_e{}^f + S_{21Z}[\Box\varphi g^{ij} - \varphi^{ij}] + S_{22Z}\,\delta^{aie}_{bdf}\,\varphi_a\varphi^b g^{dj}\xi_e{}^f + S_{23Z}\,\delta^{aie}_{bdf}\,\xi_a\xi^b g^{dj}\varphi_e{}^f +$$

$$+ S_{24Z}\,\delta^{aie}_{bdf}\,\varphi_a\xi^b g^{dj}\xi_e{}^f + S_{24Z}\,\delta^{aie}_{bdf}\,\xi_a\varphi^b g^{dj}\xi_e{}^f\,\}\,, \qquad (2.44)$$

$$\Phi^i(L_2) = g^{\frac{1}{2}}\{4S_{21X}\,\varphi^i\,\delta^{ab}_{cd}\,R_{ab}{}^{cd} + 2S_{21Z}\,\xi^i\,\delta^{ab}_{cd}\,R_{ab}{}^{cd} + 2S_{22X}\varphi^i\,\delta^{abe}_{cdf}\,\varphi_e\varphi^f R_{ab}{}^{cd} + S_{22Z}\xi^i\,\delta^{abe}_{cdf}\,\varphi_e\varphi^f R_{ab}{}^{cd} +$$

$$+ S_{22}\,\delta^{abi}_{cdf}\,\varphi^f R_{ab}{}^{cd} + S_{22}\,\delta^{abe}_{cdf}\,\varphi_e g^{fi} R_{ab}{}^{cd} + 2S_{23X}\varphi^i\,\delta^{abe}_{cdf}\,\xi_e\xi^f R_{ab}{}^{cd} + S_{23Z}\xi^i\,\delta^{abe}_{cdf}\,\xi_e\xi^f R_{ab}{}^{cd} +$$

$$+ 4S_{24X}\varphi^i\,\delta^{abe}_{cdf}\,\varphi_e\xi^f R_{ab}{}^{cd} + 2S_{24Z}\,\xi^i\,\delta^{abe}_{cdf}\,\varphi_e\xi^f R_{ab}{}^{cd} + 2S_{24}\,\delta^{abi}_{cdf}\,\xi^f R_{ab}{}^{cd} - 16S_{21XX}\varphi^i\,\delta^{ac}_{bd}\,\varphi_a{}^b\varphi_c{}^d +$$

$$- 8S_{21XZ}\,\xi^i\,\delta^{ac}_{bd}\,\varphi_a{}^b\varphi_c{}^d - 8S_{22XX}\varphi^i\,\delta^{ace}_{bdf}\,\varphi_a\varphi^b\varphi_c{}^d\varphi_e{}^f - 4S_{22XZ}\,\xi^i\,\delta^{ace}_{bdf}\,\varphi_a\varphi^b\varphi_c{}^d\varphi_e{}^f - 4S_{22X}\,\delta^{ice}_{bdf}\,\varphi^b\varphi_c{}^d\varphi_e{}^f +$$

$$- 4S_{22X}\,\delta^{ace}_{bdf}\,\varphi_a g^{bi}\varphi_c{}^d\varphi_e{}^f - 8S_{23XX}\varphi^i\,\delta^{ace}_{bdf}\,\xi_a\xi^b\varphi_c{}^d\varphi_e{}^f - 4S_{23XZ}\xi^i\,\delta^{ace}_{bdf}\,\xi_a\xi^b\varphi_c{}^d\varphi_e{}^f - 8S_{24XX}\varphi^i\,\delta^{ace}_{bdf}\,\varphi_a\xi^b\varphi_c{}^d\varphi_e{}^f +$$

$$- 4S_{24XZ}\xi^i\,\delta^{ace}_{bdf}\,\varphi_a\xi^b\varphi_c{}^d\xi_e{}^f - 4S_{24X}\,\delta^{ice}_{bdf}\,\xi^b\varphi_c{}^d\varphi_e{}^f - 8S_{24XX}\varphi^i\,\delta^{ace}_{bdf}\,\xi_a\varphi^b\varphi_c{}^d\varphi_e{}^f - 4S_{24XZ}\xi^i\,\delta^{ace}_{bdf}\,\xi_a\varphi^b\varphi_c{}^d\varphi_e{}^f +$$

$$-4S_{24X}\,\delta^{ace}_{bdf}\,\xi_a g^{bi}\varphi_c{}^d\varphi_e{}^f - 16S_{21ZX}\varphi^i\,\delta^{ac}_{bd}\,\varphi_a{}^b\xi_c{}^d - 8S_{21ZZ}\xi^i\,\delta^{ac}_{bd}\,\varphi_a{}^b\xi_c{}^d - 8S_{22XZ}\varphi^i\,\delta^{ace}_{bdf}\,\varphi_a\varphi^b\varphi_c{}^d\xi_e{}^f +$$

$$-4S_{22ZZ}\xi^i\,\delta^{ace}_{bdf}\,\varphi_a\varphi^b\varphi_c{}^d\xi_{ef} - 4S_{22Z}\,\delta^{ice}_{bdf}\,\varphi^b\varphi_c{}^d\xi_e{}^f - 4S_{22Z}\,\delta^{ace}_{bdf}\,\varphi_a g^{bi}\varphi_c{}^d\xi_e{}^f - 8S_{23ZX}\varphi^i\,\delta^{ace}_{bdf}\,\xi_a\xi^b\varphi_c{}^d\xi_e{}^f +$$

$$-4S_{23ZZ}\xi^i\,\delta^{ace}_{bdf}\,\xi_a\xi^b\varphi_c{}^d\xi_e{}^f - 8S_{24ZX}\varphi^i\,\delta^{ace}_{bdf}\,\varphi_a\xi^b\varphi_c{}^d\xi_e{}^f - 4S_{24ZZ}\,\xi^i\,\delta^{ace}_{bdf}\,\varphi_a\xi^b\varphi_c{}^d\xi_e{}^f - 4S_{24Z}\,\delta^{ice}_{bdf}\,\xi^b\varphi_c{}^d\xi_e{}^f +$$

$$-8S_{24ZX}\varphi^i\,\delta^{ace}_{bdf}\,\xi_a\varphi^b\varphi_c{}^d\xi_e{}^f - 4S_{24ZZ}\xi^i\,\delta^{ace}_{bdf}\,\xi_a\varphi^b\varphi_c{}^d\xi_e{}^f - 4S_{24Z}\,\delta^{ace}_{bdf}\,\xi_a g^{bi}\varphi_c{}^d\xi_e{}^f - 16S_{21YX}\,\varphi^i\,\delta^{ac}_{bd}\,\xi_a{}^b\xi_c{}^d +$$

$$-8S_{21YZ}\xi^{i}\,\delta^{ac}_{bd}\,\xi_{a}{}^{b}\xi_{c}{}^{d}\;-8S_{22YX}\varphi^{i}\,\delta^{ace}_{bdf}\,\varphi_{a}\varphi^{b}\xi_{c}{}^{d}\xi_{e}{}^{f}\;-4S_{22YZ}\,\xi^{i}\,\delta^{ace}_{bdf}\,\varphi_{a}\varphi^{b}\xi_{c}{}^{d}\xi_{e}{}^{f}\;-\;4S_{22Y}\,\delta^{ice}_{bdf}\,\varphi^{b}\xi_{c}{}^{d}\xi_{e}{}^{f}\;+$$

$$-4S_{22Y}\,\delta^{ace}_{bdf}\,\varphi_{a}g^{bi}\xi_{c}{}^{d}\xi_{e}{}^{f}-8S_{23YX}\varphi^{i}\,\delta^{ace}_{bdf}\,\xi_{a}\xi^{b}\xi_{c}{}^{d}\xi_{e}{}^{f}-4S_{23YZ}\xi^{i}\,\delta^{ace}_{bdf}\,\xi_{a}\xi^{b}\xi_{c}{}^{d}\xi_{e}{}^{f}-8S_{24YX}\varphi^{i}\,\delta^{ace}_{bdf}\,\varphi_{a}\xi^{b}\xi_{c}{}^{d}\xi_{e}{}^{f}+$$

$$-4S_{24YZ}\xi^{i}\,\delta^{ace}_{bdf}\,\varphi_{a}\xi^{b}\xi_{c}{}^{d}\xi_{e}{}^{f}-4S_{24Y}\,\delta^{ice}_{bdf}\,\xi^{b}\xi_{c}{}^{d}\xi_{e}{}^{f}-8S_{24YX}\,\varphi^{i}\,\delta^{ace}_{bdf}\,\xi_{a}\varphi^{b}\xi_{cd}\xi_{e}{}^{f}-4S_{24YZ}\xi^{i}\,\delta^{ace}_{bdf}\,\xi_{a}\varphi^{b}\xi_{c}{}^{d}\xi_{e}{}^{f}+$$

$$-4S_{24Y}\,\delta^{ace}_{bdf}\,\xi_{a}g^{bi}\xi_{c}{}^{d}\xi_{e}{}^{f}\},\qquad(2.45)$$

$$\Xi^{i}(L_{2})=g^{\frac{1}{2}}\{4S_{21Y}\,\xi^{i}\,\delta^{ab}_{cd}\,R_{ab}{}^{cd}+2S_{21Z}\,\varphi^{i}\,\delta^{ab}_{cd}\,R_{ab}{}^{cd}+2S_{23Y}\xi^{i}\,\delta^{abe}_{cdf}\,\varphi_{e}\varphi^{f}R_{ab}{}^{cd}+S_{23Z}\varphi^{i}\,\delta^{abe}_{cdf}\,\xi_{e}\xi^{f}R_{ab}{}^{cd}+$$

$$+\;S_{23}\,\delta^{abi}_{cdf}\,\xi^{f}R_{ab}{}^{cd}+S_{23}\,\delta^{abe}_{cdf}\,\xi_{e}g^{fi}R_{ab}{}^{cd}+2S_{22Y}\xi^{i}\,\delta^{abe}_{cdf}\,\varphi_{e}\varphi^{f}R_{ab}{}^{cd}+S_{22Z}\varphi^{i}\,\delta^{abe}_{cdf}\,\varphi_{e}\varphi^{f}R_{ab}{}^{cd}+$$

$$+4S_{24Y}\xi^{i}\,\delta^{abe}_{cdf}\,\xi_{e}\varphi^{f}R_{ab}{}^{cd}+2S_{24Z}\,\varphi^{i}\,\delta^{abe}_{cdf}\,\xi_{e}\varphi^{f}R_{ab}{}^{cd}+2S_{24}\,\delta^{abi}_{cdf}\,\varphi^{f}R_{ab}{}^{cd}-16S_{21YY}\xi^{i}\,\delta^{ac}_{bd}\,\xi_{a}{}^{b}\xi_{c}{}^{d}+$$

$$-\;8S_{21YZ}\,\varphi^{i}\,\delta^{ac}_{bd}\,\xi_{a}{}^{b}\xi_{c}{}^{d}-8S_{23YY}\xi^{i}\,\delta^{ace}_{bdf}\,\xi_{a}\xi^{b}\xi_{c}{}^{d}\xi_{e}{}^{f}-4S_{23YZ}\,\varphi^{i}\,\delta^{ace}_{bdf}\,\xi_{a}\xi^{b}\xi_{c}{}^{d}\xi_{e}{}^{f}-4S_{23Y}\,\delta^{ice}_{bdf}\,\xi^{b}\xi_{c}{}^{d}\xi_{e}{}^{f}+$$

$$-\;4S_{23Y}\,\delta^{ace}_{bdf}\,\xi_{a}g^{bi}\,\xi_{c}{}^{d}\xi_{e}{}^{f}-8S_{22YY}\xi^{i}\,\delta^{ace}_{bdf}\,\varphi_{a}\varphi^{b}\xi_{c}{}^{d}\xi_{e}{}^{f}-4S_{22YZ}\varphi^{i}\,\delta^{ace}_{bdf}\,\varphi_{a}\varphi^{b}\xi_{c}{}^{d}\xi_{e}{}^{f}-8S_{24YY}\xi^{i}\,\delta^{ace}_{bdf}\,\xi_{a}\varphi^{b}\xi_{c}{}^{d}\xi_{e}{}^{f}+$$

$$-\;4S_{24YZ}\varphi^{i}\,\delta^{ace}_{bdf}\,\xi_{a}\varphi^{b}\xi_{c}{}^{d}\xi_{e}{}^{f}-4S_{24Y}\,\delta^{ice}_{bdf}\,\varphi^{b}\xi_{c}{}^{d}\xi_{e}{}^{f}-8S_{24YY}\xi^{i}\,\delta^{ace}_{bdf}\,\varphi_{a}\xi^{b}\xi_{cd}\xi_{e}{}^{f}-4S_{24YZ}\varphi^{i}\,\delta^{ace}_{bdf}\,\varphi_{a}\xi^{b}\xi_{c}{}^{d}\xi_{e}{}^{f}+$$

$$-4S_{24Y}\,\delta^{ace}_{bdf}\,\varphi_{a}g^{bi}\xi_{c}{}^{d}\xi_{e}{}^{f}-16S_{21ZY}\xi^{i}\,\delta^{ac}_{bd}\,\varphi_{a}{}^{b}\xi_{c}{}^{d}-8S_{21ZZ}\varphi^{i}\,\delta^{ac}_{bd}\,\varphi_{a}{}^{b}\xi_{c}{}^{d}-8S_{23YZ}\xi^{i}\,\delta^{ace}_{bdf}\,\xi_{a}\xi^{b}\varphi_{c}{}^{d}\xi_{e}{}^{f}+$$

$$-4S_{23ZZ}\varphi^{i}\,\delta^{ace}_{bdf}\,\xi_{a}\xi^{b}\varphi_{c}{}^{d}\xi_{e}{}^{f}-4S_{23Z}\,\delta^{ice}_{bdf}\,\xi^{b}\varphi_{c}{}^{d}\xi_{e}{}^{f}-4S_{23Z}\,\delta^{ace}_{bdf}\,\xi_{a}g^{bi}\varphi_{c}{}^{d}\xi_{e}{}^{f}-8S_{22ZX}\xi^{i}\,\delta^{ace}_{bdf}\,\varphi_{a}\varphi^{b}\varphi_{c}{}^{d}\xi_{e}{}^{f}+$$

$$-4S_{22ZZ}\varphi^{i}\,\delta^{ace}_{bdf}\,\varphi_{a}\varphi^{b}\varphi_{c}{}^{d}\xi_{e}{}^{f}-8S_{24ZY}\xi^{i}\,\delta^{ace}_{bdf}\,\xi_{a}\varphi^{b}\varphi_{c}{}^{d}\xi_{e}{}^{f}-4S_{24ZZ}\,\varphi^{i}\,\delta^{ace}_{bdf}\,\xi_{a}\varphi^{b}\varphi_{c}{}^{d}\xi_{e}{}^{f}-4S_{24Z}\,\delta^{ice}_{bdf}\,\varphi^{b}\varphi_{c}{}^{d}\xi_{e}{}^{f}+$$

$$-8S_{24ZY}\xi^{i}\,\delta^{ace}_{bdf}\,\varphi_{a}\xi^{b}\varphi_{c}{}^{d}\xi_{e}{}^{f}-4S_{24ZZ}\varphi^{i}\,\delta^{ace}_{bdf}\,\varphi_{a}\xi^{b}\varphi_{c}{}^{d}\xi_{ef}-4S_{24Z}\,\delta^{ace}_{bdf}\,\varphi_{a}g^{bi}\varphi_{c}{}^{d}\xi_{e}{}^{f}-16S_{21YX}\,\xi^{i}\,\delta^{ac}_{bd}\,\varphi_{a}{}^{b}\varphi_{c}{}^{d}+$$

$$-8S_{21XZ}\varphi^{i}\,\delta^{ac}_{bd}\,\varphi_{a}{}^{b}\varphi_{c}{}^{d}\;-8S_{23YX}\xi^{i}\,\delta^{ace}_{bdf}\,\xi_{a}\xi^{b}\varphi_{c}{}^{d}\varphi_{e}{}^{f}\;-4S_{23XZ}\varphi^{i}\,\delta^{ace}_{bdf}\,\xi_{a}\xi^{b}\varphi_{c}{}^{d}\varphi_{e}{}^{f}\;-\;4S_{23X}\,\delta^{ice}_{bdf}\,\xi^{b}\varphi_{c}{}^{d}\varphi_{e}{}^{f}\;+$$

$$-4S_{23X}\,\delta^{ace}_{bdf}\,\xi_{a}g^{bi}\varphi_{c}{}^{d}\varphi_{e}{}^{f}-8S_{22YX}\xi^{i}\,\delta^{ace}_{bdf}\,\xi_{a}\varphi^{b}\varphi_{c}{}^{d}\varphi_{e}{}^{f}-4S_{22XZ}\varphi^{i}\,\delta^{ace}_{bdf}\,\varphi_{a}\varphi^{b}\varphi_{c}{}^{d}\varphi_{e}{}^{f}-8S_{24YX}\xi^{i}\,\delta^{ace}_{bdf}\,\xi_{a}\varphi^{b}\varphi_{c}{}^{d}\varphi_{e}{}^{f}+$$

$$-4S_{24XZ}\varphi^{i}\,\delta^{ace}_{bdf}\,\xi_{a}\varphi^{b}\varphi_{c}{}^{d}\varphi_{e}{}^{f}-4S_{24X}\,\delta^{ice}_{bdf}\,\varphi^{b}\varphi_{c}{}^{d}\varphi_{e}{}^{f}-8S_{24YX}\,\xi^{i}\,\delta^{ace}_{bdf}\,\varphi_{a}\xi^{b}\varphi_{c}{}^{d}\varphi_{e}{}^{f}-4S_{24XZ}\xi^{i}\,\delta^{ace}_{bdf}\,\varphi_{a}\xi^{b}\varphi_{c}{}^{d}\varphi_{e}{}^{f}+$$

$$-4S_{24X}\,\delta^{ace}_{bdf}\,\varphi_{a}g^{bi}\varphi_{c}{}^{d}\varphi_{e}{}^{f}\}.\qquad(2.46)$$

In deriving the above expressions I made use of the fact that $L_2$ is invariant under the transformation that interchanges $\varphi$ and $\xi$ in $L_2$ when $S_{21}$ and $S_{24}$ are left fixed while $S_{22}$ and $S_{23}$ are interchanged.

Using (2.43) and (2.45) in (2.30) it can be shown that since $S_{21}$, $S_{22}$, $S_{23}$ and $S_{24}$ are BST functions $E_\varphi(L_2)$ is in fact of second-order. Consequently $E_\xi(L_2)$ is also of second-order since it can be obtained from $E_\varphi(L_2)$ by the "$\varphi\leftrightarrow\xi$" transformation of $L_2$ mentioned above.

An unconscionably lengthy calculation using (2.30) and (2.42)-(2.46) along with repeated use of the identity

$$0=\tfrac{1}{2}\varphi\,\delta^{iabcd}_{npqrs}\,g^{nj}\varphi_a\xi_b\varphi^p\xi^q R_{cd}{}^{rs} = -2(XY-Z^2)G^{ij} - X[2g^{ij}\xi_a\xi_b R^{ab} + \xi^i\xi^j R - 2\xi^i\xi_a R^{aj} - 2\xi^j\xi_a R^{ai} + 2\xi_a\xi_b R^{aijb}] +$$

$$-Y[2g^{ij}\varphi_a\varphi_b R^{ab} + \varphi^i\varphi^j R - 2\varphi^i\varphi_a R^{aj} - 2\varphi^j\varphi_a R^{ai} + 2\varphi_a\varphi_b R^{aijb}] + Z[4g^{ij}\varphi_a\xi_b R^{ab} + (\varphi^i\xi^j+\varphi^j\xi^i)R +$$

$$-2\varphi^i\xi_a R^{aj} - 2\varphi^j\xi_a R^{ai} - 2\xi^i\varphi_a R^{aj} - 2\xi^j\varphi_a R^{ai} + 2\varphi_a\xi_b R^{aijb} + 2\varphi_a\xi_b R^{ajib}] +$$

$$+2g^{ij}\varphi_a\xi_b\varphi_c\xi_d R^{abcd} + 2\varphi_a\xi_b\xi_c R^{abci}\varphi^j + 2\varphi_a\xi_b\xi_c R^{abcj}\varphi^i + 2\xi_a\varphi_b\varphi_c R^{abci}\xi^j + 2\xi_a\varphi_b\varphi_c R^{abcj}\xi^i +$$

$$+2\varphi^i\varphi^j\xi_a\xi_b R^{ab} + 2\xi^i\xi^j\varphi_a\varphi_b R^{ab} - 2(\varphi^i\xi^j+\varphi^j\xi^i)\varphi_a\xi_b R^{ab}\,, \tag{2.47}$$

shows that $E^{ij}(L_2)$ is also of second-order and the terms in it involving $R_{abcd}$ will have the form given in (2.41) provided $S_{21}$, $S_{22}$, $S_{23}$ and $S_{24}$ are chosen so that

$$\alpha_{31} = -[-S_{21}+2XS_{21X}+2YS_{21Y}+2ZS_{21Z}] - [\tfrac{1}{2}XS_{22}+X^2S_{22X}+Z^2S_{22Y}+XZS_{22Z}] +$$

$$-[\tfrac{1}{2}YS_{23}+Z^2S_{23X}+Y^2S_{23Y}+YZS_{23Z}] - [ZS_{24}+2XZS_{24X}+2YZS_{24Y}+(XY+Z^2)S_{24Z}] \tag{2.48a}$$

$$\alpha_{32} = S_{22}+2S_{21X}+XS_{22X}-YS_{22Y}+2ZS_{24X}+YS_{24Z} \tag{2.48b}$$

$$\alpha_{33} = S_{23}+2S_{21Y}+YS_{23Y}-XS_{23X}+XS_{24Z}+2ZS_{24Y} \tag{2.48c}$$

$$\alpha_{34} = S_{24}+S_{21Z}+\tfrac{1}{2}XS_{22Z}+\tfrac{1}{2}YS_{23Z}+ZS_{22Y}+ZS_{23X}. \tag{2.48d}$$

The system of first order partial differential equations presented in (2.48) is considerably more difficult then the one we had to deal with in (2.38), which arose when we were considering the Lagrangian $L_3$ and the terms in $A_3{}^{ij}$. Nevertheless I shall demonstrate how they can be solved in general, and this will also be done for a particular example of interest in Appendix **D**.

Before trying to solve the system given in (2.48) the first thing one should do is check that when $\alpha_{31}$, $\alpha_{32,}$ $\alpha_{33}$ and $\alpha_{34}$ are given by (2.48), with $S_{21}$, $S_{22}$, $S_{23}$ and $S_{24}$ being BST functions, then they satisfy all of the conditions presented in **Lemma 5** when $\alpha_1 = \alpha_2 = 0$. A straightforward calculation shows that they do.

To solve the system (2.48) we note is that (2.48b)-(2.48d) can be solved for $S_{21X}$, $S_{21Y}$ and $S_{21Z}$ in terms of quantities which do not involve $S_{21}$. If we take these solutions and plug them into (2.48a) we discover that $S_{21}$ is given by

$$S_{21} = [\alpha_{31}+X\alpha_{32}+Y\alpha_{33}+2Z\alpha_{34}] - \tfrac{1}{2}XS_{22} - \tfrac{1}{2}YS_{23} - ZS_{24} + (XY-Z^2)[S_{22Y}+S_{23X}-S_{24Z}] . \tag{2.49}$$

Next we use (2.49) to compute $S_{21X}$, $S_{21Y}$, $S_{21Z}$ and insert these expressions back into (2.48b)-(2.48d). The end result is that if we define

$$S:= S_{22Y} + S_{23X} - S_{24Z}, \;\; D:=XY-Z^2, \;\; \sigma_1:= -2[\tfrac{1}{2}\alpha_{32}+\alpha_{31X}+X\alpha_{32X}+Y\alpha_{33X}+2Z\alpha_{34X}], \tag{2.50a}$$

$$\sigma_2:=-2[\tfrac{1}{2}\alpha_{33}+\alpha_{31Y}+X\alpha_{32Y}+Y\alpha_{33Y}+2Z\alpha_{34Y}], \text{ and } \sigma_3:=-[\alpha_{34}+\alpha_{31Z}+X\alpha_{32Z}+Y\alpha_{33Z}+2Z\alpha_{34Z}] \tag{2.50b}$$

then (2.48b)-(2.48d) become

$$\sigma_1 = YS + 2DS_X , \;\; \sigma_2 = XS + 2DS_Y \text{ and } \sigma_3 = -ZS + DS_Z . \tag{2.51}$$

It is very important to note that the three $\sigma$ functions depend only on the values of $\alpha_{31}$, $\alpha_{32}$, $\alpha_{33}$, $\alpha_{34}$ and *not* on $S_{21}$, $S_{22}$, $S_{23}$ or $S_{24}$. So we have reduced our system of four first order partial differential equations for $S_{21}$, $S_{22}$, $S_{23}$ and $S_{24}$ to an equivalent system of three second-order partial differential equations for the BST functions $S_{22}$, $S_{23}$ and $S_{24}$, which we shall now solve.

Upon multiplying the first equation in (2.51) by X, the second equation by Y and the third equation by 2Z and then adding the three equations together we find that

$$\sigma = S + XS_X + YS_Y + ZS_Z \tag{2.52a}$$

where

$$\sigma := \tfrac{1}{2} D^{-1} [X\sigma_1 + Y\sigma_2 + 2Z\sigma_3]. \tag{2.52b}$$

When (2.52a) is "t-d" in the usual way, as we did when studying $L_3$, it becomes

$$\sigma(t) = \frac{d(tS(t))}{dt}.$$

Taking an indefinite integral of the above equation and evaluating it for t=1 we obtain

$$S \equiv S_{22Y} + S_{23X} - S_{24Z} = \int \sigma(t)dt \Big|_{t=1} . \tag{2.53}$$

However at this point it is not really clear why the solution of (2.52a) should also be a solution to each of the three individual equations in (2.51). To see why it is, first note that the equations of (2.51) can be rewritten equivalently as

$$\tfrac{1}{2}D^{-1/2}\sigma_1 = (D^{1/2}S)_X \ , \ \tfrac{1}{2}D^{-1/2}\sigma_2 = (D^{1/2}S)_Y \text{ and } D^{-1/2}\sigma_3 = (D^{1/2}S)_Z \ . \tag{2.54}$$

This simple system of first order partial differential equations for $D^{1/2}S$ will be solvable if the integrability conditions $(D^{1/2}S)_{XY} = (D^{1/2}S)_{YX}$, $(D^{1/2}S)_{XZ} = (D^{1/2}S)_{ZX}$ and $(D^{1/2}S)_{YZ} = (D^{1/2}S)_{ZY}$ are satisfied. This entails demonstrating that

$$(D^{-1/2}\sigma_1)_Y = (D^{-1/2}\sigma_2)_X \ , \ (\tfrac{1}{2}D^{-1/2}\sigma_1)_Z = (D^{-1/2}\sigma_3)_X \text{ and } (\tfrac{1}{2}D^{-1/2}\sigma_2)_Z = (D^{-1/2}\sigma_3)_Y \ . \tag{2.55}$$

To see this is the case you should note that as of yet we have not employed the fact that $\alpha_{31}$, $\alpha_{32}$, $\alpha_{33}$ and $\alpha_{34}$ satisfy **Lemma 5**. Upon doing so we discover that

$$\sigma_1 = -2Y\alpha_{33X} + Y\alpha_{34Z} + Z\alpha_{32Z} - 2Z\alpha_{34X}; \ \sigma_2 = -2X\alpha_{32Y} + X\alpha_{34Z} + Z\alpha_{33Z} - 2Z\alpha_{34Y}; \tag{2.56a}$$

$$\sigma_3 = -Y\alpha_{33Z} - Z\alpha_{34Z} + 2Y\alpha_{34Y} + 2Z\alpha_{32Y} \text{ and } \sigma = -[\alpha_{32Y} + \alpha_{33X} - \alpha_{34Z}] \ . \tag{2.56b}$$

Using (2.56), along with the partial derivatives of (2.19) with respect to X, Y, and Z, it is a straightforward matter to prove that the integrability conditions (2.55) do indeed hold. Consequently the system of equations (2.51) has a solution for S, and those equations imply that any S that satisfies them must also satisfy (2.52) and hence be given by (2.53). Thus by our construction the S of (2.53) is the solution to each of the three equations in (2.51), which is equivalent to our original system of partial differential equations (2.48).

I summarize what we have done with the Lagrangian $L_2$ in the following

**Lemma 9:** The Lagrangian $L_2$ presented in (2.40) will yield second-order bi-scalar-tensor field equations. Moreover, $E^{ij}(L_2)$ will be devoid of terms involving $A_3{}^{ij}$, but will precisely yield the terms appearing in the $A_2{}^{ij}$ of (2.22c), along with terms of the form $A_1{}^{ij}$ and $A_0{}^{ij}$, provided the four BST functions $S_{21}$, $S_{22}$, $S_{23}$ and $S_{24}$ in $L_2$ are chosen so that $S_{21}$ is given by (2.48) and $S_{22}$, $S_{23}$ and $S_{24}$ are any set of three BST functions that satisfy (2.53) with $\sigma$ as given in (2.56b).■

Equation (2.53) shows that we have a great deal of latitude in the solution for $S_{22}$, $S_{23}$ and $S_{24}$. For if the triple $(S_{22}, S_{23}, S_{24})$ is one solution with $S_{21}$ given by (2.49), then it generates a family of triples $(\hat{S}_{22}, \hat{S}_{23}, \hat{S}_{24})$ where

$$\hat{S}_{22}:=S_{22}+f_2(\varphi,\xi,X)+k_2Z+c_2;\ \ \hat{S}_{23}:=S_{23}+f_3(\varphi,\xi,Y)+k_3Z+c_3;\text{ and }\ \hat{S}_{24}:=S_{24}+f_4(\varphi,\xi,X)+h_4(\varphi,\xi,Y), \tag{2.57a}$$

with $f_2$, $f_3$, $f_4$, $h_4$ being real valued functions of the indicated variables and $k_2$, $k_3$, $c_2$ and $c_3$ being real constants. Using (2.57a) in (2.49) we find that the $\hat{S}_{21}$ corresponding to the triple $(\hat{S}_{22}, \hat{S}_{23}, \hat{S}_{24})$ is

$$\hat{S}_{21} = S_{21} - \tfrac{1}{2}X[f_2(\varphi,\xi,X)+k_2Z+c_2] - \tfrac{1}{2}Y[f_3(\varphi,\xi,Y)+k_3Z+c_3] - Z[f_4(\varphi,\xi,X)+h_4(\varphi,\xi,Y)]\ . \tag{2.57b}$$

If we now replace the S's on the right hand side of (2.48) with the $\hat{S}$'s of (2.57) we find that the $\hat{S}$'s generate the same values for $\alpha_{31}$, $\alpha_{32}$, $\alpha_{33}$ and $\alpha_{34}$ as do the S's. This, of course, is no surprise, but it is reassuring. Thus we see that there is a plethora of different version of $L_2$ that can generate the second term within curly brackets in (2.22).

We now choose $\mathfrak{L}_2$ in the **Theorem** to be the $L_2$ given in (2.40). (The coefficient functions in $\mathfrak{L}_2$ and $L_2$ differ by numerical factors, but that is irrelevant.)

At this point we have found two Lagrangians, $L_3$ and $L_2$, given by (2.29) and (2.40) which are such that if L yields the $A^{ij}$ presented in (2.22) then $L-L_3-L_2$ will yield a variational derivative with respect to $g_{ij}$ which has the form $A_1{}^{ij} + A_0{}^{ij}$, where these tensor densities are given by (2.22d) and (2.22e). I shall now produce a Lagrangian $L_1$ that yields $A_1{}^{ij}$. In order to do that we shall have to examine the divergence of $A_1{}^{ij}$ in more detail, since I failed to present any restrictions on the coefficients of this block of terms in the statement of **Lemma 5**. The reason why I did not determine the restrictions we needed earlier was because those restrictions were intimately connected with the coefficients appearing in the $A_3{}^{ij}$ and $A_2{}^{ij}$ blocks of terms in (2.22a), and hence quite convoluted. The conditions we require to determine $L_1$ are presented in

**Lemma 10:** When $A^{ij} = A_1{}^{ij} + A_0{}^{ij}$, where $A_1{}^{ij}$ and $A_0{}^{ij}$ are given by (2.22d) and (2.22e) with $A^{ij}{}_{|j} \equiv 0 \bmod(\varphi^i,\xi^i)$, then its coefficient functions must satisfy

$$\beta_{81} = \beta_{91} = \beta_{86} = \beta_{96} = 0;\ \beta_{83} = \beta_{94};\ \beta_{84} = \beta_{92}; \tag{2.58a}$$

$$\beta_{82Z}=2\beta_{84X};\ \beta_{93Z}=2\beta_{83Y};\ \beta_{85}=2\beta_{83X}-2\beta_{82Y};\ \beta_{85}=\beta_{84Z}-2\beta_{83X};\ \beta_{95}=\beta_{83Z}-2\beta_{84Y};\ \beta_{95}=2\beta_{84Y}-2\beta_{93X}; \tag{2.58b}$$

$$2\beta_{82Y} = 4\beta_{83X} - \beta_{84Z} \text{ and } 2\beta_{93X} = 4\beta_{84Y} - \beta_{83Z}, \tag{2.58c}$$

where $(\beta_{85},\beta_{95})$ is a BSTCP.

When $A_1{}^{ij}=0$ then

$$\gamma_2 = -2\gamma_{1X},\ \gamma_3 = -2\gamma_{1Y} \text{ and } \gamma_4 = -\gamma_{1Z}. \tag{2.59d}$$

**Proof:** To begin (2.59c) follows immediately from (2.59b), but I include it as a separate equation because it turns out to be useful when applying the lemma. The fact that $(\beta_{85},\beta_{95})$ is a BSTCP also follows from (2.59b).

The proof of the remainder of this lemma is similar to that of **Lemma 5**, so I shall only provide a brief sketch of how it is accomplished. First of all, **Lemma 5** gives us the equations in (2.58a). Thus when you take the divergence of $A_1{}^{ij} + A_0{}^{ij}$ on the index j (working $\bmod(\varphi^i,\xi^i)$), keeping track of only those terms that involve third order derivatives of the scalar fields, you find that those third-order terms vanish. Next we shall only be interested in the terms of $(A_1{}^{ij}{}_{|j} + A_0{}^{ij}{}_{|j})$ $\bmod(\varphi^i,\xi^i)$ that are of second degree in the scalar fields, and these terms arise from $A_1{}^{ij}{}_{|j}$. Contracting that equation with $V^i$, which is taken to be a unit vector perpendicular to $\varphi^i$ and $\xi^i$, gives 0.

Miraculously that contracted equation is devoid of terms involving $\Box\varphi$, $\Box\xi$ and any contractions of the second partial derivatives of the scalar fields with second partial derivatives of the scalar fields. Consequently we can write out that contracted equation in terms of the components of $\varphi_{ab}$ and $\xi_{ab}$ contracted with the vectors of the set $\{\varphi^i,\xi^i,V^i\}$. So, for example, a typical term in the equation is $\varphi^a\varphi^b\xi^jV^i\varphi_{ab}\xi_{ij}[\beta_{85}+2\beta_{82Y}-2\beta_{83X}] = \varphi_{11}\xi_{23}[\beta_{85}+2\beta_{82Y}-2\beta_{83X}]$, using frame notation for the second covariant derivatives ($\varphi_{11}:=\varphi^a\varphi^b\varphi_{ab}$, and so on) . The remaining conditions of (2.58b-c) then follow from the independence of the frame components of the unique elements in the set $\{\varphi_{ab}\varphi_{cd}$, $\varphi_{ab}\xi_{cd}$, $\xi_{ab}\xi_{cd}\}$, and the arbitrariness of the scalar fields $\varphi$ and $\xi$.

(2.59d) is obtained without much effort in a similar manner.■

We now desire to find a Lagrangian $L_1$ which yields second-order bi-scalar-tensor field equations with $E^{ij}(L_1)$ having the form of $A_1^{\ ij} + A_0^{\ ij}$, with the first degree terms in $E^{ij}(L_1)$ identically equal to the terms in the general $A_1^{\ ij}$. As usual **Lemma 7** suggests that we look at $A_1^{\ ij}g_{ij}$. Doing so leads us to consider

$$L_1 := g^{1/2}\{S_{11}(X\Box\varphi + 2\varphi^a\varphi^b\varphi_{ab}) + S_{12}(Y\Box\varphi + 2\xi^a\xi^b\varphi_{ab}) + S_{13}(2Z\Box\varphi + 4\varphi^a\xi^b\varphi_{ab}) +$$

$$+ \ S_{13}(X\Box\xi + 2\varphi^a\varphi^b\xi_{ab}) + S_{14}(Y\Box\xi + 2\xi^a\xi^b\xi_{ab}) + S_{12}(2Z\Box\xi + 4\varphi^a\xi^b\xi_{ab}) +$$

$$+ \ S_{15}(X\xi^a\xi^b\varphi_{ab}+Y\varphi^a\varphi^b\varphi_{ab} - 2Z\varphi^a\xi^b\varphi_{ab}) + S_{16}(X\xi^a\xi^b\xi_{ab}+Y\varphi^a\varphi^b\xi_{ab} - 2Z\varphi^a\xi^b\xi_{ab})\}, \tag{2.60a}$$

where

$$S_{15}=2S_{12X}-2S_{11Y}=S_{13Z}-2S_{12X};\ S_{16}=2S_{13Y}-2S_{14X}=S_{12Z}-2S_{13Y};\ S_{11Z}=2S_{13X};\ S_{14Z}=2S_{12Y}. \tag{2.60b}$$

Note that $L_1$ is invariant under the transformation

$$S_{11}\leftrightarrow S_{14};\ S_{12}\leftrightarrow S_{13}; S_{15}\leftrightarrow S_{16};\ X\leftrightarrow Y;\ Z\leftrightarrow Z;\ \varphi\leftrightarrow\xi;\ \partial_x\leftrightarrow\partial_y;\ \partial_Z\leftrightarrow\partial_Z\ . \tag{2.61}$$

Invariance under (2.61) will prove very useful when computing various partial derivatives and Euler-Lagrange tensor densities.

In order to compute $E^{ij}(L_1)$, $E_\varphi(L_1)$ and $E_\xi(L_1)$ using (2.30) and (2.32) we shall require $L_{1\varphi}^{\ \ ;ij}$, $L_{1\xi}^{\ \ ;ij}$, $\Phi^i(L_1)$ and $\Xi^i(L_1)$. These are easily found to be given by

$$
\begin{aligned}
L_{1\varphi}{}^{;ij} = g^{1/2}\{ & S_{11}(Xg^{ij}+2\varphi^i\varphi^j) + S_{12}(Yg^{ij}+2\xi^i\xi^j) + S_{13}(2Zg^{ij}+2\varphi^i\xi^j+2\xi^i\varphi^j) + \\
& + \quad S_{15}(X\xi^i\xi^j+Y\varphi^i\varphi^j-Z\varphi^i\xi^j-Z\varphi^j\xi^i)\} \ ,
\end{aligned} \qquad (2.62a)
$$

$$
\begin{aligned}
L_{1\xi}{}^{;ij} = g^{1/2}\{ & S_{12}(2Zg^{ij}+2\varphi^i\xi^j+2\xi^i\varphi^j) + S_{13}(Xg^{ij}+2\varphi^i\varphi^j) + S_{14}(Yg^{ij}+2\xi^i\xi^j) + \\
& + \quad S_{16}(X\xi^i\xi^j+Y\varphi^i\varphi^j-Z\varphi^i\xi^j-Z\varphi^j\xi^i)\} \ ,
\end{aligned} \qquad (2.62b)
$$

$$
\begin{aligned}
& \Phi^i(L_1) = g^{1/2}\{2S_{11X}\varphi^i(X\Box\varphi + 2\varphi_{ab}\varphi^a\varphi^b) + S_{11Z}\xi^i(X\Box\varphi + 2\varphi_{ab}\varphi^a\varphi^b) + S_{11}(2\varphi^i\Box\varphi + 4\varphi^{ia}\varphi_a) + \\
& + 2S_{12X}\varphi^i(Y\Box\varphi + 2\varphi_{ab}\xi^a\xi^b + 2Z\Box\xi + 4\xi_{ab}\xi^a\varphi^b) + S_{12Z}\xi^i(Y\Box\varphi + 2\varphi_{ab}\xi^a\xi^b + 2Z\Box\xi + 4\xi_{ab}\xi^a\varphi^b) + \\
& + S_{12}(2\xi^i\Box\xi + 4\xi^{ia}\xi_a) + 2S_{13X}\varphi^i(X\Box\xi + 2\xi_{ab}\varphi^a\varphi^b + 2Z\Box\varphi + 4\varphi_{ab}\varphi^a\xi^b) + \\
& + S_{13Z}\xi^i(X\Box\xi + 2\xi_{ab}\varphi^a\varphi^b + 2Z\Box\varphi + 4\varphi_{ab}\varphi^a\xi^b) + S_{13}(2\varphi^i\Box\xi + 4\xi^{ia}\varphi_a + 2\xi^i\Box\varphi + 4\varphi^{ia}\xi_a) + \\
& + 2S_{14X}\varphi^i(Y\Box\xi + 2\xi_{ab}\xi^a\xi^b) + S_{14Z}\xi^i(Y\Box\xi + 2\xi_{ab}\xi^a\xi^b) + 2S_{15X}\varphi^i(X\xi^a\xi^b\varphi_{ab}+Y\varphi^a\varphi^b\varphi_{ab}-2Z\varphi^a\xi^b\varphi_{ab}) + \\
& + S_{15Z}\xi^i(X\xi^a\xi^b\varphi_{ab}+Y\varphi^a\varphi^b\varphi_{ab}-2Z\varphi^a\xi^b\varphi_{ab}) + S_{15}(2\varphi^i\xi^a\xi^b\varphi_{ab}+2Y\varphi_b\varphi^{ib} - 2\xi^i\varphi^a\xi^b\varphi_{ab} - 2Z\xi_b\varphi^{bi}) + \\
& + 2S_{16X}\varphi^i(X\xi^a\xi^b\xi_{ab}+Y\varphi^a\varphi^b\xi_{ab}-2Z\varphi^a\xi^b\xi_{ab}) + S_{16Z}\xi^i(X\xi^a\xi^b\xi_{ab}+Y\varphi^a\varphi^b\xi_{ab}-2Z\varphi^a\xi^b\xi_{ab}) + \\
& + S_{16}(2\varphi^i\xi^a\xi^b\xi_{ab}+2Y\varphi_b\xi^{ib} - 2\xi^i\varphi^a\xi^b\xi_{ab} - 2Z\xi_b\xi^{bi})\},
\end{aligned} \qquad (2.62c)
$$

$$
\begin{aligned}
& \Xi^i(L_1) = g^{1/2}\{2S_{11Y}\xi^i(X\Box\varphi + 2\varphi_{ab}\varphi^a\varphi^b) + S_{11Z}\varphi^i(X\Box\varphi + 2\varphi_{ab}\varphi^a\varphi^b) + \\
& + 2S_{12Y}\xi^i(Y\Box\varphi + 2\varphi_{ab}\xi^a\xi^b + 2Z\Box\xi + 4\xi_{ab}\xi^a\varphi^b) + S_{12Z}\varphi^i(Y\Box\varphi + 2\varphi_{ab}\xi^a\xi^b + 2Z\Box\xi + 4\xi_{ab}\xi^a\varphi^b) + \\
& + S_{12}(2\xi^i\Box\varphi + 4\varphi^{ia}\xi_a + 2\varphi^i\Box\xi + 4\xi^{ia}\varphi_a) + 2S_{13Y}\xi^i(X\Box\xi + 2\xi_{ab}\varphi^a\varphi^b + 2Z\Box\varphi + 4\varphi_{ab}\varphi^a\xi^b) + \\
& + S_{13Z}\varphi^i(X\Box\xi + 2\xi_{ab}\varphi^a\varphi^b + 2Z\Box\varphi + 4\varphi_{ab}\varphi^a\xi^b) + S_{13}(2\varphi^i\Box\varphi + 4\varphi^{ia}\varphi_a) + \\
& + 2S_{14Y}\xi^i(Y\Box\xi + 2\xi_{ab}\xi^a\xi^b) + S_{14Z}\varphi^i(Y\Box\xi + 2\xi_{ab}\xi^a\xi^b) + S_{14}(2\xi^i\Box\xi + 4\xi^{ia}\xi_a) + \\
& + 2S_{15Y}\xi^i(X\xi^a\xi^b\varphi_{ab}+Y\varphi^a\varphi^b\varphi_{ab}-2Z\varphi^a\xi^b\varphi_{ab}) + S_{15Z}\varphi^i(X\xi^a\xi^b\varphi_{ab}+Y\varphi^a\varphi^b\varphi_{ab}-2Z\varphi^a\xi^b\varphi_{ab}) + \\
& + S_{15}(2\xi^i\varphi^a\varphi^b\varphi_{ab}+2X\xi_b\varphi^{ib} - 2\varphi^i\varphi^a\xi^b\varphi_{ab} - 2Z\varphi_b\varphi^{bi}) + 2S_{16Y}\xi^i(X\xi^a\xi^b\xi_{ab}+Y\varphi^a\varphi^b\xi_{ab}-2Z\varphi^a\xi^b\xi_{ab}) + \\
& + S_{16Z}\varphi^i(X\xi^a\xi^b\xi_{ab}+Y\varphi^a\varphi^b\xi_{ab}-2Z\varphi^a\xi^b\xi_{ab}) + S_{16}(2\xi^i\varphi^a\varphi^b\xi_{ab}+2X\xi_b\xi^{ib} - 2\varphi^i\varphi^a\xi^b\xi_{ab} - 2Z\varphi_b\xi^{bi})\}.
\end{aligned} \qquad (2.62d)
$$

Due to (2.30), (2.32) and (2.62) it is clear that $E^{ij}(L_1)$ will be of second-order, and of first degree. Unfortunately it is not as obvious that $E_\varphi(L_1)$ or $E_\xi(L_1)$ will be of second-order. Nevertheless a lengthy calculation shows that due to (2.30), (2.62) and the conditions given in (2.60b), $E_\varphi(L_1)$ is indeed of second-order. Hence due to the aforementioned symmetry given in (2.61), the same is true for $E_\xi(L_1)$.

The next thing we need to determine is the first degree part of $E^{ij}(L_1)$, and to demonstrate that it can be made to equal the $A_1{}^{ij}$ given in (2.22d). A lengthy calculation, using (2.30) and (2.62), shows that this will be the case if

$$\beta_{82}= XS_{11X}+YS_{11Y}+ZS_{11Z};\ \beta_{83}= XS_{12X}+YS_{12Y}+ZS_{12Z};\ \beta_{84}= XS_{13X}+YS_{13Y}+ZS_{13Z}; \tag{2.63a}$$

$$\beta_{85}= S_{15}+XS_{15X}+YS_{15Y}+ZS_{15Z};\ \beta_{93}=XS_{14X}+YS_{14Y}+ZS_{14Z};\ \beta_{95}= S_{16}+XS_{16X}+YS_{16Y}+ZS_{16Z}. \tag{2.63b}$$

The first thing we need to check is that when the $\beta_{..}$'s are given by (2.63) they actually satisfy the conditions of **Lemma 10**, when the $S_{..}$'s satisfy (2.60b). A simple calculation shows that this is in fact the case. The next thing we must show is that when we are presented with $\beta_{..}$'s that satisfy **Lemma 10** we can solve the partial differential equations of (2.63) for the $S_{..}$'s. Well, we have encountered partial differential equations of this form in our previous treatment of the Lagrangians $L_2$ and $L_3$ and know from that experience that the required solutions are

$$S_{11}=\int t^{-1}\beta_{82}(t)dt\Big|_{t=1},\quad S_{12}=\int t^{-1}\beta_{83}(t)dt\Big|_{t=1},\quad S_{13}=\int t^{-1}\beta_{84}(t)dt\Big|_{t=1},\quad S_{14}=\int t^{-1}\beta_{93}(t)dt\Big|_{t=1}, \tag{2.64a}$$

$$S_{15}=\int \beta_{85}(t)dt\Big|_{t=1}\quad \text{and}\quad S_{16}=\int \beta_{95}(t)dt\Big|_{t=1}. \tag{2.64b}$$

In view of the above work we have established

**Lemma 11:** When $A^{ij} = A_1{}^{ij} + A_0{}^{ij}$, where $A_1{}^{ij}$ and $A_0{}^{ij}$ are given by (2.22d) and (2.22e) with $A^{ij}{}_{|j}\equiv 0 \bmod(\varphi^i,\xi^i)$, then the Lagrangian $L_1$ defined in (2.60) yields second-order bi-scalar-tensor field equations with $E^{ij}(L_1)$ having the form of $A^{ij}$ with the first degree terms in $E^{ij}(L_1)$ precisely equal to those in $A_1{}^{ij}$ when $S_{11}$, $S_{12}$, $S_{13}$, $S_{14}$, $S_{15}$ and $S_{16}$ are given by (2.64).■

The Lagrangian $L_1$ appears to be a bit different then the Lagrangian $\mathcal{L}_1$ of (1.11) in the **Theorem**. Nevertheless it does the job of yielding the $A_1{}^{ij}$ term given in (2.22d). However, the problem in using it is that it is difficult in practice to find $S_{11}$, $S_{12}$, $S_{13}$, $S_{14}$, $S_{15}$ and $S_{16}$ that satisfy (2.60b). So I shall now demonstrate that $L_1$ and $\mathcal{L}_1$ are indeed equivalent up to divergences and terms that can be absorbed into $\mathcal{L}_0$.. To that end we begin by noting that the Lagrangian $L_{15}$ appearing in (1.12e) has numerous terms in common with the terms in $L_1$ which involve $S_{15}$ and $S_{16}$. To capitalize on that observation let us now add 0 to $L_1$ which I take to have the form

$$0= S_{15}(XY-Z^2)\Box\varphi - S_{15}(XY-Z^2)\Box\varphi + S_{16}(XY-Z^2)\Box\xi - S_{16}(XY-Z^2)\Box\xi \,.$$

Upon doing so we obtain

$$\begin{aligned} L_1 := {} & g^{1/2}\{S_{11}(X\Box\varphi + 2\varphi^a\varphi^b\varphi_{ab}) + S_{12}(Y\Box\varphi + 2\xi^a\xi^b\varphi_{ab} + 2Z\Box\xi + 4\varphi^a\xi^b\xi_{ab}) + S_{13}(2Z\Box\varphi + 4\varphi^a\xi^b\varphi_{ab} + \\ & + X\Box\xi + 2\varphi^a\varphi^b\xi_{ab}) + S_{14}(Y\Box\xi + 2\xi^a\xi^b\xi_{ab}) + S_{15}(XY-Z^2)\Box\varphi + S_{16}(XY-Z^2)\Box\xi\} + \\ & - g^{1/2}\{ S_{15}[(XY-Z^2)\Box\varphi - X\xi^a\xi^b\varphi_{ab} - Y\varphi^a\varphi^b\varphi_{ab} + 2Z\varphi^a\xi^b\varphi_{ab}] + \\ & + S_{16}[(XY-Z^2)\Box\xi - X\xi^a\xi^b\xi_{ab} - Y\varphi^a\varphi^b\xi_{ab} + 2Z\varphi^a\xi^b\xi_{ab}) \}. \end{aligned} \tag{2.65}$$

Since we know that $(S_{15},S_{16})$ is a BSTCP the second term in curly brackets in (2.65) is the Lagangian $L_{15}$ up to sign, which is irrelevant.

The next thing that we shall do with $L_1$ is group together the terms which multiply $\Box\varphi$ and $\Box\xi$. Upon doing so we see that due to (2.60b), equation (2.65) can be rewritten as

$$L_1 = \bar{L}_{11} + L_R - L_{15} \tag{2.66a}$$

where

$$\bar{L}_{11} := g^{1/2}( \bar{S}_{11}\,\Box\varphi + \bar{S}_{12}\,\Box\xi\,), \tag{2.66b}$$

$$\bar{S}_{11} := XS_{11} + YS_{12} + 2ZS_{13} + 2S_{12X}(XY-Z^2) - 2S_{11Y}(XY-Z^2)\,, \tag{2.66c}$$

$$\bar{S}_{12} := XS_{13} + YS_{14} + 2ZS_{12} + 2S_{13Y}(XY-Z^2) - 2S_{14X}(XY-Z^2)\,, \tag{2.66d}$$

and

$$L_R := g^{1/2}\{2S_{11}\varphi^a\varphi^b\varphi_{ab} + 2S_{12}\xi^a\xi^b\varphi_{ab} + 4S_{12}\varphi^a\xi^b\xi_{ab} + 2S_{13}\varphi^a\varphi^b\xi_{ab} + 4S_{13}\varphi^a\xi^b\varphi_{ab} + 2S_{14}\xi^a\xi^b\xi_{ab}\}. \tag{2.66e}$$

A straightforward calculation, using (2.60b), (2.66c) and (2.66d), shows that the pair $(\bar{S}_{11},\bar{S}_{12})$ is a BSTCP. Hence the Lagrangian $\bar{L}_{11}$ given in (2.66b), can be taken to be the Lagrangian $L_{11}$ appearing in $\mathcal{L}_1$ of the **Theorem**. Hence due to (2.66a) all that is left for us to do to prove that the $L_1$ given in (2.66a) is the $\mathcal{L}_1$ of the **Theorem**, is to show how the remnant Lagrangian $L_R$ is expressible as the sum of Lagrangians of the form $L_{12}$, $L_{13}$ and $L_{14}$. This can be done as follows.

In order to simplify the form of the ensuing calculation I shall rewrite the Lagrangians $L_{12}$, $L_{13}$ and $L_{14}$ as

$$L_{12} := g^{1/2}\{u_1(X\Box\varphi - \varphi^a\varphi^b\varphi_{ab}) + u_2(X\Box\xi - \varphi^a\varphi^b\xi_{ab})\} \tag{2.67a}$$

$$L_{13} := g^{1/2}\{v_1(Y\Box\varphi - \xi^a\xi^b\varphi_{ab}) + v_2(Y\Box\xi - \xi^a\xi^b\xi_{ab})\} \tag{2.67b}$$

$$L_{14} := g^{1/2}\{w_1(Z\Box\varphi - \varphi^a\xi^b\varphi_{ab}) + w_2(Z\Box\xi - \varphi^a\xi^b\xi_{ab})\}\ , \tag{2.67c}$$

where $(u_1,u_2)$, $(v_1,v_2)$ and $(w_1,w_2)$ are BSTCPs. Thus we have

$$L_{12}+L_{13}+2L_{14} = g^{1/2}\{u_1X\Box\varphi + v_1Y\Box\varphi + 2w_1Z\Box\varphi + u_2X\Box\xi + v_2Y\Box\xi + 2w_2Z\Box\xi - u_1\varphi^a\varphi^b\varphi_{ab} - u_2\varphi^a\varphi^b\xi_{ab} + \\ - v_1\xi^a\xi^b\varphi_{ab} - v_2\xi^a\xi^b\xi_{ab} - 2w_1\varphi^a\xi^b\varphi_{ab} - 2w_2\varphi^a\xi^b\xi_{ab}\}. \tag{2.68}$$

Upon comparing (2.66e) with (2.68) we see that the last six terms in (2.68) correspond to terms in $L_R$, but $L_R$ is devoid of terms involving $\Box\varphi$ and $\Box\xi$ . These $\Box\varphi$ and $\Box\xi$ terms can be incorporated into divergences by noting that; *e.g.,*

$$(u_1X\varphi^a)_{|a} = u_1X\Box\varphi + 2u_1\varphi^a\varphi^b\varphi_{ab} + u_{1\varphi}X^2 + u_{1\xi}XZ + 2u_{1X}X\varphi^a\varphi^b\varphi_{ab} + 2u_{1Y}X\varphi^a\xi^b\xi_{ab} + u_{1Z}X\varphi^a\varphi^b\xi_{ab} + \\ + u_{1Z}X\varphi^a\xi^b\varphi_{ab}\ . \tag{2.69}$$

There are similar terms for $(u_2X\xi^a)_{|a}$ , $(v_1Y\varphi^a)_{|a}$ , $(v_2Y\xi^a)_{|a}$ , $(2w_1Z\varphi^a)_{|a}$ and $(2w_2Z\xi^a)_{|a}$ . Using (2.69) and its counterparts in (2.68) to replace the $\Box\varphi$ and $\Box\xi$ terms allows us to show, after a bit of effort, that if we take

$$S_{11} = -(Xu_{1X} + Yv_{1X} + 2Zw_{1X} + {}^3/_2u_1) \tag{2.70a}$$

$$S_{12} = -½(Xu_{2Z} + Yv_{2Z} + 2Zw_{2Z} + 2w_2 + v_1) \tag{2.70b}$$

$$S_{13} = -½(Xu_{1Z} + Yv_{1Z} + 2Zw_{1Z} + 2w_1 + u_2)\ \text{, and} \tag{2.70c}$$

$$S_{14} = -(Xu_{2Y} + Yv_{2Y} + 2Zw_{2Y} + {}^3/_2v_2) \tag{2.70d}$$

in the expression for $L_R$ given in (2.66e) then

$$L_{12}+L_{13}+2L_{14} =$$

$$=L_R + (u_1X\varphi^a)_{|a} + (u_2X\xi^a)_{|a} + (v_1Y\varphi^a)_{|a} + (v_2Y\xi^a)_{|a} + (2w_1Z\varphi^a)_{|a} + (2w_2Z\xi^a)_{|a} - u_{1\varphi}X^2 - u_{1\xi}XZ +$$

$$- u_{2\varphi}XZ - u_{2\xi}XY - v_{1\varphi}XY - v_{1\xi}YZ - v_{2\varphi}YZ - v_{2\xi}Y^2 - 2w_{1\varphi}XZ - 2w_{1\xi}Z^2 - 2w_{2\varphi}Z^2 - 2w_{2\xi}YZ. \qquad (2.71)$$

From (2.60b) we know that the coefficients $S_{11}$, $S_{12}$, $S_{13}$ and $S_{14}$ appearing in $L_R$ must satisfy

$$4S_{12X} = S_{13Z} + 2S_{11Y};\ \ 4S_{13Y} = S_{12Z} + 2S_{14X};\ S_{11Z} = 2S_{13X};\ S_{14Z} = 2S_{12Y}. \qquad (2.72)$$

The S's of (2.70) do satisfy these four conditions since $(u_a,u_b)$, $(v_a,v_b)$ and $(w_a,w_b)$ are BSTCPs. Thus to complete our proof that $L_R$ can be expressed in the form of $L_{12}+L_{13}+2L_{14}$ plus a divergence and a term of the form $g^{1/2}S_0(\varphi,\xi,X,Y,Z)$, all we need do is choose the three BSTCPs $(u_a,u_b)$, $(v_a,v_b)$ and $(w_a,w_b)$ to satisfy (2.70) for the given values of $S_{11}$, $S_{12}$, $S_{13}$ and $S_{14}$. I summarize what we have just done in

**Lemma 12**: The Lagrangian $L_1$ given in (2.66a) will precisely yield the $A_1^{ij}$ term in (2.22a) along with some terms of the form $A_0^{ij}$ when the coefficients $(S_{11},S_{12},S_{13},S_{14})$ appearing in the Lagrangian $L_R$ of (2.66e) are chosen in accordance with (2.70), where the triple of BSTCPs $(u_1,u_2)$, $(v_1,v_2)$ and $(w_1,w_2)$, are the coefficient function appearing in the Lagrangians $L_{12}$, $L_{13}$ and $L_{14}$ of (2.67), so that $L_R$ can be expressed as in (2.71).■

Before proceeding further I would like to point out a reformulation of (2.70). To begin, note that these equations suggest that we consider the two function $F_1$ and $F_2$ defined by

$$F_1:=Xu_1+Yv_1+2Zw_1 \text{ and } F_2:=Xu_2+Yv_2+2Zw_2. \qquad (2.73)$$

Using (2.70), along with the fact that $(u_1,u_2)$, $(v_1,v_2)$, and $(w_1,w_2)$ are BSTCPs, we easily find that

$$F_{1Y} = -S_{12}-w_2+\tfrac{1}{2}v_1,\ F_{2X} = -S_{13}-w_1+\tfrac{1}{2}u_2\,,$$

$$u_1= -2(S_{11} + F_{1X}),\ v_1=-(2S_{12}+F_{2Z}),\ w_1=-(2S_{13}+\tfrac{1}{2}F_{1Z}+F_{2X}), \qquad (2.74a)$$

$$u_2= -(2S_{13}+F_{1Z}),\ v_2= -2(S_{14} + F_{2Y}),\ w_2=-(2S_{12}+\tfrac{1}{2}F_{2Z}+F_{1Y}). \qquad (2.74b)$$

Upon combining (2.73) with (2.74) we discover that

$$F_1 = -2XF_{1X} - YF_{2Z} - ZF_{1Z} - 2ZF_{2X} - [2XS_{11} + 2YS_{12} + 4ZS_{13}], \tag{2.75a}$$

$$F_2 = -2YF_{2Y} - XF_{1Z} - ZF_{2Z} - 2ZF_{1Y} - [2XS_{13} + 2YS_{14} + 4ZS_{12}]. \tag{2.75b}$$

In addition the fact that $(u_1,u_2)$, $(v_1,v_2)$, and $(w_1,w_2)$ are BSTCPs, and that $S_{11}$, $S_{12}$, $S_{13}$ and $S_{14}$ satisfy (2.60b) implies that

$$4F_{1XY} - F_{1ZZ} = 2(S_{13Z} - 2S_{11Y}) \text{ and } 4F_{2XY} - F_{2ZZ} = 2(S_{12Z} - 2S_{14X}). \tag{2.75c}$$

So we see that if we can solve the four partial differential equations given in (2.75) for $F_1$ and $F_2$ then we can use (2.74) to determine the required BSTCPs $(u_1,u_2)$, $(v_1,v_2)$, and $(w_1,w_2)$. Thus $F_1$ and $F_2$ are the generating functions for these three BSTCPs, and hence we have a second method for finding these functions.

At this point in our proof we have produced three second-order bi-scalar tensor Lagrangians $L_3$, $L_2$ and $L_1$ presented in (2.29), (2.40) and (2.66a), which can be used to show that all of the terms appearing in the $A^{ij}$ given in (2.22a) can be obtained from a variational principle with the possible exception of $A_0^{ij}$.

**Lemma 10** tells us that that for the $A_0^{ij}$ of (2.22e), $A_0^{ij}{}_{|j} \equiv 0 \text{ mod}(\varphi^i,\xi^i)$ implies that

$$\gamma_2 = -2\gamma_{1X}\,,\ \gamma_3 = -2\gamma_{1Y} \text{ and } \gamma_4 = -\gamma_{1Z}\,.$$

With no restrictions on $\gamma_1$. Consequently we have

$$A_0^{ij} = g^{1/2}\{\gamma_1 g^{ij} - 2\gamma_{1X}\varphi^i\varphi^j - 2\gamma_{1Y}\xi^i\xi^j - \gamma_{1Z}(\varphi^i\xi^j + \varphi^j\xi^i)\}\ . \tag{2.73}$$

which in turn implies that

$$g_{ij}A_0^{ij} = g^{1/2}\{4\gamma_1 - 2X\gamma_{1X} - 2Y\gamma_{1Y} - 2Z\gamma_{1Z}\}\ . \tag{2.74}$$

After looking at the partial derivative with respect to $g_{ij}$ of various terms in (2.74) we see that if we choose $L_0 := 2g^{1/2}\gamma_1$ then $E^{ij}(L_0) = A_0^{ij}$, as given in (2.73). So we take $L_0$ to be $\mathcal{L}_0$ in the **Theorem**.

So at long last, we have peeled off all the layers of the onion, and the **Theorem** has been proved. To recapitulate, we have constructed the most general second-order bi-scalar-tensor field equations that can be derived in a space of four-dimensions from a Lagrangian of arbitrary differential order, and also shown that those equations can be derived from the second-order Lagrangian presented in (1.9).

But in fact we have proved even more, since in the course of establishing the **Theorem** we have proved the following

**Corollary:** Suppose that in a space of four-dimensions the triple of second-order bi-scalar-tensor tensorial concomitants $(A^{ij},B,C)$ are such that $A^{ij}$ is a symmeteric contravariant tensor density of rank 2, with B and C being scalar densities such that $A^{ij}{}_{|j} = \frac{1}{2}\varphi^i B + \frac{1}{2}\xi^i C$. Then $(A^{ij},B,C)$ arise as the Euler-Lagrange tensor densities of some second-order bi-scalar-tensor Lagrangian, $\mathfrak{L}$, as presented in the **Theorem**, with $A^{ij}=E^{ij}(\mathfrak{L})$, $B=E_\varphi(\mathfrak{L})$ and $C=E_\xi(\mathfrak{L})$. ■

We have always known that the condition $A^{ij}{}_{|j} = \frac{1}{2}\varphi^i B + \frac{1}{2}\xi^i C$, was necessary to prove the **Theorem**, but now we know that it is also sufficient.

## Section 3: Concluding Remarks

The **Theorem** shows us that the Lagrangian which yields the most general second-order bi-salar-tensor field equations can have as many as eleven coefficient functions that are functions of the five variables φ, ξ, X, Y and Z. We need to find some way to restrict these functions in order to apply these equations to our universe. In the scalar-tensor case we know that there are four coefficient functions of just two variables φ and X. For that case Ezquiaga and Zumalaćarregui [19], T.Baker, *et al.* [20], J.Sakstein and B.Jain [21] and P.Creminelli and F.Vernizzi [22], have shown that since GW170817 implied that the speed of gravity waves must equal the speed of light, the quintic coefficient of scalar-tensor theory must vanish, and the quartic coefficient must be independent of X. Perhaps a similar analysis of the bi-scalar-tensor field equations will show that $\mathfrak{L}_3$ must vanish, and the coefficient functions appearing in $\mathfrak{L}_2$ must only depend on φ and ξ.

It is important to note that the **Theorem** does not provide us with the form of all second-order bi-scalar-tensor Lagrangians that yield second-order field equations. It just says that the Lagrangians presented in the **Theorem** can be used to generate the most general second-order bi-scalar-tensor field equations in a space of four-dimensions. As a result one can occasionally find second-order bi-scalar-tensor Lagrangians that do not appear in the statement of the **Theorem** and yet yield field equations of the requisite type. One such Lagrangian is

$$\begin{aligned} L_{*} := g^{1/2}\{ & K\,\delta^{abcd}_{pqrs}\,\varphi_a\xi_b\varphi^p\xi^q R_{cd}{}^{rs} - 4K_X\,\delta^{abcd}_{pqrs}\,\varphi_a\xi_b\varphi^p\xi^q\varphi_c{}^r\varphi_d{}^s - 4K_Y\,\delta^{abcd}_{pqrs}\,\varphi_a\xi_b\varphi^p\xi^q\xi_c{}^r\xi_d{}^s + \\ & - 4\,K_Z\,\delta^{abcd}_{pqrs}\,\varphi_a\xi_b\varphi^p\xi^q\varphi_c{}^r\xi_d{}^s\}\,, \end{aligned} \tag{3.1}$$

where K is a BST function. The existence of this Lagrangian was pointed out to me by Professor Kobayashi, along with the Lagrangian $L_{15}$ of (1.12e), when I showed him a preliminary version of this paper Horndeski [23] in which the second-order field equations associated with these Lagrangians were absent–a big oversight on my part. $L_{15}$ and $L_{*}$ appear among the many Lagrangians discussed in Ohashi, *et al.,* [3], Kobayashi, *et al.,* [24], and Akama & Kobayashi [25]. $L_{15}$ has been incorporated into this paper while $L_{*}$ is discussed in **Appendix D**, where I show how its Euler-Lagrange tensor densitites can be obtained from a Lagrangian of the form $\mathfrak{L}_2+\mathfrak{L}_1+\mathfrak{L}_0$.

While in the process of correcting the errors in my preliminary version of this paper Horndeski [23], I encountered an interesting way of generating bi-scalar-tensor Lagrangians that yield second-order field equations from other bi-scalar-tensor Lagrangians with this property. This observation may prove useful in constructing multi-scalar-tensor Lagrangians in spaces of arbitrary dimension that generate second-order field equations. This construction method proceeds as follows.

Suppose that we have a bi-scalar-tensor Lagrangian that is expressed in terms of generalized Kronecker deltas. There are four extension operators that can act on such a Lagrangian. (The extensions that I am about to discuss here are different then those used to extend bi-scalar

Lagrangians from flat spaces to curved spaces, as is done, *e.g.,* in Akama & Kobayashi [25].) These operators essentially have the effect of replacing each k×k generalized Kronecker delta $\delta^{\cdots}_{\cdots}$ by an (1+k)×(1+k) generalized Kronecker delta $\delta^{a\cdots}_{b\cdots}$ where the indices a and b are summed with either $\varphi_a\varphi^b$, $\xi_a\xi^b$, $\varphi_a\xi^b$ or $\xi_a\varphi^b$. To illustrate let us consider the Lagrangian $L_{11}$ of (1.12a); *viz.,*

$$L_{11} := g^{\frac{1}{2}}(S_{11}\Box\varphi + S_{12}\Box\xi) = g^{\frac{1}{2}}(S_{11}\delta_q^{\ p}\varphi_p^{\ q} + S_{12}\delta_q^{\ p}\xi_p^{\ q}).$$

I now define the extensions $\mathscr{E}_{\varphi\varphi}(L_{11})$, $\mathscr{E}_{\xi\xi}(L_{11})$, $\mathscr{E}_{\varphi\xi}(L_{11})$ and $\mathscr{E}_{\xi\varphi}(L_{11})$ by

$$\mathscr{E}_{\varphi\varphi}(L_{11}) := g^{\frac{1}{2}}\{\kappa_1\,\delta^{ap}_{bq}\,\varphi_a\varphi^b\varphi_p^{\ q} + \kappa_2\,\delta^{ap}_{bq}\,\varphi_a\varphi^b\xi_p^{\ q}\} \tag{3.2}$$

$$\mathscr{E}_{\xi\xi}(L_{11}) := g^{\frac{1}{2}}\{\lambda_1\,\delta^{ap}_{bq}\,\xi_a\xi^b\varphi_p^{\ q} + \lambda_2\,\delta^{ap}_{bq}\,\xi_a\xi^b\xi_p^{\ q}\} \tag{3.3}$$

$$\mathscr{E}_{\varphi\xi}(L_{11}) := g^{\frac{1}{2}}\{\mu_1\,\delta^{ap}_{bq}\,\varphi_a\xi^b\varphi_p^{\ q} + \mu_2\,\delta^{ap}_{bq}\,\varphi_a\xi^b\xi_p^{\ q}\} \tag{3.4}$$

$$\mathscr{E}_{\xi\varphi}(L_{11}) := g^{\frac{1}{2}}\{\nu_1\,\delta^{ap}_{bq}\,\xi_a\varphi^b\varphi_p^{\ q} + \nu_2\,\delta^{ap}_{bq}\,\xi_a\varphi^b\xi_p^{\ q}\} \tag{3.5}$$

where the pairs $(\kappa_1,\kappa_2)$, $(\lambda_1,\lambda_1)$, $(\mu_1,\mu_2)$ and $(\nu_1,\nu_2)$ are BSTCPs that replace the BSTCP $(S_{11},S_{12})$ each time we do an extension of $L_{11}$. Note that $\mathscr{E}_{\varphi\xi}(L_{11})$ and $\mathscr{E}_{\xi\varphi}(L_{11})$ are essentially the same, and hence I shall not include $\mathscr{E}_{\xi\varphi}(L_{11})$ in what follows. This redundancy will occur frequently and I shall disregard the duplicate Lagrangian. There is only one unique extension of the four Lagrangians presented above, which I shall take to be $\mathscr{E}_{\varphi\varphi}(\mathscr{E}_{\xi\xi}(L_{11})$, which can be expressed as

$$\mathscr{E}_{\varphi\varphi}(\mathscr{E}_{\xi\xi}(L_{11}) = \{\tau_1\,\delta^{abp}_{cdq}\,\varphi_a\xi_b\varphi^c\xi^d\varphi_p^{\ q} + \tau_2\,\delta^{abp}_{cdq}\,\varphi_a\xi_b\varphi^c\xi^d\xi_p^{\ q}\} \tag{3.6}$$

where $(\tau_1,\tau_2)$ is a BSTCP. The Lagrangians given in (3.2)-(3.6) are just the Lagrangians $L_{12}$, $L_{13}$, $L_{14}$ and $L_{15}$ given in (1.12b)-(1.12e). Hence by applying the extension operators to the Lagrangian $L_{11}$ we have essentially generated all of the Lagrangians appearing in $\mathscr{L}_1$, which (when added to a Lagrangian of type $\mathcal{L}_0$) generate the second-order, bi-scalar-tensor field equations that are of first degree.

This extension process can now be applied to the Lagrangians appearing in $\mathcal{L}_2$ and $\mathcal{L}_3$ of the **Theorem**. As I mention in the paragraph after (2.57b), the Lagrangian $\mathcal{L}_2$ and $L_2$ are essentially the same up to numerical coefficients, where $L_2$ is presented in (2.40). Using this fact we see that the Lagrangian $L_{21}$ of (1.14a) is given by

$$L_{21} := \tfrac{1}{2} g^{1/2}\{S_{21}\,\delta^{pq}_{rs}\,R_{pq}{}^{rs} - 4S_{21X}\,\delta^{pq}_{rs}\,\varphi_p{}^r\varphi_q{}^s - 4S_{21Y}\,\delta^{pq}_{rs}\,\xi_p{}^r\xi_q{}^s - 4S_{21Z}\,\delta^{pq}_{rs}\,\varphi_p{}^r\xi_q{}^s\}. \tag{3.7}$$

The four extensions of $L_{21}$ would be (where I am using the notation of the **Theorem** for the coefficient functions in the first three Lagrangians below)

$$\mathcal{E}_{\varphi\varphi}(L_{21}) := \tfrac{1}{2} g^{1/2}\{S_{22}\,\delta^{apq}_{brs}\,\varphi_a\varphi^b R_{pq}{}^{rs} - 4S_{22X}\,\delta^{apq}_{brs}\,\varphi_a\varphi^b\varphi_p{}^r\varphi_q{}^s - 4S_{22Y}\,\delta^{apq}_{brs}\,\varphi_a\varphi^b\xi_p{}^r\xi_q{}^s + -4S_{22Z}\,\delta^{apq}_{brs}\,\varphi_a\varphi^b\varphi_p{}^r\xi_q{}^s\}, \tag{3.8}$$

$$\mathcal{E}_{\xi\xi}(L_{21}) := \tfrac{1}{2} g^{1/2}\{S_{23}\,\delta^{apq}_{brs}\,\xi_a\xi^b R_{pq}{}^{rs} - 4S_{22X}\,\delta^{apq}_{brs}\,\xi_a\xi^b\varphi_p{}^r\varphi_q{}^s - 4S_{23Y}\,\delta^{apq}_{brs}\,\xi_a\xi^b\xi_p{}^r\xi_q{}^s + -4S_{24Z}\,\delta^{apq}_{brs}\,\xi_a\xi^b\varphi_p{}^r\xi_q{}^s\}, \tag{3.9}$$

$$\mathcal{E}_{\varphi\xi}(L_{21}) := \tfrac{1}{2} g^{1/2}\{S_{24}\,\delta^{apq}_{brs}\,\varphi_a\xi^b R_{pq}{}^{rs} - 4S_{24X}\,\delta^{apq}_{brs}\,\varphi_a\xi^b\varphi_p{}^r\varphi_q{}^s - 4S_{24Y}\,\delta^{apq}_{brs}\,\varphi_a\xi^b\xi_p{}^r\xi_q{}^s + -4S_{24Z}\,\delta^{apq}_{brs}\,\varphi_a\xi^b\varphi_p{}^r\xi_q{}^s\}, \tag{3.10}$$

$$\mathcal{E}_{\varphi\varphi}(\mathcal{E}_{\xi\xi}(L_{21})) := \tfrac{1}{2} g^{1/2}\{K\,\delta^{abpq}_{cdrs}\,\varphi_a\xi_b\varphi^c\xi^d R_{pq}{}^{rs} - 4K_X\,\delta^{abpq}_{cdrs}\,\varphi_a\xi_b\varphi^c\xi^d\varphi_p{}^r\varphi_q{}^s - 4K_Y\,\delta^{abpq}_{cdrs}\,\varphi_a\xi_b\varphi^c\xi^d\xi_p{}^r\xi_q{}^s + - 4K_Z\,\delta^{abpq}_{cdrs}\,\varphi_a\xi_b\varphi^c\xi^d\varphi_p{}^r\xi_q{}^s\} \tag{3.11}$$

where $S_{22}$, $S_{23}$, $S_{24}$ and K are BST functions. The Lagrangian $\mathcal{E}_{\varphi\varphi}(\mathcal{E}_{\xi\xi}(L_{21}))$ is the Lagrangian $L_*$ appearing in (3.1).

Lastly, the Lagrangian $\mathcal{L}_3$ of the **Theorem** given in (1.15) is related to the $L_3$ of (2.29) by $\mathcal{L}_3 = -\tfrac{1}{4}L_3$. This Lagrangian will admit three extensions all of which use the 4x4 generalized Kronecker delta and hence are not of the form of $\mathcal{L}_3$. Due to the **Theorem** none of these extensions can produce field equations that differ from those provided by the Lagrangians of the **Theorem**, and so I won't bother to present them.

The above analysis suggests how to find Lagrangians that yield second-order field equations when dealing with multi-scalar tensor field theories in spaces of dimension n=2m. We know from what we have done in **Section 2** that a Lagrangian that can yield such field theories will be the sum of Lagrangians that are algebraically of $0^{th}$, $1^{st}$, $2^{nd}$, ..., $n-1^{st}$ order in the second covariant derivatives of the multi-scalar fields. Now just guess the "simplest" Lagrangian for each of those orders based upon what we have seen for the bi-scalar-tensor case in the **Theorem**. Once this has been done, determine what constraints the coefficient functions in those ansatz Lagrangians must satisfy if they are to yield second-order multi-scalar-tensor field theories. Lastly, extend each of these simplest Lagrangians as far as you can in the space that you are in, using the obvious extension operators applicable to your multi-scalar-tensor theory. In this case you should arrive at a Lagrangian that yields a fairly general second-order multi-scalar tensor field theory in a space of dimension n=2m. Proving that you have actually found a Lagrangian that yields the most general such theory will be quite a formidable and odious task.

**Acknowledgements**

I would like to thank Dr. Katzufumi Takahashi for numerous discussions on scalar-tensor field theories over the years, and for exposing me to the formidable work of Ohashi, Tanahashi, Kobayashi and Yamaguchi that appears in reference [3]. I also want to thank Professor Emeritus John Wainwright for discussions on the wave equation in Minkowski Space, Dr. Stephen J. Aldersley for discussions concerning the inverse Euler-Lagrange problem, that involves going from field equation candidates back to possible Lagrangian generators of those equations, and Dr. James

S. Barber for discussions on the possible application of bi-scalar theory to Higgs fields. Lastly I owe an immense debt of gratitude to Prof. Tsutomu Kobayashi for many discussions on the **Theorem**, its formulation and proof. Without Prof. Kobayashi's observations this paper would not have been possible.

## Appendix A: Constructing Scalar Concomitants of $g_{ij}$, $\varphi$, $\xi$, $\varphi_{,i}$ and $\xi_{,i}$

The purpose of this section is to prove

**Lemma A:** Suppose that $\theta$ is a class $C^1$ scalar concomitant with functional form $\theta = \theta(g_{ij}, \varphi, \varphi_{,i}, \xi, \xi_{,i})$. Then in a space of dimension $n \geq 3$, there exists a function F of $\varphi$, $\xi$, X, Y and Z, where

$$X := g^{ij}\varphi_{,i}\varphi_{,j}\,, \quad Y := g^{ij}\xi_{,i}\xi_{,j} \text{ and } Z := g^{ij}\varphi_{,i}\xi_{,j} \tag{A.1}$$

for which $\theta = F(\varphi, \xi, X, Y, Z)$.

**Proof:** We could appeal to Weyl [15] for this result, but he actually requires his invariants to be polynomials in their arguments (*see*, section 5, page 23 in [15]) and I rather not do that. In fact the construction I shall present only requires $\theta$ to be of class $C^1$, and proceeds as follows.

Since $\theta$ is a scalar concomitant we know that if P is an arbitrary point in an n-dimensional manifold M, with x and x' charts at P then we may write

$$\theta(g_{lm}B^l{}_iB^m{}_j, \varphi, \varphi_{,l}B^l{}_i, \xi, \xi_{,l}B^l{}_i) = \theta(g_{ij}, \varphi, \varphi_{,i}, \xi, \xi_{,i})\,, \tag{A.2}$$

where $B^r{}_s := \dfrac{\partial x^r}{\partial x'^s}$. Since $B^r{}_s$ is essentially an arbitrary invertible matrix when evaluated at P, we can differentiate (A.2) with respect to $B^r{}_s$ and evaluate that result for the identity coordinate transformation, where $B^r{}_s = \delta^r{}_s$, to obtain, after some manipulation, that

$$\theta_\varphi{}^{;s}\varphi^t + \theta_\xi{}^{;s}\xi^t + 2\theta^{;st} = 0 \tag{A.3}$$

where I am using the notation introduced in (2.4b). Since $\theta^{;st} = \theta^{;ts}$, (A.3) tells us that

$$\theta_\varphi{}^{;s}\varphi^t + \theta_\xi{}^{;s}\xi^t = \theta_\varphi{}^{;t}\varphi^s + \theta_\xi{}^{;t}\xi^s\ . \tag{A.4}$$

Next we multiply (A.4) by $\varphi_{,t}$ and by $\xi_{,t}$ to obtain the following two equations

$$X\theta_{\varphi}{}^{;s} + Z\theta_{\xi}{}^{;s} = \varphi_{,t}\theta_{\varphi}{}^{;t}\varphi^{s} + \varphi_{,t}\theta_{\xi}{}^{;t}\xi^{s}\,, \tag{A.5a}$$

$$Z\theta_{\varphi}{}^{;s} + Y\theta_{\xi}{}^{;s} = \xi_{,t}\theta_{\varphi}{}^{;t}\varphi^{s} + \xi_{,t}\theta_{\xi}{}^{;t}\xi^{s}\,. \tag{A.5b}$$

If we confine ourselves to the open (and dense) subset of X, Y, Z space where $D := XY - Z^2 \neq 0$, then (A.5) can be solved for $\theta_{\varphi}{}^{;s}$ and $\theta_{\xi}{}^{;s}$. Thus we can write

$$\theta_{\varphi}{}^{;s} = \Phi_1\varphi^{s} + \Phi_2\xi^{s} \;\text{ and }\; \theta_{\xi}{}^{;s} = \Xi_1\varphi^{s} + \Xi_2\xi^{s}\,, \tag{A.6}$$

where $\Phi_1$, $\Phi_2$, $\Xi_1$ and $\Xi_2$ are scalar concomitants of class $C^0$ and have the same functional form as $\theta$. (A.6) is the key equation in showing that $\theta$ must actually be a funtcion $\varphi$, $\xi$, X, Y and Z, as I shall presently demonstrate.

We can use the defining equations $X=g^{ij}\varphi_{,i}\varphi_{,j}$ and $Y=g^{ij}\xi_{,i}\xi_{,j}$ to solve for $\varphi_{,1}$ in terms of $g_{ij}$, X and $\varphi_{,\alpha}$ ($\alpha=2,\ldots,n$) and for $\xi_{,1}$ in terms of $g_{ij}$, Y and $\xi_{,\alpha}$. Since these equations are quadratic the solutions will have two branches, just choose one, and it will be obvious from my construction that what is done will also be valid for the other branches. Use these solutions for $\varphi_{,1}$ and $\xi_{,1}$ in the equation $Z=g^{ij}\varphi_{,i}\xi_{,j}$ to solve for $\xi_{,2}$ in terms of $g_{ij}$, X, $\varphi_{,\alpha}$, Y, Z, and $\xi_{,A}$ where $A=3,\ldots,n$. So at present we can say that there exists a scalar function $F = F(g_{ij}, \varphi, \xi, X, \varphi_{,\alpha}, Y, Z, \xi_{,A})$ for which

$$\theta = F(g_{ij}, \varphi, \xi, X, \varphi_{,\alpha}, Y, Z, \xi_{,A})\,, \tag{A.7}$$

built from $\theta$ in the obvious way. The partial derivatives of $\theta$ are related to those of F by

$$\frac{\partial\theta}{\partial\varphi_1} = 2\frac{\partial F}{\partial X}\varphi^{1} + \frac{\partial F}{\partial Z}\xi^{1} \tag{A.8}$$

$$\frac{\partial\theta}{\partial\varphi_\alpha} = 2\frac{\partial F}{\partial X}\varphi^{\alpha} + \frac{\partial F}{\partial Z}\xi^{\alpha} + \frac{\partial F}{\partial\varphi_{,\alpha}} \tag{A.9}$$

$$\frac{\partial\theta}{\partial\xi_1} = 2\frac{\partial F}{\partial Y}\xi^{1} + \frac{\partial F}{\partial Z}\varphi^{1} \tag{A.10}$$

$$\frac{\partial\theta}{\partial\xi_2} = 2\frac{\partial F}{\partial Y}\xi^{2} + \frac{\partial F}{\partial Z}\varphi^{2} \tag{A.11}$$

$$\frac{\partial\theta}{\partial\xi_{A}} = 2\frac{\partial F}{\partial Y}\xi^{A} + \frac{\partial F}{\partial Z}\varphi^{A} + \frac{\partial F}{\partial\xi_{,A}} \quad . \tag{A.12}$$

(A.6) also provide us with expressions for the partial derivatives of $\theta$ with respect to $\varphi_{,a}$ and $\xi_{,a}$. Upon comparing those expressions we can conclude that if $n \geq 3$ then

$$\frac{\partial F}{\partial\varphi_{,\alpha}} = 0 \quad \text{and} \quad \frac{\partial F}{\partial\xi_{,A}} = 0$$

and hence F is independent of $\varphi_{,\alpha}$ and $\xi_{,A}$. Note that if we were in a 2-dimensional space, the fact that $\varphi_{,1}$, $\xi_{,1}$ and $\xi_{,2}$ were functions of X, Y, Z and $\varphi_{,2}$ would prevent us from concluding that the partial of F with respect to $\varphi_{,2}$ must vanish. In fact the result we are trying to establish is not true when n=2, as I shall elaborate about below.

From what we have done so far we can use (A.7) to conclude that $\theta = F(g_{ij}, \varphi, \xi, X, Y, Z)$ when X, Y and Z are such that $XY - Z^2 \neq 0$, which is required for (A.6) to hold. Since that set is open and dense in the set of all possible values of X, Y and Z, we can use the continuity of F to deduce that $\theta = F(g_{ij}, \varphi, \xi, X, Y, Z)$ for all X, Y and Z. If we now examine the effect of a coordinate transformation at an arbitrary point P on the scalar function F we obtain the equation

$$F(g_{lm}B^{l}{}_{i}B^{m}{}_{j}, \varphi, \xi, X, Y, Z) = F(g_{ij}, \varphi, \xi, X, Y, Z) \ .$$

When this equation is differentiated with respect to $B^{r}{}_{s}$ the resulting expression can be shown to imply that the partial derivative of F with respect to $g_{ij}$ vanishes, and hence $\theta = F(\varphi, \xi, X, Y, Z)$ as desired.■

In a space of 2-dimension **Lemma A** is false. For in a 2-space we also have the scalar invariant $W := g^{-\frac{1}{2}}\varepsilon^{ij}\varphi_{,i}\xi_{,j}$. So the defining equations for W, X, Y and Z can be solved for $\varphi_{,1}$, $\varphi_{,2}$, $\xi_{,1}$ and $\xi_{,2}$ to show that if $\theta=\theta(g_{ij},\varphi,\varphi_{,i}, \xi, \xi_{,i})$ then there exists a function $F = F(\varphi, \xi, W, X, Y, Z)$ for which $\theta = F$.

**Appendix B: Construction of Covariant Tensors with Property S in a space of 4-Dimensions**

We need to construct in a space of 4-dimensions all tensorial concomitants (of weight 0) $\theta_{i_1 \cdots i_{2k}} = \theta_{i_1 \cdots i_{2k}}(g_{hi}, \varphi, \varphi_{,h}, \xi, \xi_{,h})$ with property S, when k=1,2,3,4. To obtain the corresponding results for tensor densities just multiply the results derived here by $g^{½}$. We begin with k=1.

Let $\theta_{ab} = \theta_{ab}(g_{hk}, \varphi, \varphi_{,h}, \xi, \xi_{,h})$ be a symmetric tensorial concomitant. If P is an arbitrary point of our differentiable manifold M, with x and x' charts at P, then the $\theta_{ab}$ version of (A.2) at P is

$$\theta_{ab}(g_{rs}B^r{}_h B^s{}_k, \varphi, \varphi_{,r}B^r{}_h, \xi, \xi_{,r}B^r{}_h) = \theta_{rs}B^r{}_a B^s{}_b\,. \tag{B.1}$$

Upon differentiating this equation with respect to $B^l{}_m$ and then evaluating the result for the identity coordinate transformation we find that, after some manipulation,

$$2\theta_{ab}{}^{;lm} = \theta^l{}_b\delta^m{}_a + \theta_a{}^l\delta^m{}_b - \frac{\partial\theta_{ab}}{\partial\varphi_{,m}}\varphi^l - \frac{\partial\theta_{ab}}{\partial\xi_{,m}}\xi^l\,. \tag{B.2}$$

Since the left-hand side of (B.2) is symmetric in l,m, the right-hand side must also enjoy this symmetry. Consequently (B.2) implies that

$$\theta^l{}_b\delta^m{}_a + \theta_{al}\delta^m{}_b - \frac{\partial\theta_{ab}}{\partial\varphi_{,m}}\varphi^l - \frac{\partial\theta_{ab}}{\partial\xi_{,m}}\xi^l = \theta^m{}_b\delta^l{}_a + \theta_a{}^m\delta^l{}_b - \frac{\partial\theta_{ab}}{\partial\varphi_{,l}}\varphi^m - \frac{\partial\theta_{ab}}{\partial\xi_{,l}}\xi^m\,. \tag{B.3}$$

What we want to do is solve (B.3) for $\theta_{ab}$ in terms of $g_{ij}$, $\varphi_{,i}$, $\xi_{,i}$ and vector tensorial concomitants of the same functional form as $\theta_{ab}$. To that end we begin by contracting on l and b to obtain

$$\theta\delta_a{}^m + \theta_a{}^m - \frac{\partial\theta_{ab}}{\partial\varphi_{,m}}\varphi^b - \frac{\partial\theta_{ab}}{\partial\xi_{,m}}\xi^b = \theta^m{}_a + n\theta_a{}^m - \frac{\partial\theta_{ab}}{\partial\varphi_{,b}}\varphi^m - \frac{\partial\theta_{ab}}{\partial\xi_{,b}}\xi^m\,, \tag{B.4}$$

where $\theta := \theta_b{}^b$ and n is the dimension of the space, which we shall eventually choose to be 4. To rewrite the third and fourth terms on the left-hand side of (B.4) in a more useful form note that

$$\frac{\partial(\theta_{ab}\varphi^b)}{\partial\varphi_{,m}} = \frac{\partial\theta_{ab}}{\partial\varphi_{,m}}\varphi^b + \theta_a{}^m \quad\text{and}\quad \frac{\partial(\theta_{ab}\xi^b)}{\partial\xi_{,m}} = \frac{\partial\theta_{ab}}{\partial\xi_{,m}}\xi^b + \theta_a{}^m\,. \tag{B.5}$$

Combining (B.4) and (B.5) we easily find that when $\theta_a{}^m = \theta^m{}_a$

$$(n-2)\theta_a{}^m = \theta\delta_a{}^m - \frac{\partial(\theta_{ab}\varphi^b)}{\partial\varphi_{,m}} - \frac{\partial(\theta_{ab}\xi^b)}{\partial\xi_{,m}} + \frac{\partial\theta_{ab}}{\partial\varphi_{,b}}\varphi^m + \frac{\partial\theta_{ab}}{\partial\xi_{,b}}\xi^m\,. \tag{B.6}$$

Now if we knew the most general $\theta_a=\theta_a(g_{ij},\varphi,\varphi_{,i},\xi,\xi_{,i})$ we could employ it to compute the right-hand side of (B.6). Well by proceeding in the footsteps taken to get from (B.1) to (B.6) it is easy to demonstrate for $\theta_a$ we have

$$(n-3)\theta^m = \frac{\partial\theta_l}{\partial\varphi_{,l}}\varphi^m + \frac{\partial\theta_l}{\partial\xi_{,l}}\xi^m - \frac{\partial(\theta_l\varphi^l)}{\partial\varphi_{,m}} - \frac{\partial(\theta_l\xi^l)}{\partial\xi_{,m}} . \qquad \text{(B.7)}$$

We can employ **Lemma A** to give form to the four scalar concomitants on the right-hand side of (B.7) and thereby establish

**Lemma B.1:** In a space of dimension $n>3$ the most general tensorial concomitant of the form $\theta_a = \theta_a(g_{ij}, \varphi, \varphi_{,i}, \xi, \xi_{,i})$ is given by $\theta_a = \theta_1\varphi_{,a} + \theta_2\xi_{,a}$ where $\theta_1$ and $\theta_2$ are arbitrary scalar concomitants of $\varphi$, $\xi$, X, Y and Z.■

Note that when n=3 there is another vector that could be added to $\theta^m$; *viz.,* $\theta_3 g^{-\frac{1}{2}}\varepsilon^{abc}\varphi_{,b}\xi_{,c}$ where $\theta_3$ is an arbitrary scalar concomitant of $\varphi$, $\xi$, X, Y and Z.

Using **Lemma A** and **Lemma B.1** in (B.6) yields, after some work, the proof of

**Lemma B.2:** In a space of dimension $n>3$ the most general symmetric tensorial concomitant of the form $\theta_{ab} = \theta_{ab}(g_{ij}, \varphi, \varphi_{,i}, \xi, \xi_{,i})$ is given by

$$\theta_{ab}= \theta_1 g_{ab} + \theta_2\varphi_{,a}\varphi_{,b} + \theta_3\xi_{,a}\xi_{,b} + \theta_4(\varphi_a\xi_b+\varphi_b\xi_a) , \qquad \text{(B.8)}$$

where for A=1,...,4, $\theta_A=\theta_A(\varphi,\xi,X,Y,Z)$ are arbitrary.■

**Lemma B.2** is also valid in a space of 2-dimensions, but the proof that I have provided does not show that.

Before I sketch how one goes about constructing the 4 index tensorial concomitant with Property S that we shall need in **Lemma 4**, I shall demonstrate how to build the 8 and 6 index tensorial concomitants with Property S, since their form surprisingly follows from **Lemma A** and **Lemma B.2**.

In **Section 2** I point out that Lovelock [13] has shown that the 10 index tensorial concomitant which has Property S in a space of 4-dimensions vanishes. Let $\theta_{i_1 \cdots i_8}$ be a tensorial concomitant of $g_{ij}$, $\varphi$, $\varphi_{,i}$, $\xi$ and $\xi_{,i}$ that has Property S. Then it only has one independent component; *viz.,* $\theta_{11223344}$. Scalar concomitants also only have one independent component which suggests that there might be a relationship between the 8 index tensorial concomitants with Property S and the scalar concomitants.

Now a 6 index tensorial concomitant $\theta_{i_1 \cdots i_6}$ with Property S has 10 independent components in a space of 4-dimensions; *viz.,* $\theta_{112233}$, $\theta_{112244}$, $\theta_{113344}$, $\theta_{223344}$, $\theta_{123344}$, $\theta_{132244}$, $\theta_{142233}$, $\theta_{231144}$, $\theta_{241133}$ and $\theta_{341122}$, which is precisely the same number of independent components that a symmetric 2 index tensorial concomitant in a space of 4-dimensions would have. This is not a coincidence. To establish the connection between these spaces of tensors I have to introduce an 8 index tensorial concomitant with property S that will provide us with maps between these tensor concomitant spaces.

In [13] Lovelock introduced an 8 index tensor with Property S. To define it let $\Sigma_{a,b}$ denote the operator which has the following action when applied to a tensor with components $T^{\ldots a \ldots b \ldots}$

$$\Sigma_{a,b}(T^{\ldots a \ldots b \ldots}) := \tfrac{1}{2}(T^{\ldots a \ldots b \ldots} + T^{\ldots b \ldots, a \ldots}). \tag{B.9}$$

I now define the 8-index Lovelock tensor (in a space of 4-dimensions) by

$$\Theta_{L\, i_1 \ldots i_8} := g\Sigma_{i_1 i_2} \ldots \Sigma_{i_7 i_8}\, \varepsilon_{i_1 i_3 i_5 i_7}\, \varepsilon_{i_2 i_4 i_6 i_8}\ . \tag{B.10}$$

$\Theta_L$ is not identically zero since we have $\Theta_{L11223344}=g$. In addition if $\theta_{a \ldots b}$ is an 8 index tensor with Property S, then $\theta_{a \ldots b} = g^{-1}\theta_{11223344}\Theta_{La \ldots b}$, since they both have Property S and the same non-zero component. This establishes the relationship between scalar concomitants of $g_{ij}$, $\varphi$, $\varphi_{,i}$, $\xi$, $\xi_{,i}$ and 8 index tensorial concomitants of the same arguments with Property S.

For k=0,1,2,3,4 I let $\mathbf{PS_{2k}}$ denote the vector space of covariant tensorial concomitants of $g_{ij}$,

$\varphi$, $\varphi_{,i}$, $\xi$ and $\xi_{,i}$ which have Property S and rank 2k. I define the Lovelock map, $\Theta_L$, mapping **PS₂** into **PS₆** by

$$\Theta_L\colon \theta_{ab} \rightarrow \theta^{ab}\Theta_{Labcdefhi}\,.$$

We shall also let $\Theta_L$ denote the map from **PS₆** into **PS₂** defined by

$$\Theta_L\colon \theta_{abcdef} \rightarrow \theta^{abcdef}\Theta_{Labcdefhi}\,.$$

Clearly the map $\Theta_L$ is a linear map between these two vector spaces. It is a straightforward matter (relying heavily on Property S) to show that

$$\Theta_L(\Theta_L(\theta_{ab})) = \pm 12\,\theta_{ab} \quad\text{and}\quad \Theta_L(\Theta_L(\theta_{abcdef})) = \pm 18\theta_{abcdef} \qquad \text{(B.11)}$$

where the minus sign applies for Lorentzian signatures, and the plus sign to all other signatures of the metric tensor. Due to (B.11) we see that $\Theta_L$ maps **PS₂** onto **PS₆** and hence due to **Lemma B.2** we can easily establish

**Lemma B.3:** In a space of 4-dimensions the most general 8 index and 6 index tensorial concomitants of $g_{ij}$, $\varphi$, $\varphi_{,i}$, $\xi$ and $\xi_{,i}$ are given by

$$\theta_{abcdefhi} = g\theta\{\varepsilon_{aceh}\varepsilon_{bdfi} + \varepsilon_{bceh}\varepsilon_{adfi} + \varepsilon_{adeh}\varepsilon_{bcfi} + \varepsilon_{bdeh}\varepsilon_{acfi} + \varepsilon_{acfh}\varepsilon_{bdei} + \varepsilon_{bcfh}\varepsilon_{adei} + \varepsilon_{adfh}\varepsilon_{bcei} + \varepsilon_{bdfh}\varepsilon_{acei}\}, \qquad \text{(B.12)}$$

$$\begin{aligned}
\theta^{abcdef} = {} & \alpha_1\{\delta^{ace}_{pqr}\,g^{pb}g^{qd}g^{rf} + \delta^{bce}_{pqr}\,g^{pa}g^{qd}g^{rf} + \delta^{ade}_{pqr}\,g^{pb}g^{qc}g^{rf} + \delta^{bde}_{pqr}\,g^{pa}g^{qc}g^{rf}\} + \\
& + \alpha_2\{\delta^{aceh}_{pqrs}\,\varphi_h\varphi^s g^{pb}g^{qd}g^{rf} + \delta^{bceh}_{pqrs}\,\varphi_h\varphi^s g^{pa}g^{qd}g^{rf} + \delta^{adeh}_{pqrs}\,\varphi_h\varphi^s g^{pb}g^{qc}g^{rf} + \delta^{bdeh}_{pqrs}\,\varphi_h\varphi^s g^{pa}g^{qc}g^{rf}\} + \\
& + \alpha_3\{\delta^{aceh}_{pqrs}\,\xi_h\xi^s g^{pb}g^{qd}g^{rf} + \delta^{bceh}_{pqrs}\,\xi_h\xi^s g^{pa}g^{qd}g^{rf} + \delta^{adeh}_{pqrs}\,\xi_h\xi^s g^{pb}g^{qc}g^{rf} + \delta^{bdeh}_{pqrs}\,\xi_h\xi^s g^{pa}g^{qc}g^{rf}\} + \\
& + \alpha_4\{\delta^{aceh}_{pqrs}\,\varphi_h\xi^s g^{pb}g^{qd}g^{rf} + \delta^{bceh}_{pqrs}\,\varphi_h\xi^s g^{pa}g^{qd}g^{rf} + \delta^{adeh}_{pqrs}\,\varphi_h\xi^s g^{pb}g^{qc}g^{rf} + \delta^{bdeh}_{pqrs}\,\varphi_h\xi^s g^{pa}g^{qc}g^{rf} + \\
& + \delta^{aceh}_{pqrs}\,\xi_h\varphi^s g^{pb}g^{qd}g^{rf} + \delta^{bceh}_{pqrs}\,\xi_h\varphi^s g^{pa}g^{qd}g^{rf} + \delta^{adeh}_{pqrs}\,\xi_h\varphi^s g^{pb}g^{qc}g^{rf} + \delta^{bdeh}_{pqrs}\,\xi_h\varphi^s g^{pa}g^{qc}g^{rf}\}\,, \qquad \text{(B.13)}
\end{aligned}$$

where $\theta$, $\alpha_1$, $\alpha_2$, $\alpha_3$ and $\alpha_4$ are scalar concomitants of $\varphi$, $\xi$, X, Y and Z.

**Proof:** (B.12) is obvious, and (B.13) follows from (B.8) and (B.10) since $\delta^{abcd}_{pqrs} = \varepsilon^{abcd}\,\varepsilon_{pqrs}$.■

At this juncture one should note that $\Theta_L$ also provides us with a map from $\mathbf{PS_4}$ into itself defined in the obvious way. This map is such that $\Theta_L(\Theta_L(\theta_{abcd})) = \pm 9\theta_{abcd}$ , and hence $\Theta_L$ is an isomorpism of $\mathbf{PS_4}$ onto itself. Unfortunately $\Theta_L$ an can not be used to build the elements of $\mathbf{PS_4}$. I shall show how that is done shortly. But first some remarks on how to build concomitants with Property S in spaces of even dimension n=2m greater than 4.

The above work has shown us that in a space of four-dimensions we have a "duality" (*i.e.,* a canonical isomorphism) between the spaces $\mathbf{PS_{2k}}$ and $\mathbf{PS_{2(4\text{-}k)}}$ for k=0,1 which enables us to build elements of $\mathbf{PS_{2(4\text{-}k)}}$ from elements of $\mathbf{PS_{2k}}$ when k=0,1. I suspect that this duality persists when n=2m (m>2). Let me elaborate.

For k=0,1,...,n=2m, set

$\mathbf{PS_{2k}(n)}$:= {tensorial concomitants of $g_{ij}$, $\varphi$, $\varphi_{,i}$, $\xi$ and $\xi_{,i}$ , with Property S and covariant rank 2k in a space of dimension n}, (B.14)

and let

$$\Theta_L{}^{i_1 j_1 \ldots i_n j_n} := g^{-1}\, \Sigma_{i_1 j_1} \ldots \Sigma_{i_n j_n}\, \varepsilon^{i_1 i_2 \ldots i_n}\, \varepsilon^{j_1 j_2 \ldots j_n} \quad . \tag{B.15}$$

$\Theta_L{}^{i_1 \ldots j_n}$ gives rise to a Lovelock map, $\Theta_L$, from $\mathbf{PS_{2k}(n)}$ into $\mathbf{PS_{2(n\text{-}k)}(n)}$ in the obvious way. My conjecture is that this map is in fact an isomorphism. If that can be demonstrated then the duality between these spaces can easily be employed to build the elements of $\mathbf{PS_{2(n-k)}(n)}$ from the elements of $\mathbf{PS_{2k}(n)}$ when k=0,1,...,m−1, m≥2. This would prove to be immensely useful to those trying to generalize the **Theorem** to spaces of dimension >4, or to more than two scalar fields, or to both.

We shall now return to the problem of building the most general element, $\theta_{abcd}$ of $\mathbf{PS_4}$.

As usual we begin with the basic transformation which $\theta_{abcd}$ must satisfy since it is to be a tensorial concomitant; *viz*., at an arbitrary point P

$$\theta_{abcd}(g_{pq}B^p{}_r B^q{}_s,\ \varphi,\ \varphi_{,p}B^p{}_r,\ \xi,\ \xi_{,p}B^p{}_r) = \theta_{pqrs}B^p{}_a B^q{}_b B^r{}_c B^s{}_d\ . \tag{B.16}$$

After differentiating (B.16) with respect to $B^l_{\ m}$, setting $B^l_{\ m}=\delta^l_{\ m}$ and raising the index l we find after some rearrangement that

$$2\theta_{abcd}{}^{;lm} + \theta_{\varphi abcd}{}^{;m}\varphi^l + \theta_{\xi abcd}{}^{;m}\xi^l = \delta^m_{\ a}\theta^l_{\ bcd} + \delta^m_{\ b}\theta_a{}^l{}_{cd} + \delta^m_{\ c}\theta_{ab}{}^l{}_d + \delta^m_{\ d}\theta_{abc}{}^l\,. \tag{B.17}$$

The first term on the left-hand side of (B.17) is symmetric in l and m. Thus if we take the remaining terms on the left-hand side over to the right-hand side, the new right-hand side will have to be symmetric in l and m. So upon taking this new right-hand side and equating it to itself after interchanging l and m, we discover after contracting on a and m and lowering the index l that (where I have yet to assume that $\theta_{abcd}$ has any symmetries)

$$(n-3)\theta_{lbcd} + \theta_{blcd} + \theta_{cbld} + \theta_{dbcl} = g_{lb}\theta_a{}^a{}_{cd} + g_{lc}\theta_{ab}{}^a{}_d + g_{ld}\theta_{abc}{}^a + \frac{\partial\theta_{abcd}}{\partial\varphi_{,a}}\varphi_{,l} + \frac{\partial\theta_{abcd}}{\partial\xi_{,a}}\xi_{,l} - \frac{\partial(\theta_{abcd}\varphi^a)}{\partial\varphi_{,p}}g_{pl} + $$
$$- \frac{\partial(\theta_{abcd}\xi^a)}{\partial\xi_{,p}}g_{pl}\,. \tag{B.18}$$

If we now choose n=4 and invoke Property S, then (B.18) will give us

$$\theta_{lbcd} = g_{lb}\theta_a{}^a{}_{cd} - \tfrac{1}{2}\, g_{lc}\theta_a{}^{ab}{}_d - \tfrac{1}{2}\, g_{ld}\theta_a{}^a{}_{bc} + \frac{\partial\theta_{abcd}}{\partial\varphi_{,a}}\varphi_{,l} + \frac{\partial\theta_{abcd}}{\partial\xi_{,a}}\xi_{,l} - \frac{\partial(\theta_{abcd}\varphi^a)}{\partial\varphi_{,p}}g_{pl} - \frac{\partial(\theta_{abcd}\xi^a)}{\partial\xi_{,p}}g_{pl}\,, \tag{B.19}$$

which expresses $\theta_{lbcd}$ in terms of two index concomitants with Property S, and three index concomitants, $t_{bcd}$, which have Property T; *viz.,*

$$t_{bcd} = t_{bdc} \text{ and } t_{(bcd)} = 0. \tag{B.20}$$

**Lemma B.2** takes care of the two index concomitants which are symmetric, so I now have to demonstrate how to build the three index concomitants with Property T.

To construct the 3 index concomitants $\theta_{abc}$ with Property T we begin, as usual, with the transformation equation

$$\theta_{abc}(g_{pq}B^p_{\ r}B^q_{\ s}, \varphi, \varphi_{,p}B^p_{\ r}, \xi, \xi_{,p}B^p_{\ r}) = \theta_{pqr}B^p_{\ a}B^q_{\ b}B^r_{\ c}\,.$$

Working with this equation as we have done previously we find that when $\theta_{abc}$ has Property T

$$\frac{\partial\theta_{abc}}{\partial\varphi_a}\varphi_d + \frac{\partial\theta_{abc}}{\partial\xi_a}\xi_d - g_{dl}\frac{\partial(\theta_{abc}\varphi^a)}{\partial\varphi_l} - g_{dl}\frac{\partial(\theta_{abc}\xi^a)}{\partial\xi_l} = (n-4)\theta_{dbc} - \theta_{aac}g_{bd} - \theta_{ab}{}^a g_{cd}\,. \tag{B.21}$$

Thus we see that in a space of 4-dimensions our usual approach to constructing concomitants has failed catastrophically! To see how it might still be possible to compute $\theta_{abc}$ let us go back and examine the 2 index concomitant $\theta_{ab}$ that we constructed under the assumption that it was symmetric.

If we do not assume that $\theta_{ab}$ is symmetric, then the analysis that led from (B.4) to (B.6) would yield

$$\theta_{ad} + \theta_{da} = \theta g_{md} - \frac{\partial(\theta_{ab}\varphi^b)}{\partial\varphi_{,m}} g_{md} - \frac{\partial(\theta_{ab}\xi^b)}{\partial\xi_{,m}} g_{md} + \frac{\partial\theta_{ab}}{\partial\varphi_{,b}}\varphi_d + \frac{\partial\theta_{ab}}{\partial\xi_{,b}}\xi_d \ . \qquad \text{(B.22)}$$

So to compute the arbitrary 2 index concomitant of $g_{ij}$, $\varphi$, $\varphi_{,i}$, $\xi$ and $\xi_{,i}$ we would need to know the form of $\theta_{ab}$ when it is antisymmetric, because (B.22) won't help. This can be done as follows.

Since $\theta_{ab}\varphi^b$ and $\theta_{ab}\xi^b$ are vector concomitants we know that there exist scalar functions M, N, P and Q for which

$$\theta_{ab}\varphi^b = M\varphi_a + N\xi_a \quad \text{and} \quad \theta_{ab}\xi^b = P\varphi_a + Q\xi_a \ .$$

Since $\theta_{ab}$ is assumed to be antisymmetric it is easy to show by looking at $\theta_{ab}\varphi^a\varphi^b$, $\theta_{ab}\varphi^a\xi^b$, $\theta_{ab}\xi^a\xi^b$ and $\theta_{ab}\xi^a\varphi^b$ that

$$\theta_{ab}\varphi^b = M\varphi_a - MXZ^{-1}\xi_a \ \text{and} \ \theta_{ab}\xi^b = MYZ^{-1}\varphi_a - M\xi_a \ . \qquad \text{(B.23)}$$

Similarly for the dual of $\theta_{ab}$ we would have

$$*\theta_{ab}\varphi^b = M'\varphi_a - M'XZ^{-1}\xi_a \ \text{and} \ *\theta_{ab}\xi^b = M'YZ^{-1}\varphi_a - M'\xi_a \ , \qquad \text{(B.24)}$$

for a scalar concomitant M'. Now let $\{V^a, W^a\}$ be orthogonal unit vectors which span the plane perpendicular to the one spanned by $\{\varphi^a, \xi^a\}$. We can expand the two form $\theta_{ab}$ as

$$\theta_{ab} = \alpha_1(\varphi_a\xi_b - \varphi_b\xi_a) + \alpha_2(\varphi_a V_b - \varphi_b V_a) + \alpha_3(\varphi_a W_b - \varphi_b W_a) + \alpha_4(\xi_a V_b - \xi_b V_a) + \alpha_5(\xi_a W_b - \xi_b W_a) +$$
$$+ \alpha_6(V_a W_b - V_b W_a), \qquad \text{(B.25)}$$

where $\alpha_1, \ldots, \alpha_6$ are scalar concomitants. By examining the contraction of $\theta_{ab}$ from (B.25) with $\varphi^b$ and $\xi^b$, and comparing that result with (B.23) we find that on the set $\Delta := \{(X,Y,Z)|\ XY - Z^2 \neq 0\}$

$$\theta_{ab} = \alpha_1(\varphi_a\xi_b - \varphi_b\xi_a) + \alpha_6(V_aW_b - V_bW_a)\ . \tag{B.26}$$

Using (B.26) along with (B.24) it is possible to show that the second term on the right-hand side of (B.26) is a scalar multiple of the dual of $(\varphi_a\xi_b - \varphi_b\xi_b)$ and hence we have established

**Lemma B.3:** The most general skew symmetric 2 index concomitant of $g_{ij}$, $\varphi$, $\varphi_{,i}$, $\xi$ and $\xi_{,i}$ in a space of 4-dimensions is

$$\theta_{ab} = \alpha_1\omega_{ab} + \alpha_2 *\omega_{ab}$$

where $\omega_{ab} := \frac{1}{2}(\varphi_a\xi_b - \varphi_b\xi_a)$, $*\omega_{ab} := \frac{1}{2}g^{-\frac{1}{2}}\varepsilon_{abcd}\omega^{cd}$, with $\alpha_1$ and $\alpha_2$ being scalar concomitants of $\varphi$, $\xi$, X, Y and Z.■

Now one can employ arguments similar to the one we used above to prove **Lemma B.3** to convince oneself that our coveted 3 index tensorial concomitant with Property T can be built as a polynomial using $g_{ij}$, $\varphi_{,i}$ , $\xi_{,i}$ and the Levi-Civita symbol along with scalar coefficients of $\varphi$, $\xi$, X, Y and Z. You can then use that quantity with **Lemma B.2** and (B.1) to try to build the 4 index concomitant with Property S. It would be a mess. The most important thing to come out of that approach is the realization that $\theta_{abcd}$ must be a polynomial in $g_{ij}$, $\varphi_{,i}$, $\xi_{,i}$ and $\varepsilon_{hijk}$, and so we essentially established Wehl's [15] result on how to build that concomitant. The end result of using that approach is presented in

**Lemma B.4:** In a space of 4-dimensions the most general 3 index tensorial concomitant with Property T and the most general 4 index tensorial concomitant with Property S which are functions of $g_{ij}$, $\varphi$, $\varphi_{,i}$, $\xi$ and $\xi_{,i}$ are given by

$$\theta_{abc} = \alpha_1(g_{ac}\varphi_b + g_{ab}\varphi_c - 2g_{bc}\varphi_a) + \alpha_2(g_{ac}\xi_b + g_{ab}\xi_c - 2g_{bc}\xi_a) + \alpha_3(\varphi_a\xi_b\varphi_c + \varphi_a\varphi_b\xi_c - 2\xi_a\varphi_b\varphi_c) +$$
$$+ \alpha_4(\xi_a\varphi_b\xi_c + \xi_a\xi_b\varphi_c - 2\varphi_a\xi_b\xi_c) + \alpha_5(*\omega_{ab}\varphi_c + *\omega_{ac}\varphi_b) + \alpha_6(*\omega_{ab}\xi_c + *\omega_{ac}\xi_b) \tag{B.27}$$

$$\theta_{abcd} = \beta_1(g_{ac}g_{bd} + g_{ad}g_{bc} - 2g_{ab}g_{cd}) + \beta_2(g_{ac}\varphi_b\varphi_d + g_{ad}\varphi_b\varphi_c + g_{bc}\varphi_a\varphi_d + g_{bd}\varphi_a\varphi_c - 2g_{ab}\varphi_c\varphi_d - 2g_{cd}\varphi_a\varphi_b) +$$

$$+ \beta_3(g_{ac}\xi_b\xi_d + g_{ad}\xi_b\xi_c + g_{bc}\xi_a\xi_d + g_{bd}\xi_a\xi_c - 2g_{ab}\xi_c\xi_d - 2g_{cd}\xi_a\xi_b) + \beta_4(g_{ac}\varphi_b\xi_d + g_{ac}\xi_b\varphi_d + g_{bd}\varphi_a\xi_c + g_{bd}\xi_a\varphi_c +$$

$$+ g_{ad}\varphi_b\xi_c + g_{ad}\xi_b\varphi_c + g_{bc}\varphi_a\xi_d + g_{bc}\xi_a\varphi_d - 2g_{ab}\varphi_c\xi_d - 2g_{ab}\xi_c\varphi_d - 2g_{cd}\varphi_a\xi_b - 2g_{cd}\xi_a\varphi_b) +$$

$$+ \beta_5(\varphi_a\varphi_c\xi_b\xi_d + \varphi_b\varphi_c\xi_a\xi_d + \varphi_a\varphi_d\xi_b\xi_c + \varphi_b\varphi_d\xi_a\xi_c - 2\varphi_a\varphi_b\xi_c\xi_d - 2\varphi_c\varphi_d\xi_a\xi_b) +$$

$$+ \beta_6({*\omega_{bc}}\varphi_d\xi_a + {*\omega_{da}}\varphi_b\xi_c + {*\omega_{ac}}\varphi_d\xi_b + {*\omega_{ca}}\varphi_b\xi_d + {*\omega_{bd}}\varphi_c\xi_a + {*\omega_{db}}\varphi_a\xi_c + {*\omega_{ad}}\varphi_c\xi_b + {*\omega_{cb}}\varphi_a\xi_d) \qquad \text{(B.28)}$$

where $\alpha_1,..., \alpha_6$, $\beta_1,..., \beta_6$ are scalar concomitants of $\varphi$, $\xi$, X, Y and Z with ${*\omega_{ab}} := \frac{1}{2} g^{-\frac{1}{2}} \varepsilon_{abcd}\varphi^c\xi^d$.■

**Appendix C: Remarks on the Proof of Lemma 5**

In **Lemma 5** I present some of the conditions which the coefficients appearing in the $A^{ij}$ of **Lemma 4** must satisfy if $A^{ij}{}_{|j}$ is to be of second-order and such that $A^{ij}{}_{|j} \equiv 0 \text{mod}(\varphi^i,\xi^i)$. I shall now give an indication of how I arrived at those conditions.

To begin one writes out $A^{ij}{}_{|j}$ and while doing so drops all second-order terms that are proportional to either $\varphi^i$ or $\xi^i$. The terms obtained can be placed into 22 separate groups involving R....R...., $(\varphi..)^2$R...., $(\xi..)^2$R...., $(\varphi..)(\xi..)$R...., $\varphi$..R...., $\xi$..R...., R...., $(\varphi..)^4$, $(\varphi..)^3(\xi..)$, $(\varphi..)^2(\xi..)^2$, $(\varphi..)(\xi..)^3$, $(\xi..)^4$, $(\varphi..)^3$, $(\varphi..)^2(\xi..)$, $(\varphi..)(\xi..)^2$, $(\xi..)^3$, $(\varphi..)^2$, $(\varphi..)(\xi..)$, $(\xi..)^2$, $(\varphi..)$, $(\xi..)$ and terms not involving $\varphi$.., $\xi$.. or R.... , where $\varphi$.., and $\xi$.. denote terms involving the second derivatives of $\varphi$ and $\xi$. Each of these groups must be $\equiv 0 \text{mod}(\varphi^i,\xi^i)$. For the purposes of **Lemma 5**, I did not bother to examine the blocks of terms in $A^{ij}{}_{|j}$ that were of second, first or zero degree in $\{\varphi.., \xi..\}$. These terms are examined in **Lemma 10**.

To begin our analysis of the terms in $A^{ij}{}_{|j}$, let P be an arbitrary point in a 4-dimensional manifold with $\varphi$ and $\xi$ arbitrary scalar fields on a neighborhood of P, and $g_{ij}$ a pseudo-Riemannian metric tensor on a neighborhood of P. So in general $\{\varphi^a,\xi^a\}$ spans a 2-plane at P, and let $V^a$ be any unit vector perpendicular to the $\{\varphi^a, \xi^a\}$ plane. The results we establish under these assumptions will be true for the case when $\varphi^a$ and $\xi^a$ are parallel by continuity. The condition that $A^{ij}{}_{|j} \equiv 0 \text{mod}(\varphi^i,\xi^i)$ is equivalent to $V_iA^{ij}{}_{|j}=0$. To assist in our analysis of the blocks of terms in $V_iA^{ij}{}_{|j}=0$, it is often convenient to evaluate the terms in a block for special choices of the metric tensor, such as a space of constant curvature, or a space with vanishing Ricci tensor. When such choices are made, that does not really effect the coefficient functions in the block since they will be functions of φ, ξ, X, Y and Z, which can still assume all possible values subject to $XY - Z^2 \neq 0$. I let $W^a$ be another vector at P, which is perpendicular to the $\{\varphi^a, \xi^a\}$ plane and require $g_{ij}V^iV^j = \varepsilon_V$, $g_{ij}W^iW^j = \varepsilon_W$, $g_{ij}V^iW^j = 0$, with $\varepsilon_V = \pm 1$, and $\varepsilon_W = \pm 1$ depending upon the signature of the metric and the nature of the $\{\varphi^a, \xi^a\}$ plane. So the ordered collection of four contravariant vectors $\{\varphi^a, \xi^a, V^a, W^a\}$, numbered 1,2,3,4, forms a basis for $T_PM$ (the tangent space to M at P) that we can employ to expand the components of the curvature tensor when the need arises. In a 4-dimensional space the curvature tensor has 20 independent components, but when we require $R_{ij}=0$, only 10 of those components are independent. In those instances we can take the independent components to be

$$R_{1213}, R_{1214}, R_{1223}, R_{1224}, R_{1313}, R_{1314}, R_{1324}, R_{1414}, R_{2323}, R_{2324} \tag{C.1}$$

with the dependent components being

$$R_{1212}, R_{1234}, R_{1323}, R_{1334}, R_{1424}, R_{1434}, R_{2334}, R_{2424}, R_{2434}, R_{3434} \tag{C.2}$$

where, *e.g.,* $R_{1324} := R_{abcd}\varphi^aV^b\xi^cW^d$. (Note that some of the dependent components that could be in the list (C.2), such as $R_{1423}$, have been left out since they are obtainable from other elements in (C.1) and(C.2).)

With these preliminaries disposed of let us now begin our examination of the various blocks of terms in $V_iA^{ij}{}_{|j}=0$. The term that is quadratic in the curvature tensor is given by

$$0= g^{1/2}V_i\{\alpha_1\,\delta^{abei}_{cdfq}\,R_{ab}{}^{cd}\varphi_mR_e{}^{mfq} + \alpha_2\,\delta^{abei}_{cdfq}\,R_{ab}{}^{cd}\xi_mR_e{}^{mfq}\}.$$

This term vanishes identically since in a four-dimensional space

$$0= \delta^{iabcd}_{pqrst}\,\varphi^pR_{ab}{}^{qr}R_{cd}{}^{st} \text{ and } 0 = \delta^{iabcd}_{pqrst}\,\xi^pR_{ab}{}^{qr}R_{cd}{}^{st}\,,$$

and hence provides no constraint on $\alpha_1$ and $\alpha_2$.

Other dimensionally dependent identities you will require in analyzing $V_i A^{ij}{}_{|j}=0$ are

$$0 = \delta^{iabcd}_{mpqrs}\, \xi^m \xi_a{}^p \xi_b{}^q R_{cd}{}^{rs} \quad \text{and} \quad 0 = \delta^{iabcd}_{mpqrs}\, \varphi^m \xi_a{}^p \xi_b{}^q R_{cd}{}^{rs},$$

along with similar ones obtained from these by replacing $\xi_a{}^p$ by $\varphi_a{}^p$, and the pair $\xi_a{}^p \xi_b{}^q$ by $\varphi_a{}^p \varphi_b{}^q$.

An analysis of the terms involving R....φ..φ.., R....ξ..ξ.. and R....φ..ξ.., in $V_i A^{ij}{}_{|j}=0$ shows

$$\alpha_{1X} = -{}^3/_4\beta_1,\ \alpha_{1Z} = -\tfrac{1}{2}\beta_2,\ \alpha_{2Y} = -{}^3/_4\beta_4,\ \alpha_{2Z} = -\tfrac{1}{2}\beta_3,\ \alpha_{2X} = -\tfrac{1}{4}\beta_2,\ \alpha_{1Y} = -\tfrac{1}{4}\beta_3\,. \tag{C.3}$$

Next examination of the terms involving $(\varphi..)^4$, $(\varphi..)^3\xi..$, $(\varphi..)^2\ (\xi..)^2$, $\varphi..(\xi..)^3$ and $(\xi..)^4$ all yield noting new.

The term involving R.... in $V_i A^{ij}{}_{|j}=0$ is given by

$$\begin{aligned}0=\ &\alpha_{31\varphi} V_i\, \delta^{abi}_{cdq}\, \varphi^q R_{ab}{}^{cd} + \alpha_{31\xi} V_i\, \delta^{abi}_{cdq}\, \xi^q R_{ab}{}^{cd} + \alpha_{32\xi} V_i\, \delta^{abei}_{cdfq}\, \xi^q \varphi_e \varphi^f R_{ab}{}^{cd} + \alpha_{33\varphi} V_i\, \delta^{abei}_{cdfq}\, \varphi^q \xi_e \xi^f R_{ab}{}^{cd} + \\ &+ \alpha_{34\varphi} V_i\, \delta^{abei}_{cdfq}\, \varphi^q \varphi_e \xi^f R_{ab}{}^{cd} + \alpha_{34\xi} V_i\, \delta^{abei}_{cdfq}\, \xi^q \xi_e \varphi^f R_{ab}{}^{cd} - \beta_{81} V_a \varphi_b R^{ab} + \beta_{83} \xi_a \xi_b \varphi_c V_i R^{cabi} + \beta_{84} \varphi_a \xi_b \varphi_c V_i R^{cbai} + \\ &- \beta_{86} V_i {}^*\omega^{ia} \varphi^b \xi^c \varphi^d R_{adbc} - \beta_{91} V_a \xi_b R^{ab} + \beta_{92} \varphi_a \varphi_b \xi_c V_i R^{cabi} + \beta_{94} \xi_a \xi_b \varphi_c V_i R^{acbi} - \beta_{96} V_i {}^*\omega^{ia} \varphi^b \xi^c \xi^d R_{adbc}\end{aligned} \tag{C.4}$$

where ${}^*\omega^{ij} := \tfrac{1}{2} g^{-1/2} \varepsilon^{ijab} \varphi_a \xi_b$. We examine (C.4) for a geometry in which $R^{ab}=0$. After expanding all of the generalized Kronecker deltas in (C.4) we find that (C.4) becomes

$$\begin{aligned}0=\ &4\alpha_{32\xi} R_{1213} + 4\alpha_{33\varphi} R_{2123} + 4\alpha_{34\varphi} R_{2113} + 4\alpha_{34\xi} R_{1223} + \beta_{83} R_{1223} + \beta_{84} R_{1213} + \beta_{92} R_{2113} + \beta_{94} R_{2123} + \\ &- {}^*\omega^{32} R_{2112} - \beta_{86} {}^*\omega^{34} R_{4112} - \beta_{96} {}^*\omega^{31} R_{1212} - \beta_{96} {}^*\omega^{34} R_{4212}\,,\end{aligned} \tag{C.5}$$

where it turns out that ${}^*\omega^{31} = {}^*\omega^{32} = 0$ and ${}^*\omega^{34} = \tfrac{1}{2} g^{-1/2} D$, with $D := XY - Z^2$. Due to (C.1) we know that all of the R.... terms in (C.5) are independent and thus we have obtained the following constraints

$$4\alpha_{32\xi} - 4\alpha_{34\varphi} + \beta_{84} - \beta_{92} = 0,\quad 4\alpha_{34\xi} - 4\alpha_{33\varphi} + \beta_{83} - \beta_{94} = 0, \text{ and } \beta_{86} = \beta_{96} = 0. \tag{C.6}$$

The results presented in (C.6) can now be used back in (C.4), where we shall no longer assume that $R_{ab}= 0$. This produces two more constraints which are given by

$$\beta_{81} = 4[Z(\alpha_{32\xi} - \alpha_{34\varphi}) + Y(\alpha_{34\xi} - \alpha_{33\varphi}) - \alpha_{31\varphi}] \text{ and } \beta_{91} = 4[Z(\alpha_{33\varphi} - \alpha_{34\xi}) + X(\alpha_{34\varphi} - \alpha_{32\xi}) - \alpha_{31\xi}]. \tag{C.7}$$

Thus from the work done so far we have established (2.14), (2.15), (2.20) and (2.21) of **Lemma 5** without much effort. (Note (2.15) follows from(2.14).) Things are now about to take a turn for the worst.

The remaining blocks of terms in $V_iA^{ij}{}_{|j}=0$ to be examined are the six blocks invoving R....φ.., R....ξ.., $(φ..)^3$, $(φ..)^2$ξ.., φ..$(ξ..)^2$ and $(ξ..)^3$ . We really only need to perform detailed analysis of the 1st, 3rd, and 4th blocks in that list due to invariance of various blocks under the φ↔ξ transformation I mention in several places in the paper. To assist in analyzing the first block one will have to differentiate with respect to $φ_{rs}$ to obtain a two index tensorial concomitant. This concomitant can be evaluated for geometries with $R_{ab}=0$, then taking various contractions after expanding all of the generalized Kronecker deltas. To assist in this analysis you need to know the form of the dependent components of R.... listed in (C.2) in terms of the independent components of R.... given in (C.1). These are given by

$R_{1212}= -DX^{-1}[ε_V R_{1313}+ε_W R_{1414}]$; $R_{1234}=2R_{1324}-YZ^{-1}R_{1314}-XZ^{-1}R_{2324}$; (C.8a)

$R_{1323}=-½D(XZ)^{-1}ε_Vε_W R_{1414}+½XZ^{-1}R_{2323}+ ½ZX^{-1}R_{1313}$ ; $R_{1334}=ε_V ZD^{-1}R_{1214} -ε_V XD^{-1}R_{1224}$; (C.8b)

$R_{1424}=½(XY+Z^2)/(XZ)R_{1414}-½ε_Vε_W XZ^{-1}R_{2323}+½ε_Vε_W ZX^{-1}R_{1313}$; (C.8c)

$R_{1434}=ε_W XD^{-1}R_{1223}-ε_W ZD^{-1}R_{1213}$; $R_{2334}=ε_V YD^{-1}R_{1214}-ε_V ZD^{-1}R_{1224}$; (C.8d)

$R_{2434}=ε_W ZD^{-1}R_{1223}-ε_W YD^{-1}R_{1213}$; $R_{3434}=-ε_V X^{-1}R_{1313}-ε_W X^{-1}R_{1414}$; (C.8e)

$R_{2424}=ε_W YX^{-1}[ε_V R_{1313}+ε_W R_{1414}]-ε_Vε_W R_{2323}$ . (C.8f)

After an incredible amount of effort you should be able to obtain most of (2.16)-(2.19). To do this you shall need to incorporate the results that you get from the φ↔ξ transformation of the results that you did obtain, which would correspond to what you would have got from analyzing the R....ξ.. term in $V_iA^{ij}{}_{|j}=0$. But you won't be able to show that $α_{31}$, $α_{32}$, $α_{33}$ and $α_{34}$ are BST functions when $α_1=α_2=0$. To obtain that result you need to look at the third degree terms in $V_iA^{ij}{}_{|j}=0$. The first such term comes from the terms $(\varphi..)^3$, and these yield nothing new. The next block to analyze is the one involving the $(\varphi..)^2\xi..$ terms. Unfortunately to do this you will have to differentiate the $(\varphi..)^2\xi..$ block with respect to $\varphi_{lm}$, $\varphi_{pq}$ and $\xi_{rs}$, and then expand all of the generalized Kronecker deltas in that expression. By looking at various contractions of the resulting 6 index contravariant tensor you have obtained with elements of the tetrad $\{\varphi^a,\xi^a,V^a,W^a\}$ you will arrive at the remaining conditions of **Lemma 5.** It is a miserable calculation, but necessary.

**Appendix D: Relating the Lagrangian $L_*$ to those of the Theorem**

In **Section 3** I pointed out that when K is a BST function the Lagrangian $L_*$ presented below

$$L_* := g^{1/2}\{K\,\delta^{abcd}_{pqrs}\,\varphi_a\xi_b\varphi^p\xi^q R_{cd}{}^{rs} - 4K_X\,\delta^{abcd}_{pqrs}\,\varphi_a\xi_b\varphi^p\xi^q\varphi_c{}^r\varphi_d{}^s - 4K_Y\,\delta^{abcd}_{pqrs}\,\varphi_a\xi_b\varphi^p\xi^q\xi_c{}^r\xi_d{}^s +$$
$$-4\,K_Z\,\delta^{abcd}_{pqrs}\,\varphi_a\xi_b\varphi^p\xi^q\varphi_c{}^r\xi_d{}^s\}\,, \qquad \text{(D.1)}$$

yields second-order bi-scalar-tensor field equations in a space of four-dimensions. This was originally pointed out in Akama and Kobayashi [25]. Evidently $L_*$ is not one of the Lagrangians presented in the statement of the **Theorem**. What I shall do here is show how it can be represented in terms of Lagrangians that appear in the **Theorem**. This will provide us with a nice illustration of how the proof of the **Theorem** can be used in practice.

To begin, from (2.30), (2.32) and (2.22b) we can tell that $E^{hi}(L_*)$ will be devoid of an $A_3{}^{ij}$ term. Another lengthy calculation, like many of those performed in the proof of the **Theorem** to compute Euler-Lagrange tensor densities, shows that in the $A_2{}^{ij}$ part of $E^{ij}(L_*)$ (*c.f.* (2.22c))

$$\alpha_{31} = 0;\ \alpha_{32} = -[\tfrac{1}{2}YK + Z^2K_X + Y^2K_Y + YZK_Z];\ \alpha_{33} = -[\tfrac{1}{2}XK + X^2K_X + Z^2K_Y + XZK_Z] \qquad \text{(D.2a)}$$

$$\alpha_{34} = \tfrac{1}{2}ZK + XZK_X + YZK_Y + \tfrac{1}{2}(XY + Z^2)K_Z\,. \qquad \text{(D.2b)}$$

It is easily seen that $\alpha_{31}$, $\alpha_{32}$, $\alpha_{33}$ and $\alpha_{34}$ satisfy the conditions of **Lemma 5**, as they must, but it is a nice check on the accuracy of the computation of them.

What we wish to do next is use these $\alpha_{3\cdot}$'s to build a Lagrangian $\mathscr{L}_2$ of the form $\mathscr{L}_2 = L_{21}+L_{22}+L_{23}+L_{24}$, which is such that $E^{ij}(\mathscr{L}_2)$ will have the same $A_2^{ij}$ part that $L_*$ has. The proof of the **Theorem** tells us precisely how to do this. To that end we need to compute the coefficient functions $S_{21}$, $S_{22}$, $S_{23}$ and $S_{24}$ that appear in (2.40). From (2.49) we know that

$$S_{21} = [\alpha_{31}+X\alpha_{32}+Y\alpha_{33}+2Z\alpha_{34}] - \tfrac{1}{2}XS_{22} - \tfrac{1}{2}YS_{23} - ZS_{24} + (XY-Z^2)[S_{22Y}+S_{23X}-S_{24Z}]\ .$$

Using (D.2) this becomes, upon setting $D := XY - Z^2$,

$$S_{21} = -D[K+XK_X+YK_Y+ZK_Z] - \tfrac{1}{2}XS_{22} - \tfrac{1}{2}YS_{23} - ZS_{24} + D[S_{22Y}+S_{23X}-S_{24Z}]\ , \qquad \text{(D.3)}$$

where due to (2.53)

$$S_{22Y} + S_{23X} - S_{24Z} = \int \sigma(t)dt\Big|_{t=1} \qquad \text{(D.4)}$$

with (2.56b) telling us that

$$\sigma = -[\alpha_{32Y}+\alpha_{33X}-\alpha_{34Z}]\ . \qquad \text{(D.5)}$$

Since K is a BST function we can use (D.2) and (D.5) to deduce that

$$\sigma = {}^{3}/_{2}K + {}^{7}/_{2}[XK_X + YK_Y + ZK_Z] + X^2K_{XX} + Y^2K_{YY} + Z^2K_{ZZ} + 2XYK_{XY} + 2YZK_{YZ} + 2XZK_{XZ}\ .$$

Using this expression for $\sigma$ in (D.4) we discover that

$$S_{22Y} + S_{23X} - S_{24Z} = \int \{ {}^{3}/_{2}\frac{d}{dt}(tK(t)) + \frac{d}{dt}(t^2XK_X(t) + t^2YK_Y(t) + t^2ZK_Z(t))\} dt\Big|_{t=1}$$

and hence

$$S_{22Y} + S_{23X} - S_{24Z} = {}^{3}/_{2}K + XK_X + YK_Y + ZK_Z\ . \qquad \text{(D.6)}$$

Combining (D.3) and (D.6) tells us that

$$S_{21} = \tfrac{1}{2}[KD - XS_{22} - YS_{23} - 2ZS_{24}]\ . \qquad \text{(D.7)}$$

Now we know that $S_{21}$ must be a BST function, and it turns out that (D.7) implies that is so due to (D.6). So we are consistent!

In summary, the Lagrangians $L_*$ and $L_2$ will yield the same $A_2^{ij}$ terms when the coefficient functions $S_{21}$, $S_{22}$, $S_{23}$ and $S_{24}$ in the Lagrangian $L_2$ presented in (2.40) are chosen in accordance with (D.6) and (D.7). We shall now take a closer look at the connection between $L_*$ and $L_2$ in view of his new information.

To that end we need to first expand the generalized Kronecker deltas in $L_{\ast}$. Upon doing so we find that

$$L_{\ast} = g^{1/2}\{2KDR - 4XK\xi_a\xi_b R^{ab} - 4YK\varphi_a\varphi_b R^{ab} + 8ZK\varphi_a\xi_b R^{ab} + 4K\varphi_a\xi_b\varphi_c\xi_d R^{abcd} +$$

$$-4K_X[D((\Box\varphi)^2 - \varphi^{ab}\varphi_{ab}) + 2\Box\varphi(2Z\varphi^a\xi^b\varphi_{ab} - X\xi^a\xi^b\varphi_{ab} - Y\varphi^a\varphi^b\varphi_{ab}) + 2X\xi^a\xi^b\varphi_{ac}\varphi_b{}^c + 2Y\varphi^a\varphi^b\varphi_{ac}\varphi_b{}^c +$$

$$-4Z\varphi^a\xi^b\varphi_{ac}\varphi_b{}^c + 2\varphi^a\varphi^b\varphi_{ab}\xi^c\xi^d\varphi_{cd} - 2\varphi^a\xi^b\varphi_{ab}\varphi^c\xi^d\varphi_{cd}\,] +$$

$$-4K_Y[D((\Box\xi)^2 - \xi^{ab}\xi_{ab}) + 2\Box\xi(2Z\varphi^a\xi^b\xi_{ab} - X\xi^a\xi^b\xi_{ab} - Y\varphi^a\varphi^b\xi_{ab}) + 2X\xi^a\xi^b\,\xi_{ac}\xi_b{}^c + 2Y\varphi^a\varphi^b\xi_{ac}\xi_b{}^c +$$

$$-4Z\varphi^a\xi^b\xi_{ac}\xi_b{}^c + 2\varphi^a\varphi^b\xi_{ab}\xi^c\xi^d\xi_{cd} - 2\varphi^a\xi^b\xi_{ab}\varphi^c\xi^d\xi_{cd}\,] +$$

$$-4K_Z[D(\Box\varphi\;\Box\xi - \varphi^{ab}\xi_{ab}) + \Box\xi(2Z\varphi^a\xi^b\varphi_{ab} - X\xi^a\xi^b\varphi_{ab} - Y\varphi^a\varphi^b\varphi_{ab}) + \Box\varphi(2Z\varphi^a\xi^b\xi_{ab} - X\xi^a\xi^b\xi_{ab} - Y\varphi^a\varphi^b\xi_{ab}) +$$

$$+2X\xi^a\xi^b\varphi_{ac}\xi_b{}^c + 2Y\varphi^a\varphi^b\varphi_{ac}\xi_b{}^c - 2Z(\varphi^a\xi^b\varphi_{ac}\xi_b{}^c + \varphi^a\xi^b\xi_{ac}\varphi_b{}^c) + \varphi^a\varphi^b\varphi_{ab}\xi^c\xi^d\xi_{cd} + \xi^a\xi^b\varphi_{ab}\varphi^c\varphi^d\xi_{cd} +$$

$$-2\varphi^a\xi^b\varphi_{ab}\varphi^c\xi^d\xi_{cd}\,]\,\}. \tag{D.9}$$

For the Lagrangian $L_2$ of (2.40) we have

$$L_2 = g^{1/2}\{[4S_{21} + 2XS_{22} + 2YS_{23} + 4ZS_{24}]R - 4S_{22}\varphi_a\varphi_b R^{ab} - 4S_{23}\xi_a\xi_b R^{ab} - 8S_{24}\varphi_a\xi_b R^{ab} +$$

$$-[(\Box\varphi)^2 - \varphi^{ab}\varphi_{ab}](8S_{21X} + 4XS_{22X} + 4YS_{23X} + 8ZS_{24X}) - [(\Box\xi)^2 - \xi^{ab}\xi_{ab}](8S_{21Y} + 4XS_{22Y} + 4YS_{23Y} + 8ZS_{24Y}) +$$

$$-[\Box\varphi\Box\xi - \varphi^{ab}\xi_{ab}](8S_{21Z} + 4XS_{22Z} + 4YS_{23Z} + 8ZS_{24Z}) + \Box\varphi(8S_{22X}\varphi^a\varphi^b\varphi_{ab} + 8S_{23X}\xi^a\xi^b\varphi_{ab} + 16S_{24X}\varphi^a\xi^b\varphi_{ab}) +$$

$$+\Box\varphi(4S_{22Z}\varphi^a\varphi^b\xi_{ab} + 4S_{23Z}\xi^a\xi^b\xi_{ab} + 8S_{24Z}\varphi^a\xi^b\xi_{ab}) + \Box\xi(8S_{22Y}\varphi^a\varphi^b\xi_{ab} + 8S_{23Y}\xi^a\xi^b\xi_{ab} + 16S_{24Y}\varphi^a\xi^b\xi_{ab}) +$$

$$+\Box\xi(4S_{22Z}\varphi^a\varphi^b\varphi_{ab} + 4S_{23Z}\xi^a\xi^b\varphi_{ab} + 8S_{24Z}\varphi^a\xi^b\varphi_{ab}) - 8S_{22X}\varphi^a\varphi^b\varphi_{ac}\varphi_b{}^c - 8S_{23X}\xi^a\xi^b\varphi_{ac}\varphi_b{}^c - 16S_{24X}\varphi^a\xi^b\varphi_{ac}\varphi_b{}^c +$$

$$-8S_{22Y}\varphi^a\varphi^b\xi_{ac}\xi_b{}^c - 8S_{23Y}\xi^a\xi^b\xi_{ac}\xi_b{}^c - 16S_{24Y}\varphi^a\xi^b\xi_{ac}\xi_b{}^c - 8S_{22Z}\varphi^a\varphi^b\varphi_{ac}\xi_b{}^c - 8S_{23Z}\xi^a\xi^b\varphi_{ac}\xi_b{}^c +$$

$$-8S_{24Z}\varphi^a\xi^b\varphi_{ac}\xi_b{}^c - 8S_{24Z}\varphi^a\xi^b\xi_{ac}\varphi_b{}^c\}\,. \tag{D.10}$$

I shall let $L_{2\ast}$ denote what $L_2$ becomes when evaluated for the coefficient functions $S_{21}$, $S_{22}$, $S_{23}$ and $S_{24}$ which satisfy (D.6) and (D.7). We wish to determine what $L_{\ast} - L_{2\ast}$ is equal to. To that end we need to examine some of the implications of (D.6) and (D.7). We begin by rewriting (D.6) as

$$DK = 2S_{21}+XS_{22}+YS_{23}+2ZS_{24}. \tag{D.11}$$

Upon differentiation with respect to X, Y and Z this equation gives us

$$DK_X+YK = 2S_{21X}+XS_{22X}+YS_{23X}+2ZS_{24X}+S_{22} \tag{D.12a}$$

$$DK_Y+XK = 2S_{21Y}+XS_{22Y}+YS_{23Y}+2ZS_{24Y}+S_{23} \tag{D.12b}$$

$$DK_Z-2ZK = 2S_{21Z}+XS_{22Z}+YS_{23Z}+2ZS_{24Z}+2S_{24} \tag{D.12c}$$

If we multiply (D.12a) by $4\,[(\Box\varphi)^2-\varphi^{ab}\varphi_{ab}]$, (D12b) by $4\,[(\Box\xi)^2-\xi^{ab}\xi_{ab}]$ and (D.12c) by $4\,[(\Box\varphi\Box\xi+\varphi^{ab}\xi_{ab}]$, then we see that when $L_{2*}$ is subtracted from $L_{*}$ we shall eliminate numerous terms involving $[(\Box\varphi)^2-\varphi^{ab}\varphi_{ab}]$, $[(\Box\xi)^2-\xi^{ab}\xi_{ab}]$ and $[(\Box\varphi\Box\xi-\varphi^{ab}\xi_{ab}]$. But we shall also introduce the terms

$$4YK\,[(\Box\varphi)^2-\varphi^{ab}\varphi_{ab}],\ 4XK\,[(\Box\xi)^2-\xi^{ab}\xi_{ab}] \text{ and } -8ZK\,[(\Box\varphi\Box\xi-\varphi^{ab}\xi_{ab}].$$

These terms can be dealt with by noticing that

$$YK\,[(\Box\varphi)^2-\varphi^{ab}\varphi_{ab}] = (YK\,[\Box\varphi\,\varphi^a-\varphi_b\varphi^{ab}])_{|a} + YK\varphi^a\varphi^b R_{ab} - 2K\Box\varphi\varphi^a\xi^b\xi_{ab} + 2K\varphi^a\xi^b\varphi_{ac}\xi_b{}^c - XYK_\varphi\Box\varphi +$$
$$+YK_\varphi\,\varphi^a\varphi^b\varphi_{ab} - YZK_\xi\Box\varphi + YK_\xi\varphi^a\xi^b\varphi_{ab} - 2YK_X\varphi^c\varphi_{ca}\,[\Box\varphi\,\varphi^a-\varphi_b\varphi^{ab}] - 2YK_Y\xi^c\xi_{ca}\,[\Box\varphi\,\varphi^a-\varphi_b\varphi^{ab}] +$$
$$-YK_Z\varphi^c\xi_{ca}\,[\Box\varphi\,\varphi^a-\varphi_b\varphi^{ab}] - YK_Z\xi^c\varphi_{ca}\,[\Box\varphi\,\varphi^a-\varphi_b\varphi^{ab}]\,, \tag{D.13}$$

along with similar expressions for $XK\,[(\Box\xi)^2-\xi^{ab}\xi_{ab}]$, $ZK\,[\Box\varphi\Box\xi-\varphi^{ab}\xi_{ab}]$, $S_{22}[(\Box\varphi)^2-\varphi^{ab}\varphi_{ab}]$, $S_{23}[(\Box\xi)^2-\xi^{ab}\xi_{ab}]$ and $S_{24}[\Box\varphi\Box\xi-\varphi^{ab}\xi_{ab}]$.

Now we notice that $L_{*}$ has the term $K\varphi^a\xi^b\varphi^c\xi^d R_{abcd}$ in it while $L_{2*}$ does not have such a term. This can be addressed using the fact that

$$K\varphi^a\xi^b\varphi^c\xi^d R_{abcd} = K\varphi^a\varphi^c\xi^d(\xi_{adc}-\xi_{acd}) \text{ and } K\varphi^a\xi^b\varphi^c\xi^d R_{abcd} = K\xi^b\varphi^c\xi^d(\varphi_{bcd}-\varphi_{bdc})$$

and so

$$K\varphi^a\xi^b\varphi^c\xi^d R_{abcd} = \tfrac{1}{2}K\varphi^a\varphi^c\xi^d(\xi_{adc}-\xi_{acd}) + \tfrac{1}{2}K\xi^b\varphi^c\xi^d(\varphi_{bcd}-\varphi_{bdc}) =$$
$$= [\tfrac{1}{2}K\varphi^a\varphi^c\xi^d\xi_{ad}]_{|c} - [\tfrac{1}{2}K\varphi^a\varphi^c\xi^d]_{|c}\,\xi_{ad} - [\tfrac{1}{2}K\varphi^a\varphi^c\xi^d\xi_{ac}]_{|d} + [\tfrac{1}{2}K\varphi^a\varphi^c\xi^d]_{|d}\,\xi_{ac} +$$
$$+ [\tfrac{1}{2}K\xi^b\varphi^c\xi^d\varphi_{bc}]_{|d} - [\tfrac{1}{2}K\xi^b\varphi^c\xi^d]_{|d}\,\varphi_{bc} - [\tfrac{1}{2}K\xi^b\varphi^c\xi^d\varphi_{bd}]_{|c} + [\tfrac{1}{2}K\xi^b\varphi^c\xi^d]_{|c}\,\varphi_{bd}\,. \tag{D.14}$$

I chose to represent $K\varphi^a\xi^b\varphi^c\xi^d R_{abcd}$ in the manner I did to preserve the invariance of $L_{*}-L_{2*}$ under the interchange of $\varphi$ and $\xi$. If we had selected to use only one of the two possible representations of $K\varphi^a\xi^b\varphi^c\xi^d R_{abcd}$ we would have broken the $\varphi\leftrightarrow\xi$ symmetry and thereby lost an important method of testing the accuracy of our calculations. I also used this technique to handle terms like $[\Box\varphi\Box\xi-\varphi^{ab}\xi_{ab}]$ when decomposing it into divergences.

Employing (D.11)-(D.14) we discover, after some rearrangment that

$$\begin{aligned}
L_{*}-L_{2*}=g^{1/2}\{&[(8S_{22Y}-4S_{24Z}-6K-8YK_Y-4ZK_Z)\Box\varphi\varphi^a\xi^b\xi_{ab}+(8S_{23X}-4S_{24Z}-6K-8XK_X-4ZK_Z)\Box\xi\varphi^a\xi^b\varphi_{ab}]+\\
&+[(4S_{22Z}-8S_{24X}-4YK_Z-8ZK_X)\Box\varphi\varphi^a\xi^b\varphi_{ab}+(4S_{23Z}-8S_{24Y}-4XK_Z-8ZK_Y)\Box\xi\varphi^a\xi^b\xi_{ab}]+\\
&+[(4S_{24Z}-8S_{23X}+6K+8XK_X+4ZK_Z)\Box\varphi\xi^a\xi^b\varphi_{ab}+(4S_{24Z}-8S_{22Y}+6K+8YK_Y+4ZK_Z)\Box\xi\varphi^a\varphi^b\xi_{ab}]+\\
&+[(8S_{24Y}-4S_{23Z}+8ZK_Y+4XK_Z)\Box\varphi\xi^a\xi^b\xi_{ab}+(8S_{24X}-4S_{22Z}+8ZK_X+4YK_Z)\Box\xi\varphi^a\varphi^b\varphi_{ab}]+\\
&-4K_Z\varphi^a\varphi^b\varphi_{ab}\xi^c\xi^d\xi_{cd}+6K_Z\varphi^a\xi^b\varphi_{ab}\varphi^c\xi^d\xi_{cd}-4K_Z\xi^a\xi^b\varphi_{ab}\varphi^c\varphi^d\xi_{cd}+\\
&+[(8S_{24X}-4S_{22Z}+8ZK_X+4YK_Z)\varphi^a\xi^b\varphi_{ac}\varphi_b{}^c+(8S_{24Y}-4S_{23Z}+8ZK_Y+4XK_Z)\varphi^a\xi^b\xi_{ac}\xi_b{}^c]+\\
&+[(8S_{23X}-4S_{24Z}-6K-8XK_X-4ZK_Z)\xi^a\xi^b\varphi_{ac}\varphi_b{}^c+(8S_{22Y}-4S_{24Z}-6K-8YK_Y-4ZK_Z)\varphi^a\varphi^b\xi_{ac}\xi_b{}^c]+\\
&+[(4S_{22Z}-8S_{24X}-8ZK_X-4YK_z)\varphi^a\varphi^b\varphi_{ac}\xi_b{}^c+(4S_{23Z}-8S_{24Y}-8ZK_Y-4XK_z)\xi^a\xi^b\varphi_{ac}\xi_b{}^c]+\\
&-[4K_X\varphi^a\varphi^b\varphi_{ab}\varphi^c\xi^d\xi_{cd}+4K_Y\varphi^a\xi^b\varphi_{ab}\xi^c\xi^d\xi_{cd}]+[4K_Y\varphi^a\xi^b\xi_{ab}\varphi^c\xi^d\xi_{cd}+4K_X\varphi^a\xi^b\varphi_{ab}\varphi^c\xi^d\varphi_{cd}]+\\
&+[4K_X\varphi^a\xi^b\varphi_{ab}\varphi^c\varphi^d\xi_{cd}+4K_Y\xi^a\xi^b\varphi_{ab}\varphi^c\xi^d\xi_{cd}]+4K_Z\xi^a\xi^b\varphi_{ab}\varphi^c\varphi^d\xi_{cd}-2K_Z\varphi^a\xi^b\varphi_{ab}\varphi^c\xi^d\xi_{cd}+\\
&-[4K_X\varphi^a\varphi^b\varphi_{ab}\xi^c\xi^d\varphi_{cd}+4K_Y\varphi^a\varphi^b\xi_{ab}\xi^c\xi^d\xi_{cd}]\}+g^{1/2}\{4XS_{22\varphi}\Box\varphi+4ZS_{22\xi}\Box\varphi+4YK_\varphi\varphi^a\varphi^b\varphi_{ab}+\\
&+4YK_\xi\varphi^a\xi^b\varphi_{ab}+4ZS_{23\varphi}\Box\xi+4YS_{23\xi}\Box\xi-4DK_\xi\Box\xi-4S_{23\varphi}\varphi^a\xi^b\xi_{ab}-4S_{23\xi}\xi^a\xi^b\xi_{ab}+4XK_\varphi\varphi^a\xi^b\xi_{ab}+4XK_\xi\xi^a\xi^b\xi_{ab}\\
&-4DK_\varphi\Box\varphi-4S_{22\varphi}\varphi^a\varphi^b\varphi_{ab}-4S_{22\xi}\varphi^a\xi^b\varphi_{ab}+4ZS_{24\varphi}\Box\varphi+4YS_{24\xi}\Box\varphi-4ZK_\varphi\varphi^a\xi^b\varphi_{ab}-4ZK_\xi\xi^a\xi^b\xi_{ab}-4S_{24\varphi}\varphi^a\xi^b\varphi_{ab}\\
&-4S_{24\xi}\xi^a\xi^b\varphi_{ab}+4XS_{24\varphi}\Box\xi+4ZS_{24\xi}\Box\xi-4ZK_\varphi\varphi^a\varphi^b\xi_{ab}-4ZK_\xi\varphi^a\xi^b\xi_{ab}-4S_{24\varphi}\varphi^a\varphi^b\xi_{ab}-4S_{24\xi}\varphi^a\xi^b\xi_{ab}-2XK_\varphi\varphi^a\xi^b\xi_{ab}\\
&-2ZK_\xi\varphi^a\xi^b\xi_{ab}+2ZK_\varphi\varphi^a\varphi^b\xi_{ab}+2YK_\xi\varphi^a\varphi^b\xi_{ab}-2ZK_\varphi\varphi^a\xi^b\varphi_{ab}-2YK_\xi\varphi^a\xi^b\varphi_{ab}+2XK_\varphi\xi^a\xi^b\varphi_{ab}+2ZK_\xi\xi^a\xi^b\varphi_{ab}\}+\\
&+g^{1/2}\{[4(YK-S_{22})(\Box\varphi\varphi^a-\varphi_b\varphi^{ba})]_{|a}+[4(XK-S_{23})(\Box\xi\xi^a-\xi_b\xi^{ba})]_{|a}-[4(ZK+S_{24})(\Box\varphi\xi^a-\xi_b\varphi^{ba})]_{|a}+\\
&-[4(ZK+S_{24})(\Box\xi\varphi^a-\varphi_b\xi^{ba})]_{|a}+[2K\varphi^a\varphi^c\xi^d\xi_{ad}]_{|c}-[2K\varphi^a\varphi^c\xi^d\xi_{ac}]_{|d}+[2K\xi^b\varphi^c\xi^d\varphi_{bc}]_{|d}-[2K\xi^b\varphi^c\xi^d\varphi_{bd}]_{|c}\}. \quad (D.15)
\end{aligned}$$

You should note that the first term in curly brackets on the right-hand side of (D.15) is of second degree in the scalar fields. I originally hoped that this term would not be present, so that $L_{*} - L_{2*}$ would have the form of $\mathscr{L}_0+\mathscr{L}_1$, and we would have shown how $L_{*}$ could be expressed in terms of Lagrangians of the form $\mathscr{L}_0+\mathscr{L}_1+\mathscr{L}_2$. Unfortunately, things are never simple in bi-scalar-tensor field theory. What we can say though is that since $L_{*}$ and $L_{2*}$ yield the same $A_2^{ij}$ term, the variational derivative of the right-hand side of (D.15) with respect to $g_{ij}$ must have the form of $A_1^{ij}+A_0^{ij}$. Thus we can employ the methods used to establish **Lemma 12** and to finish the proof of the **Theorem** to show how to build Lagrangians of the $\mathscr{L}_1$ and $\mathscr{L}_0$ type that are equivalent to the Lagrangian on the right-hand side of (D.15). I leave that daunting task to the interested reader.

As an example of how (D.15) can be used let us suppose that K is a constant. In this case (D.9) tells us that $L_{*}$ becomes $L_{*}'$ which is given by

$$L_{*}' := g^{1/2}\{2KDR - 4XK\xi_a\xi_b R^{ab} - 4YK\varphi_a\varphi_b R^{ab} + 8ZK\varphi_a\xi_b R^{ab} + 4K\varphi_a\xi_b\varphi_c\xi_d R^{abcd}\}. \tag{D.16}$$

For this situation we can take as a solution to (D.6) and (D.7)

$$S_{22} = 0,\ S_{23} = 0,\ S_{24} = -{}^3/_2 KZ \ \text{ and } S_{21} = \tfrac{1}{2}(XY+2Z^2)K\,. \tag{D.17}$$

Using (D.17) in (D.15) we see that if we let $L_{2*}'$ denote what $L_{2*}$ becomes when the $S_{..}$'s are chosen as above then

$$\begin{aligned} L_{*}' - L_{2*}' = g^{1/2}\{&[4YK(\Box\varphi\varphi^a - \varphi_b\varphi^{ba})]_{|a} + [4XK(\Box\xi\xi^a - \xi_b\xi^{ba})]_{|a} + [2ZK(\Box\varphi\xi^a - \xi_b\varphi^{ba})]_{|a} + \\ &+[2ZK(\Box\xi\varphi^a - \varphi_b\xi^{ba})]_{|a} + [2K\varphi^a\varphi^c\xi^d\xi_{ad}]_{|c} - [2K\varphi^a\varphi^c\xi^d\xi_{ac}]_{|d} + [2K\xi^b\varphi^c\xi^d\varphi_{bc}]_{|d} - [2K\xi^b\varphi^c\xi^d\varphi_{bd}]_{|c}\}. \end{aligned} \tag{D.18}$$

Now using (D.10) we find that

$$\begin{aligned} L_{2*}' = g^{1/2}\{&2DKR + 12ZK\varphi^a\varphi^b R_{ab} - 4YK[(\Box\varphi)^2 - \varphi^{ab}\varphi_{ab}] - 4XK[(\Box\xi)^2 - \xi^{ab}\xi_{ab}] - 4ZK[\Box\varphi\,\Box\xi - \varphi^{ab}\xi_{ab}] + \\ &- 12K\,\Box\varphi\varphi^a\xi^b\xi_{ab} - 12K\,\Box\xi\varphi^a\xi^b\varphi_{ab} + 12K\varphi^a\xi^b\varphi_{ac}\xi_b{}^c + 12K\varphi^a\xi^b\xi_{ac}\varphi_b{}^c\}\,. \end{aligned} \tag{D.19}$$

So we can combine (D.18) and (D.19) to deduce that that the $L_{*}'$ of (D.16) can be expressed as

$$\begin{aligned} L_{*}' = g^{1/2}\{&2DKR + 12ZK\varphi^a\varphi^b R_{ab} - 4YK[(\Box\varphi)^2 - \varphi^{ab}\varphi_{ab}] - 4XK[(\Box\xi)^2 - \xi^{ab}\xi_{ab}] - 4ZK[\Box\varphi\,\Box\xi - \varphi^{ab}\xi_{ab}] + \\ &- 12K\,\Box\varphi\varphi^a\xi^b\xi_{ab} - 12K\,\Box\xi\varphi^a\xi^b\varphi_{ab} + 12K\varphi^a\xi^b\varphi_{ac}\xi_b{}^c + 12K\varphi^a\xi^b\xi_{ac}\varphi_b{}^c\} + \end{aligned}$$

$$+\ g^{1/2}\{[4YK(\Box\varphi\varphi^{a}-\varphi_{b}\varphi^{ba})]_{|a}+[4XK(\Box\xi\xi^{a}-\xi_{b}\xi^{ba})]_{|a}+[2ZK(\Box\varphi\xi^{a}-\xi_{b}\varphi^{ba})]_{|a}+$$

$$+[2ZK(\Box\xi\varphi^{a}-\varphi_{b}\xi^{ba})]_{|a}+[2K\varphi^{a}\varphi^{c}\xi^{d}\xi_{ad}]_{|c}-[2K\varphi^{a}\varphi^{c}\xi^{d}\xi_{ac}]_{|d}+[2K\xi^{b}\varphi^{c}\xi^{d}\varphi_{bc}]_{|d}-[2K\xi^{b}\varphi^{c}\xi^{d}\varphi_{bd}]_{|c}\}\,. \quad (D.20)$$

If you expand all of the divergences in (D.20) you will find that the right-hand side of (D.20) does in fact reduce to the $L_{*}'$ of (D.16). This provides us with a direct proof that $L_{*}'$ is indeed equal to a Lagrangian in the $\mathscr{L}_2$ class up to a divergence.

In terms of generalized Kronecker deltas the Lagrangian $L_{*}'$ can be expressed as

$$L_{*}' = g^{1/2}K\,\delta^{abcd}_{pqrs}\,\omega_{ab}\omega^{pq}R_{cd}{}^{rs} \quad (D.21)$$

where $\omega_{ab}:=\frac{1}{2}(\varphi_a\xi_b-\varphi_b\xi_a)$. This Lagrangian bears a remarkable resemblance to another Lagrangian that occurs in vector-tensor field theory. In Horndeski [26] I show that in a space of four-dimensions the most general second-order vector-tensor field equations which are compatible with the conservation of charge and for which the vector equation reduces to Maxwell's equations in a flat space can be obtained from the usual Einstein-Maxwell Lagrangian with cosmological term, plus

$$L_{+} := g^{1/2}\tau\,\delta^{abcd}_{pqrs}\,F_{ab}F^{pq}R_{cd}{}^{rs} \quad (D.22)$$

where $\tau$ is a constant and $F_{ab}$ is the electromagnetic field tensor defined by $F_{ab}:=A_{a,b}-A_{b,a}$ with $A_a$ being the electromagnetic vector potential. The similarity between the two Lagrangians (D.21) and (D.22) is quite striking.

In concluding this appendix I would like to say that although I have shown that the Lagrangian $L_{*}$ can be exprssed in terms of Lagrangians of the form $\mathscr{L}_2+\mathscr{L}_1+\mathscr{L}_0$ appearing in the **Theorem**, in practice it would probably be more useful to use it as it is defined, instead of writing it as a sum of other Lagrangians. This will probably also be the case for other bi-scalar-tensor Lagrangians one may encounter that yield second-order field equations, but do not ostensibly have the form of Lagrangians appearing in the **Theorem** (but with a sufficient amount of effort can be shown to be equivalent to such a Lagrangian) .■